\documentclass[a4paper,10pt]{article}
\usepackage{amsthm}
\usepackage{listings}
\usepackage{verbatim}
\usepackage{amsmath,amssymb}
\usepackage{amsfonts}
\usepackage{graphicx}
\usepackage{pdfpages}
\usepackage{caption}
\usepackage{subcaption}
\usepackage[english]{babel}
\usepackage{hyperref}
\usepackage{float}

\title{\begin{center}
Convergence and completeness for square-well Stark resonant state expansions \\
\end{center}}
\author{%
\text{David Juhasz}$^{1}$,\text{ Miroslav Kolesik}$^{2}$,\text{ Per K. Jakobsen}$^{1}$}

\date{}

\begin{document}
\maketitle
\begin{center}
\noindent\begin{minipage}[b]{0.8\hsize}
In this paper we  investigate the completeness of the Stark resonant eigenstates for a particle in a square-well potential. We find that the resonant state expansions for target functions converge inside the potential well and that the existence of this convergence does not depend on the depth of the potential well. By analyzing the asymptotic form of the terms in these expansions we prove some results on the relation between smoothness of target functions and the rate of convergence of the corresponding resonant state expansion.
\end{minipage}
\end{center}

\setcounter{page}{1}
\section[Introduction]{Introduction}

Decaying quantum states were first introduced in 1928 by Gamow\cite{Gamow1}
\cite{Gamow2} and independently by Gurney and Condon\cite{Gurney1}
\cite{Gurney2} in the context of nuclear physics. Decaying quantum states are
often called Gamow states.

This was however not the first time decaying eigenstates were used in physics.
As early as 1884 J. J. Thomson\cite{Thomson} used such states to describe
decay phenomena in electromagnetics. Characterization of the decaying states
using the absence of incoming waves was first introduced by
Siegert\cite{Siegert} in the context of the nuclear scattering matrix. The
Siegert characterization of decaying states was taken up by
Peierls\cite{Peierls},Couteur\cite{Couteur} and Humblet\cite{Humblet} and
developed into a powerful tool for characterizising the nuclear scattering
matrix. The decaying states characterized by Siegert became known as resonant
states and these states and their properties has been under  investigation for many years (Refs \cite{Goto}-\cite{Lind2}). \nocite{Berggren}\nocite{More}\nocite{Calderon}\nocite{Lind1} 

The fact that resonant states decay exponentially in time implies that it is  more likely to find the electron far from the nucleus than closer to it since it is more likely to have been released by an earlier time than a later one. The resonant states are thus not normalizable and therefore can not be described by the usual mathematical formalism of quantum mechanics consisting of Hilbert spaces for the states and Hermitian operators for the observables.  Two different extensions of the mathematical formalism that can include the resonant states has been developed over the years. In the first approach the notion of a rigged Hilbert space has been introduced as a model for the quantum mechanical state space while still modelling observables using Hermitian operators. In the second approach one keeps the Hilbert space as a model of state spaces but model observables using non-Hermitian operators. In particular the Hamiltonian is a non-Hermitian operator. The approach used to create these non-Hermitian energy operators proceed by analytically continuing the space variables in the Schr\o dinger equation onto a contour in the complex plane. This contour is designet in such a way that states satisfying  the Sommerfeld radiation condition at infinity are turned into exponentially decaying states on the complex contour (Refs \cite{Oleg}-\cite{Brody}). \nocite{Moiseyev1}\nocite{Bohm}\nocite{Bender}\nocite{Moiseyev2} In our paper we use this second extension of the mathematical formalism of quantum mechanics.

The purpose of this paper is to investigate the convergence and completeness of expansions based on the resonant eigenstates for a particle in a square well which is also under the influence of an external field. These are the so-called Stark resonant states. This paper thus addresses and fully answer one of the challenges posed in\cite{Brown}, where the convergence for Stark  resonant state expansions was investigated for the case of a zero range Dirac delta potential.

In section 2 we set up the problem by introducing the resonant states of a square-well potential and the positions of energy eigenvalues in the complex plane, these results are known from the research litterature\cite{Agapi}. In section 3 and 4 we investigate to what extent this expansion can be used to represent a wave function or any other function with a compact support. In particular we present two versions of a proof stating that the expansion converges on the negative side and inside of the square-well. This result is consistent with what was obtained for the case of a Dirac delta potential in \cite{Brown}. In section 5 we investigate the asymptotic form of the terms in the resonant state expansion and make precise statements about the rate of convergence and how this rate relate to the smoothness of the function that is being expanded. We also give some high precision  numerical results to verify the correctness of our asymptotic formulas and statements on convergence rates.
In a summary at the end we discuss what we have achieved and use these results to pose a conjecture on what one could expect for the case of a general potential well.

\section[Hamiltonian eigenstates]{Hamiltonian eigenstates}\label{chap:chap3}
\begin{figure}[t]
\begin{center}
\includegraphics[scale=0.3]{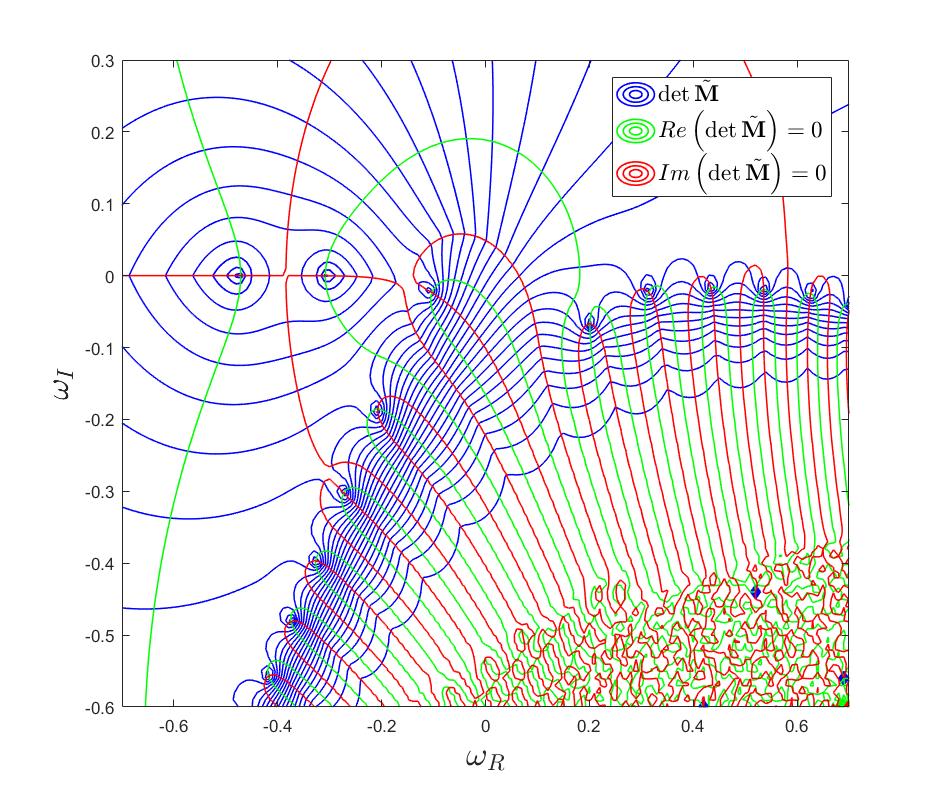}
\end{center}
\caption{Contour plot of $\det\mathbf{M}(\omega_R+i\omega_I)$. The green and red lines are the zero contours of the real and imaginary part of $\det\mathbf{M}(\omega_R+i\omega_I)$, where the parameters were $\varepsilon=0.03,V_0=0.5,d=4$. The blue lines are contours of various values of the modified eigenvalue formula $\det\tilde{\mathbf{M}}(\omega_R+i\omega_I,\rho)=\left(1-\left(1+|\det\mathbf{M}|^{0.3}\right)^{-1}+\rho\right)^{-1}$ for $\rho=0.4$. \label{fig:fig1}}
\end{figure}

Let us consider the following Hamiltonian
\begin{align}
H&=-\frac{1}{2}\partial_{xx}+V(x)-\varepsilon x,\label{eq:eq82}
\end{align}
where $\varepsilon>0$ is the strength of the external field and since we assume that it is positive, the particle is pulled to the right. $V(x)$ is the potential having the form
\begin{equation}
V(x)=\left\{
\begin{array}{l l}
-V_0 &  |x|< d,\\
0 &  |x|>d.
\end{array}\right.\label{eq:eq81}
\end{equation}
The eigenfunctions should satisfy the continuity  and the derivative continuity conditions at $x=-d,d$.
\begin{align}
\psi_\omega(x)=\chi\left\{
\begin{array}{l l}
\frac{2}{\pi^2|\det\mathbf{M}(\omega)|}\mathrm{Ai}\left(\mu\left(x+\frac{\omega}{\varepsilon}\right)\right) & x<-d,\\
\frac{2}{\pi|\det\mathbf{M}(\omega)|}\left[(B_1'A_0-B_1A_0')\mathrm{Ai}\left(\mu\left(x+\frac{\omega+V_0}{\varepsilon}\right)\right) \right. \\
\left.\quad\quad\quad\quad\quad+(A_1A_0'-A_1'A_0)\mathrm{Bi}\left(\mu\left(x+\frac{\omega+V_0}{\varepsilon}\right)\right) \right]& -d< x< d,\\\\
i\left(\frac{\overline{\det\mathbf{M}}(\omega)}{\det\mathbf{M}(\omega)}\right)^{\frac{1}{2}}\mathrm{Ci}^+\left(\mu\left(x+\frac{\omega}{\varepsilon}\right)\right)\\
-i\left(\frac{\det\mathbf{M}(\omega)}{\overline{\det \mathbf{M}}(\omega)}\right)^{\frac{1}{2}}\mathrm{Ci}^-\left(\mu\left(x+\frac{\omega}{\varepsilon}\right)\right)   & d<x,\\
\end{array}\right.\label{eq:eq132}
\end{align}
where $\mu=-(2\varepsilon)^{\frac{1}{3}}$ and $\mathrm{Ci}^{\pm}=\mathrm{Bi}(x)\pm i\mathrm{Ai}(x)$. The normalization constant can be chosen to be $\chi=2^\frac{2}{3}\varepsilon^\frac{1}{6}$ to normalize the integral
\begin{align}
\int_{-\infty}^\infty\psi_\omega(x)\psi_\omega(x')\mathrm{d}\omega=\delta(x-x').\label{eq:eq191}
\end{align}
The determinant appearing in the energy eigenfunction is given as
\begin{align}
\det\mathbf{M}(\omega)=(A_0A'_1-A'_0A_1)(B_2C'_3-B'_2C_3)-(A_0B'_1-A'_0B_1)(A_2C'_3-A'_2C_3)\label{eq:eq109}
\end{align}
together with
{\medmuskip=0mu
\thinmuskip=0mu
\thickmuskip=0mu
\begin{align}
A_0&=\mathrm{Ai}\left(\mu\left(-d+\frac{\omega}{\varepsilon}\right)\right) & A_1=\mathrm{Ai}\left(\mu\left(-d+\frac{\omega+V_0}{\varepsilon}\right)\right) & B_1=\mathrm{Bi}\left(\mu\left(-d+\frac{\omega+V_0}{\varepsilon}\right)\right)\\[\jot]
A_2&=\mathrm{Ai}\left(\mu\left(d+\frac{\omega+V_0}{\varepsilon}\right)\right) & B_2=\mathrm{Bi}\left(\mu\left(d+\frac{\omega+V_0}{\varepsilon}\right)\right) & C_3=\mathrm{Ci}^{+}\left(\mu\left(d+\frac{\omega}{\varepsilon}\right)\right)
\label{eq:eq106}
\end{align}}
The energy eigenvalues $\omega_i$ are obtained by letting $\det\mathbf{M}(\omega_i)=0$. Their positions can be seen on Figure  \ref{fig:fig1} for $\varepsilon=0.03,V_0=0.5$ and $d=4$. It contains different types of contours. Together with the zero-contours is showing also the contours of the modified determinant for better visual results. This kind of modification was first used in \cite{Kolesik}.

Figure \ref{fig:fig1} shows the shifted boundstate eigenvalues lying near to the real axis. There exist two other infnite families of eigenvalues. The family on the right side of the imaginary axis has zeros located along the real axis and they correspond to longer living states, while the "left" family have eigenvalues sitting along the ray $\arg(z)=-\frac{2\pi}{3}$ correspond to fast decaying states. This kind of structure is most likely related to case of shortranged attractive potentials, that can be observed in molecular interactions of gel phases or at percolation phenomena in colloidal systems.

\section{Resonant state expansion I}
This section contains a weaker proof of the completeness of the resonant states (\ref{eq:eq132}). First, we introduce a closed contour on which we evaluate the completness relation. On the inside of the contour we use the residue theorem to evaluate the integral while on the boundary, we use the asymptotic formulas for the resonant states. This will yield conditions where to expect convergence.

\begin{figure}[H]
\begin{center}
\includegraphics[scale=0.3]{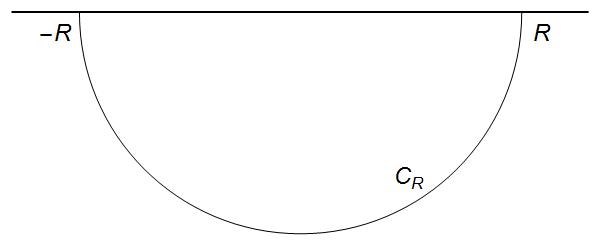}
\end{center}
\caption{Closed contour $\Gamma_R$ on the lower frequency half-plane. \label{fig:fig17}}
\end{figure}

Our staring point is the completeness relation for the scattering states.
\begin{align}
\int_{-\infty}^\infty\psi_\omega(x)\psi_\omega(x')\mathrm{d}\omega=\delta(x-x')\label{eq:eq191.1}
\end{align}
Define a function for each $R$
\begin{align}
\mathcal{F}_R(x,x')=\int_{-R}^R\psi_\omega(x)\psi_\omega(x')\mathrm{d}\omega\label{eq:eq192}
\end{align}
that converges to $\delta(x-x')$ as $R$ goes to infinity. We introduce a closed contour $\Gamma_R$ as on Figure (\ref{fig:fig17}) on the lower complex frequency half-plane.

\noindent
Note, that the curve $\Gamma_R$ we integrate along, must be negatively oriented, that is the interior of the curve is on the right side when travelling along the curve, because we chose $\mathcal{F}_R(x,x')$ to be an integral from $-R$ to $R$. This contour can be divided into two parts and using the Residue theorem, we get
\begin{align}
\int_{\Gamma_R}\psi_\omega(x)\psi_\omega(x')\mathrm{d}\omega&=\int_{C_R}\psi_\omega(x)\psi_\omega(x')\mathrm{d}\omega+\mathcal{F}_R(x,x')\nonumber\\
-2\pi i\sum_{j}\mathrm{Res}(\psi_\omega(x)\psi_\omega(x'),\omega_j)&=\int_{C_R}\psi_\omega(x)\psi_\omega(x')\mathrm{d}\omega+\mathcal{F}_R(x,x')\nonumber\\
\mathcal{F}_R(x,x')&=\frac{2\pi}{i}\sum_{j}\mathrm{Res}(\psi_\omega(x)\psi_\omega(x'),\omega_j)\nonumber\\
&-\int_{C_R}\psi_\omega(x)\psi_\omega(x')\mathrm{d}\omega\label{eq:eq193}
\end{align}
where $\omega_j$ are the poles of the integrand located on the lower half of the complex frequency plane. Observe, that the poles $\omega_j$ confined inside the contour $\Gamma_R$ are the ones for which $\det\mathbf{M}(\omega)$ is zero. The sum over residues in (\ref{eq:eq193}) can be written as
\begin{align}
&\frac{2\pi}{i}\mathrm{Res}(\psi_\omega(x)\psi_\omega(x'),\omega_j)=\psi_j(x)\psi_j(x')\label{eq:eq203}
\end{align}
with
{\medmuskip=0mu
\thinmuskip=0mu
\thickmuskip=0mu
\begin{align}
\psi_j(x)=\chi\left(\frac{2\pi}{i}\lim_{\omega\to\omega_j}\left\{\frac{\omega-\omega_j}{\det\mathbf{M}(\omega)}\right\}\right)^\frac{1}{2}\left\{
\begin{array}{l l}
i\left(\overline{\det\mathbf{M}}_j\right)^\frac{1}{2}\mathrm{Ci}^+_j\left(\mu\left(-d+\frac{\omega}{\varepsilon}\right)\right) & x>d\\
\frac{2}{\pi\left(\overline{\det\mathbf{M}}_j\right)^\frac{1}{2}} & \\
\left[a_j\mathrm{Ai}_j\left(\mu\left(-d+\frac{\omega+V_0}{\varepsilon}\right)\right)\right. & \\
\left.+b_j\mathrm{Bi}_j\left(\mu\left(-d+\frac{\omega+V_0}{\varepsilon}\right)\right)\right] & -d<x<d\\
\frac{2}{\pi^2\left(\overline{\det\mathbf{M}}_j\right)^\frac{1}{2}}\mathrm{Ai}_j\left(\mu\left(-d+\frac{\omega}{\varepsilon}\right)\right) & -d<x
\end{array}\right.\label{eq:eq204}
\end{align}}
where $a_j=(B_1'A_0-B_1A_0')\Big|_{\omega=\omega_j},b_j=(A_1A_0'-A_1'A_0)\Big|_{\omega=\omega_j}$ and \\$\mathrm{Ai}_j\left(\mu\left(-d+\frac{\omega}{\varepsilon}\right)\right)=\mathrm{Ai}\left(\mu\left(-d+\frac{\omega}{\varepsilon}\right)\right)\Big|_{\omega=\omega_j}$ etc. It is easy to verify that the functions $\psi_j(x)$ are proportional to the resonant states (\ref{eq:eq132}). We can thus rewrite (\ref{eq:eq193}) using (\ref{eq:eq204}) in the form
\begin{align}
\delta(x-x')&=\sum_{j}\psi_j(x)\psi_j(x')-\lim_{R\to\infty}\int_{C_R}\psi_\omega(x)\psi_\omega(x')\mathrm{d}\omega\label{eq:eq205}
\end{align}
The completeness of the resonant states depends on the limit of the integral. Let us perform the variable transformation $\omega=Re^{i\theta}$ so we get $\mathrm{d}\omega=Re^{i\theta}i\mathrm{d}\theta$. We use the asymptotic expressions for Airy functions from \cite{Stegun}. For the sector $-\frac{2\pi}{3}<\theta<0$ the asymptotic expression is
{\medmuskip=0mu
\thinmuskip=0mu
\thickmuskip=0mu
\begin{align}
\psi_\omega(x)=\chi\left\{
\begin{array}{l l}
\frac{2(\kappa R)^\frac{1}{4}}{i\pi^\frac{1}{2}\sigma^\frac{1}{2}}e^{-i\beta\varrho R^\frac{1}{2}}e^{\varpi xR^{\frac{1}{2}}} & x<-d\\
\frac{2(\kappa R)^\frac{1}{4}}{i\pi^\frac{1}{2}\sigma^{\frac{1}{2}}}e^{-i(\beta+\sigma)\varrho R^\frac{1}{2}}e^{\varpi R^{\frac{1}{2}}\left(x+\frac{V_0}{\varepsilon}\right)} & -d<x<d\\
-\frac{\sigma^\frac{1}{2}}{2(\kappa R)^\frac{3}{4}\pi^\frac{1}{2}}e^{i\varrho R^\frac{1}{2}(\sigma-3\beta)}e^{-\varpi R^{\frac{1}{2}}\left(x+\frac{V_0}{\varepsilon}\right)} &\\
i\frac{2(\kappa R)^\frac{1}{4}}{\pi^\frac{1}{2}\sigma^\frac{1}{2}}e^{-i\beta\varrho R^\frac{1}{2}}e^{\varpi xR^{\frac{1}{2}}}-i\frac{\sigma^\frac{1}{2}}{2\pi^\frac{1}{2}(\kappa R)^\frac{3}{4}}e^{i\beta\varrho R^\frac{1}{2}}e^{-\varpi xR^{\frac{1}{2}}} & d<x
\end{array}\right.\label{eq:eq205.1}
\end{align}}
where the quantities $\kappa,\varpi,\varrho$ are complex numbers depending on $\theta$. For the sector $-\pi<\theta<-\frac{2\pi}{3}$ the resonant states decay independently of $x$. Now let $f(x)$ be any function with a compact support. We will multiply (\ref{eq:eq205}) with $f(x')$ and integrate the equation with respect to $x'$.
\begin{align}
\int_{-\infty}^{\infty}f(x')&\delta(x-x')\mathrm{d}x'=\sum_{j}\psi_j(x)\int_{-\infty}^{\infty}f(x')\psi_j(x')\mathrm{d}x'\nonumber\\
&+\lim_{R\to\infty}\int_{-\infty}^{\infty}\int_{-\frac{2\pi}{3}}^0f(x')\psi_\omega(x)\psi_\omega(x')Re^{i\theta}i\mathrm{d}\theta\mathrm{d}x'\nonumber\\
f(x)&=\sum_{j}c_j\psi_j(x)+\lim_{R\to\infty}\int_{-\frac{2\pi}{3}}^0F(R,\theta)\psi_\omega(x)Re^{i\theta}i\mathrm{d}\theta\label{eq:eq286}
\end{align}
where we used $\omega=R e^{i\theta}$ and 
\begin{align}
F(R,\theta)&=\int_{-\infty}^{\infty}f(x')\psi_\omega(x')\mathrm{d}x'\label{eq:eq287}\\
c_j&=\int_{-\infty}^{\infty}f(x')\psi_j(x')\mathrm{d}x'\label{eq:eq288}
\end{align}
Next, we will look at all the three regions $x<-d,-d<x<d,d<x$ separately and decide whether we get convergence or not. The function $f(x)$ will have its support confined in a particular region which means it is non-zero only in this region. Therefore, we do not need to consider the cross-terms in the product $\psi_\omega(x)\psi_\omega(x')$.

Let $f(x)$ have its support in the region $x<-d$, so for $x,x'<-d$ we have
\begin{align}
f(x)&=\sum_{j}c_j\psi_j(x)+\frac{\chi 2}{\pi^\frac{1}{2}\sigma^\frac{1}{2}}\lim_{R\to\infty}R^\frac{5}{4}\nonumber\\
&\int_{-\frac{2\pi}{3}}^0F(R,\theta)\kappa^\frac{1}{4}e^{-i\beta\varrho_r R^\frac{1}{2}}e^{(2\mu)^\frac{1}{2}\left|\sin\left(\frac{1}{2}\theta\right)\right| R^{\frac{1}{2}}\left(x\varepsilon 2\mu-\beta\right)}e^{i\varpi_i xR^{\frac{1}{2}}}e^{i\theta}\mathrm{d}\theta\label{eq:eq289}
\end{align}
where $\mu=(2\varepsilon)^{-\frac{2}{3}}$, $\beta=2\mu\varepsilon d$, $\sigma=2\mu V_0$, $\kappa=2\mu e^{i\theta}$. The function $F(R,\theta)$ in (\ref{eq:eq289}) becomes
\begin{align}
F(R,\theta)&\approx\frac{2(\kappa R)^\frac{1}{4}}{i\pi^\frac{1}{2}\sigma^\frac{1}{2}}\int_{-\infty}^{\infty}f(x')e^{-i\beta\varrho_r R^\frac{1}{2}}e^{(2\mu)^\frac{1}{2}\left|\sin\left(\frac{1}{2}\theta\right)\right| R^{\frac{1}{2}}\left(x'\varepsilon 2\mu-\beta\right)}e^{i\varpi_i x'R^{\frac{1}{2}}}\mathrm{d}x'\label{eq:eq292}
\end{align}
This expression decays for $x'\varepsilon 2\mu-\beta<0$ which can be also written as $x'<d$. Essentially the same goes for the $x$ variable in (\ref{eq:eq289}), which means that in the current region $x,x'<-d$ we have a point-wise convergence for $f(x)$.

Let us move to the next region. Consider $f(x)$ now  having its support in the region $-d<x<d$, so for $-d<x,x'<d$ (\ref{eq:eq286}) becomes
\begin{align}
f(x)&=\sum_{j}c_j\psi_j(x)-\frac{\chi 2}{\pi^\frac{1}{2}\sigma^{\frac{1}{2}}}\lim_{R\to\infty} R^\frac{5}{4}\int_{-\frac{2\pi}{3}}^0 F(R,\theta)\kappa^\frac{1}{4}\nonumber\\
&e^{-i(\beta+\sigma)\varrho_r R^\frac{1}{2}}e^{(2\mu)^\frac{1}{2}\left|\sin\left(\frac{1}{2}\theta\right)\right|R^{\frac{1}{2}}\left[\varepsilon2\mu\left(x+\frac{V_0}{\varepsilon}\right)-(\beta+\sigma)\right]}e^{i\varpi_i R^{\frac{1}{2}}\left(x+\frac{V_0}{\varepsilon}\right)}e^{i\theta}\mathrm{d}\theta\nonumber\\
&+\frac{\sigma^\frac{1}{2}\chi i}{2\pi^\frac{1}{2}}\lim_{R\to\infty}R^{\frac{1}{4}}\int_{-\frac{2\pi}{3}}^0 F(R,\theta)\kappa^{-\frac{3}{4}}\nonumber\\
&e^{i\varrho_r R^\frac{1}{2}(\sigma-3\beta)}e^{(2\mu)^\frac{1}{2}\left|\sin\left(\frac{1}{2}\theta\right)\right|R^{\frac{1}{2}}\left[-\varepsilon 2\mu\left(x+\frac{V_0}{\varepsilon}\right)+(\sigma-3\beta)\right]}e^{-i\varpi_i R^{\frac{1}{2}}\left(x+\frac{V_0}{\varepsilon}\right)}e^{i\theta}\mathrm{d}\theta\label{eq:eq293}
\end{align}
The function $F(R,\theta)$ in (\ref{eq:eq287}) in this case becomes
\begin{align}
F(R,\theta)&\approx\frac{2(\kappa R)^\frac{1}{4}}{i\pi^\frac{1}{2}\sigma^{\frac{1}{2}}}\int_{-\infty}^{\infty}f(x')\nonumber\\
&e^{-i(\beta+\sigma)\varrho_r R^\frac{1}{2}}e^{(2\mu)^\frac{1}{2}\left|\sin\left(\frac{1}{2}\theta\right)\right|R^{\frac{1}{2}}\left[\varepsilon2\mu\left(x'+\frac{V_0}{\varepsilon}\right)-(\beta+\sigma)\right]}e^{i\varpi_i R^{\frac{1}{2}}\left(x'+\frac{V_0}{\varepsilon}\right)}\mathrm{d}x'\nonumber\\
&-\frac{\sigma^\frac{1}{2}}{2(\kappa R)^\frac{3}{4}\pi^\frac{1}{2}}\int_{-\infty}^{\infty}f(x')\nonumber\\
&e^{i\varrho_r R^\frac{1}{2}(\sigma-3\beta)}e^{(2\mu)^\frac{1}{2}\left|\sin\left(\frac{1}{2}\theta\right)\right|R^{\frac{1}{2}}\left[-\varepsilon 2\mu\left(x'+\frac{V_0}{\varepsilon}\right)+(\sigma-3\beta)\right]}e^{-i\varpi_i R^{\frac{1}{2}}\left(x'+\frac{V_0}{\varepsilon}\right)}\mathrm{d}x'\label{eq:eq294}
\end{align}
where $\mu=(2\varepsilon)^{-\frac{2}{3}}$, $\beta=2\mu\varepsilon d$, $\sigma=2\mu V_0$ and $\kappa=2\mu e^{i\theta}$. We can see that we have two integrals both in (\ref{eq:eq293}) and (\ref{eq:eq294}). Let us investigate under what conditions each of them vanishes. In both expressions we get the same results for $x$ and $x'$. The first integral vanishes if
\begin{align}
\varepsilon2\mu\left(x+\frac{V_0}{\varepsilon}\right)-(\beta+\sigma)&<0\nonumber\\
&\Downarrow\nonumber\\
x&<d\label{eq:eq295}
\end{align}
and the second one vanishes under the condition
\begin{align}
-\varepsilon 2\mu\left(x+\frac{V_0}{\varepsilon}\right)+(\sigma-3\beta)&<0\nonumber\\
&\Downarrow\nonumber\\
x&>-3d\label{eq:eq296}
\end{align}
One can see, that both of them are obeyed, since we are considering the region $-d<x<d$.

Finally, suppose that $f(x)$ has a support confined in $d<x$. For $d<x,x'$ the equation (\ref{eq:eq286}) reads
\begin{align}
f(x)&=\sum_{j}c_j\psi_j(x)+\frac{2}{\pi^\frac{1}{2}\sigma^\frac{1}{2}}\lim_{R\to\infty}R^\frac{5}{4}\nonumber\\
&\int_{-\frac{2\pi}{3}}^0F(R,\theta)\kappa^\frac{1}{4}e^{-i\beta\varrho_r R^\frac{1}{2}}e^{(2\mu)^\frac{1}{2}\left|\sin\left(\frac{1}{2}\theta\right)\right|R^{\frac{1}{2}}\left(\varepsilon2\mu x-\beta\right)}e^{i\varpi_i xR^{\frac{1}{2}}}e^{i\theta}\mathrm{d}\theta\nonumber\\
&-\frac{\sigma^\frac{1}{2}}{2\pi^\frac{1}{2}}\lim_{R\to\infty}R^\frac{1}{4}\nonumber\\
&\int_{-\frac{2\pi}{3}}^0F(R,\theta)\kappa^{-\frac{3}{4}}e^{i\beta\varrho_r R^\frac{1}{2}}e^{(2\mu)^\frac{1}{2}\left|\sin\left(\frac{1}{2}\theta\right)\right|R^{\frac{1}{2}}\left(-\varepsilon2\mu x+\beta\right)}e^{-i\varpi_i xR^{\frac{1}{2}}}e^{i\theta}\mathrm{d}\theta\label{eq:eq296.2}
\end{align}
The function $F(R,\theta)$ in (\ref{eq:eq287}) in this case becomes
\begin{align}
&F(R,\theta)\approx\frac{2i(\kappa R)^\frac{1}{4}}{\pi^\frac{1}{2}\sigma^\frac{1}{2}}\int_{-\infty}^{\infty}f(x')e^{-i\beta\varrho_r R^\frac{1}{2}}e^{(2\mu)^\frac{1}{2}\left|\sin\left(\frac{1}{2}\theta\right)\right|R^{\frac{1}{2}}\left(\varepsilon2\mu x'-\beta\right)}e^{i\varpi_i x'R^{\frac{1}{2}}}\mathrm{d}x'\nonumber\\
&-\frac{\sigma^\frac{1}{2}i}{2\pi^\frac{1}{2}(\kappa R)^\frac{3}{4}}\int_{-\infty}^{\infty}f(x')e^{i\beta\varrho_r R^\frac{1}{2}}e^{(2\mu)^\frac{1}{2}\left|\sin\left(\frac{1}{2}\theta\right)\right|R^{\frac{1}{2}}\left(-\varepsilon2\mu x'+\beta\right)}e^{-i\varpi_i x'R^{\frac{1}{2}}}\mathrm{d}x'\label{eq:eq296.3}
\end{align}
where as before $\mu=(2\varepsilon)^{-\frac{2}{3}}$, $\beta=2\mu\varepsilon d$, $\sigma=2\mu V_0$ and $\kappa=2\mu e^{i\theta}$. We have again two integrals both in (\ref{eq:eq296.2}) and (\ref{eq:eq296.3}). This time we study the conditions when these integrals grow exponentially in the limit $R\to\infty$. In both expressions we get the same results for $x$ and $x'$. The first integral grows if $\varepsilon2\mu x-\beta>0$ which can be also written as $x>d$. Similarly, for the second one we have growth if $x<d$.

At this point we can conclude that in region $d<x$, the first integral in (\ref{eq:eq296.2}) and (\ref{eq:eq296.3}) grows while the second decays, so overall we get a divergence in this region. Summing it up, we showed, that $f(x)$ can be represented by the resonant states (\ref{eq:eq132}) if $f(x)$ has its support confined in $x<d$. We would like to emphasize at this point, that the fact the expansion converges without any dependence of the depth of square-well is unexpected. To strengthen this result, we are going to complete another version of this proof similar to what has been done in \cite{Brown}.

\section{Resonant states expansion II}
This chapter contains a different version of the weaker proof presented in the previous section. From a theoretical point of view, this version provides a different way with more specific conclusions. We start by splitting the scattering form of the resonant states into in- and out-going parts. This allows us to split the test-function in the same way. Both of these part are then represented by integrals which we then compute using the same method as in the weak version. However, the integrand is going to be  more complicated, so we expect more conditions to deal with.

The Airy functions $\mathrm{Ci}^\pm(x)$ can be interpreted as in- and outgoing waves, which means, we can define in- and outgoing waves $\psi^\pm_\omega(x)$.
{\medmuskip=0mu
\thinmuskip=0mu
\thickmuskip=0mu
\begin{align}
&\psi^+_\omega(x)=\chi\left\{
\begin{array}{l l}
i\left(\frac{\overline{\det\mathbf{M}}(\omega)}{\det\mathbf{M}(\omega)}\right)^{\frac{1}{2}}\mathrm{Ci}^+\left(\mu\left(d+\frac{\omega}{\varepsilon}\right)\right)\frac{\mathrm{Ai}\left(\mu\left(x+\frac{\omega}{\varepsilon}\right)\right)}{\pi p(\omega)} & x<-d\\
i\left(\frac{\overline{\det\mathbf{M}}(\omega)}{\det\mathbf{M}(\omega)}\right)^{\frac{1}{2}}\mathrm{Ci}^+\left(\mu\left(d+\frac{\omega}{\varepsilon}\right)\right)\frac{1}{p(\omega)} & \\
\left[(B_1'A_0-B_1A_0')\mathrm{Ai}\left(\mu\left(x+\frac{\omega+V_0}{\varepsilon}\right)\right)\right. & \\
\left.+(A_1A_0'-A_1'A_0)\mathrm{Bi}\left(\mu\left(x+\frac{\omega+V_0}{\varepsilon}\right)\right)\right] & -d< x< d\\
i\left(\frac{\overline{\det\mathbf{M}}(\omega)}{\det\mathbf{M}(\omega)}\right)^{\frac{1}{2}}\mathrm{Ci}^+\left(\mu\left(x+\frac{\omega}{\varepsilon}\right)\right) & d<x
\end{array}\right.\label{eq:eq303}\\
&\psi^-_\omega(x)=\chi\left\{
\begin{array}{l l}
-i\left(\frac{\det\mathbf{M}(\omega)}{\overline{\det \mathbf{M}}(\omega)}\right)^{\frac{1}{2}}\mathrm{Ci}^-\left(\mu\left(d+\frac{\omega}{\varepsilon}\right)\right)\frac{\mathrm{Ai}\left(\mu\left(x+\frac{\omega}{\varepsilon}\right)\right)}{\pi p(\omega)} & x<-d\\
-i\left(\frac{\det\mathbf{M}(\omega)}{\overline{\det \mathbf{M}}(\omega)}\right)^{\frac{1}{2}}\mathrm{Ci}^-\left(\mu\left(d+\frac{\omega}{\varepsilon}\right)\right)\frac{1}{p(\omega)} & \\
\left[(B_1'A_0-B_1A_0')\mathrm{Ai}\left(\mu\left(x+\frac{\omega+V_0}{\varepsilon}\right)\right)\right. & \\
\left.+(A_1A_0'-A_1'A_0)\mathrm{Bi}\left(\mu\left(x+\frac{\omega+V_0}{\varepsilon}\right)\right)\right] & -d< x< d\\
-i\left(\frac{\det\mathbf{M}(\omega)}{\overline{\det \mathbf{M}}(\omega)}\right)^{\frac{1}{2}}\mathrm{Ci}^-\left(\mu\left(x+\frac{\omega}{\varepsilon}\right)\right) & d<x
\end{array}\right.\label{eq:eq304}
\end{align}}
where $p(\omega)=(B_1'A_0-B_1A_0')A_2+(A_1A_0'-A_1'A_0)B_2$. Observe, that the functions $\psi^\pm_\omega(x)$ are continuous at $x=-d,d$, also the derivative at $x=-d$ is continuous, but their derivative at $x=d$ is not continuous. Also observe, that by this construction we have
\begin{align}
\psi_\omega(x)=\psi^+_\omega(x)+\psi^-_\omega(x)\label{eq:eq305}
\end{align}
The completeness relation for scattering states as before is
\begin{align}
\int_{-\infty}^\infty\psi_\omega(x)\psi_\omega(x')\mathrm{d}\omega=\delta(x-x')\label{eq:eq306}
\end{align}
Let $f(x)$ be any wave-function. Then using the completeness relation (\ref{eq:eq306}) we get
\begin{align}
f(x)=\int_{-\infty}^\infty\delta(x-s)f(s)\mathrm{d}s=\int_{-\infty}^\infty\int_{-\infty}^\infty\psi_\omega(x)\psi_\omega(s)\mathrm{d}\omega f(s)\mathrm{d}s\label{eq:eq307}
\end{align}
We use this relation second times as we multiply (\ref{eq:eq306}) with $f(x)$ and integrate over the whole space:
\begin{align}
\int_{-\infty}^\infty\int_{-\infty}^\infty\psi_{\omega'}(x)\psi_{\omega'}(s)f(x)\mathrm{d}\omega'\mathrm{d}x&=\int_{-\infty}^\infty\delta(x-s)f(x)\mathrm{d}x\nonumber\\
\int_{-\infty}^\infty a(\omega')\psi_{\omega'}(s)\mathrm{d}\omega'&=f(s)\label{eq:eq308}
\end{align}
where $a(\omega')=\int_{-\infty}^\infty\psi_{\omega'}(x)f(x)\mathrm{d}x$, which is the enery representation of $f(x)$. We substitute (\ref{eq:eq308}) back into (\ref{eq:eq307}) and we get
\begin{align}
f(x)&=\int_{-\infty}^\infty\int_{-\infty}^\infty\psi_\omega(x)\psi_\omega(s)\mathrm{d}\omega\int_{-\infty}^\infty a(\omega')\psi_{\omega'}(s)\mathrm{d}\omega'\mathrm{d}s\nonumber\\
&=\int_{-\infty}^\infty a(\omega')\int_{-\infty}^\infty\psi_\omega(x)\int_{-\infty}^\infty\psi_\omega(s)\psi_{\omega'}(s)\mathrm{d}s\mathrm{d}\omega\mathrm{d}\omega'\label{eq:eq309}
\end{align}
Using the separation of the resonant states into outgoing and ingoing waves (\ref{eq:eq305}) we can rewrite (\ref{eq:eq309}) as
\begin{align}
f(x)&=\int_{-\infty}^\infty a(\omega')\int_{-\infty}^\infty\psi_\omega(x)\Upsilon^+(\omega,\omega')\mathrm{d}\omega\mathrm{d}\omega'\nonumber\\
&+\int_{-\infty}^\infty a(\omega')\int_{-\infty}^\infty\psi_\omega(x)\Upsilon^-(\omega,\omega')\mathrm{d}\omega\mathrm{d}\omega'=f^+(x)+f^-(x)\label{eq:eq310}
\end{align}
where we have defined the following quantities
\begin{align}
\Upsilon^\pm(\omega,\omega')&=\int_{-\infty}^\infty\psi_\omega(s)\psi_{\omega'}^\pm(s)\mathrm{d}s\label{eq:eq311}\\
f^\pm(x)&=\int_{-\infty}^\infty a(\omega')\int_{-\infty}^\infty\psi_\omega(x)\Upsilon^\pm(\omega,\omega')\mathrm{d}\omega\mathrm{d}\omega'\label{eq:eq312}
\end{align}
This splitting of the function $f(x)$ allows us to write the expansion of this function with resonant states in two sums: one out-going (+) and one in-going (-). It does not tell us however, how one determines whether a function is of out- or in-going type. By manipulating how we split $f(x)$ into these two part, we can set up two infinite sums for both types where the associated sums converge.

Continuing with the proof, we observe, that the integrand in (\ref{eq:eq311}) does not converge. It can be seen from the standard asymptotic expression of Airy functions, where we see that $\mathrm{Ci}^\pm(x)$ is decaying algebraically, but not fast enough to converge. That is why we introduce the following correction in the variable $\omega'$
\begin{align}
\Upsilon_{\xi}^\pm(\omega,\omega')=\int_{-\infty}^\infty\psi_\omega(s)\psi_{\omega'\pm i\xi}^\pm(s)\mathrm{d}s\label{eq:eq313}
\end{align}
We will perform calculations with a finite $\xi$ and in the end we remove this correction by letting it go to 0 in a limit. Let us now focus on the integrals $\Upsilon^\pm$. A detailed derivation is provided in Appendix \ref{App:Appendix}.
\begin{align}
\Upsilon_{\xi}^\pm(\omega,\omega')&=\frac{\chi\varepsilon}{\rho^2}\frac{\psi_\omega(d^+)}{\omega-\omega'\mp i\xi}\left(\mp i\frac{|\det\mathbf{M}(\omega')|}{p(\omega')}\right)\label{eq:eq326.1}
\end{align}
It is easy to verify that the following statement is true.
\begin{align}
\lim_{\xi\to0}\left(\Upsilon_{\xi}^+(\omega,\omega')+\Upsilon_{\xi}^-(\omega,\omega')\right)=\delta(\omega-\omega')\label{eq:eq327}
\end{align}
Let us go back to (\ref{eq:eq312}). Using (\ref{eq:eq326.1}) for to express $f^\pm(x)$ we get the following expression
\begin{align}
f^\pm(x)&=\int_{-\infty}^\infty a(\omega')\int_{-\infty}^\infty\psi_\omega(x)\Upsilon^\pm(\omega,\omega')\mathrm{d}\omega\mathrm{d}\omega'\nonumber\\
&=\frac{\chi\varepsilon}{\rho^2}\int_{-\infty}^\infty a(\omega')\left[\mp i\frac{|\det\mathbf{M}(\omega')|}{p(\omega')}\right]\int_{-\infty}^\infty\psi_\omega(x)\frac{\psi_\omega(d^+)}{\omega-\omega'\mp i\xi}\mathrm{d}\omega\mathrm{d}\omega'\label{eq:eq328}
\end{align}
Our whole focus will be on this integral
\begin{align}
P_{\xi}(\omega')&=\int_{-\infty}^\infty\psi_\omega(x)\frac{\psi_{\omega}(d^+)}{\omega-\omega'\mp i\xi}\mathrm{d}\omega\label{eq:eq329}
\end{align}
using the residue theorem. As we did in (\ref{eq:eq193}) on a closed contour $\Gamma_R$, we express $P_{\xi}(\omega')$ with the help of $\Gamma_R$. In order to do this we define
\begin{align}
P_{\xi}^R(\omega')&=\int_{-R}^R\psi_\omega(x)\frac{\psi_{\omega}(d^+)}{\omega-\omega'\mp i\xi}\mathrm{d}\omega\label{eq:eq330}
\end{align}
which converges to (\ref{eq:eq329}) as $R\to\infty$. We use also the same notation for this contour $\Gamma_R$ and we get
\begin{align}
&\int_{\Gamma_R}\psi_\omega(x)\frac{\psi_{\omega}(d^+)}{\omega-\omega'\mp i\xi}\mathrm{d}\omega=P_{\xi}^R(\omega')+\int_{C_R}\psi_\omega(x)\frac{\psi_{\omega}(d^+)}{\omega-\omega'\mp i\xi}\mathrm{d}\omega\nonumber\\
&\frac{2\pi}{i}\sum_j\mathrm{Res}\left(\psi_\omega(x)\frac{\psi_{\omega}(d^+)}{\omega-\omega'\mp i\xi},\omega_j\right)=P_{\xi}^R(\omega')+\int_{C_R}\psi_\omega(x)\frac{\psi_{\omega}(d^+)}{\omega-\omega'\mp i\xi}\mathrm{d}\omega\label{eq:eq331}
\end{align}
from which we can see, that if the the integral on the right hand side vanishes as $R$ approaches infinity, then our function $P_{\xi}(\omega')$ will equal to the sum of residues. In the following, we examine the integral along $C_R$ using the asymptotic formulas of Airy functions. They give us two separate sectors for the angle of the argument. From the asymptotic behaviours of Airy functions along rays in the lower half of the complex frequency plane $\omega=Re^{i\theta}$, we have two sectors $-\frac{2\pi}{3}<\theta<0$ and $-\pi<\theta<-\frac{2\pi}{3}$. The asymptotic expression for the sector $-\pi<\theta<-\frac{2\pi}{3}$ decays in the limit $R\to\infty$ independently of $x$. In the first sector we have the following asymptotic formula
{\medmuskip=0mu
\thinmuskip=0mu
\thickmuskip=0mu
\begin{align}
\psi_\omega(x)=\chi\left\{
\begin{array}{l l}
\frac{2(\kappa R)^\frac{1}{4}}{i\pi^\frac{1}{2}\sigma^\frac{1}{2}}e^{-i\beta\varrho R^\frac{1}{2}}e^{\varpi xR^{\frac{1}{2}}} & x<-d\\
\frac{2(\kappa R)^\frac{1}{4}}{i\pi^\frac{1}{2}\sigma^{\frac{1}{2}}}e^{-i(\beta+\sigma)\varrho R^\frac{1}{2}}e^{\varpi R^{\frac{1}{2}}\left(x+\frac{V_0}{\varepsilon}\right)} & -d<x<d\\
-\frac{\sigma^\frac{1}{2}}{2(\kappa R)^\frac{3}{4}\pi^\frac{1}{2}}e^{i\varrho R^\frac{1}{2}(\sigma-3\beta)}e^{-\varpi R^{\frac{1}{2}}\left(x+\frac{V_0}{\varepsilon}\right)} &\\
i\frac{2(\kappa R)^\frac{1}{4}}{\pi^\frac{1}{2}\sigma^\frac{1}{2}}e^{-i\beta\varrho R^\frac{1}{2}}e^{\varpi xR^{\frac{1}{2}}}-i\frac{\sigma^\frac{1}{2}}{2\pi^\frac{1}{2}(\kappa R)^\frac{3}{4}}e^{i\beta\varrho R^\frac{1}{2}}e^{-\varpi xR^{\frac{1}{2}}} & d<x
\end{array}\right.\label{eq:eq332}
\end{align}}
where $\varpi,\varrho$ are complex numbers depending on $\theta$ and $\mu=(2\varepsilon)^{-\frac{2}{3}}$, $\beta=2\mu\varepsilon d$, $\sigma=2\mu V_0$, $\kappa=2\mu e^{i\theta}$. Let us start analysing the various regions in $x$. For $x<-d$ The integrand on the right hand side in (\ref{eq:eq331}) becomes
\begin{align}
\psi_\omega(x)\frac{\psi_{\omega}(d^+)}{\omega-\omega'\mp i\xi}&\approx-\chi^2\frac{1}{\pi(\kappa R)^\frac{1}{2}\left(\omega-\omega'\mp i\xi\right)}\nonumber\\
&e^{\left|\sin\left(\frac{1}{2}\theta\right)\right|R^{\frac{1}{2}}\varepsilon (2\mu)^\frac{3}{2}(x-d)}e^{i\varpi_i R^{\frac{1}{2}}(x-d)}\nonumber\\
&+\chi^2\frac{4(\kappa R)^\frac{1}{2}}{\pi\sigma\left(\omega-\omega'\mp i\xi\right)}\nonumber\\
&e^{-i2\beta\varrho_r R^\frac{1}{2}}e^{(2\mu)^\frac{1}{2}\left|\sin\left(\frac{1}{2}\theta\right)\right| R^{\frac{1}{2}}\left(\varepsilon 2\mu(x+d)-2\beta\right)}e^{i\varpi_i R^{\frac{1}{2}}(x+d)}\label{eq:eq333}
\end{align}
It is easy to see that the first part of (\ref{eq:eq333}) decays in the limit $R\to\infty$ if $x<d$ and for the second part if $\varepsilon 2\mu(x+d)-2\beta<0$ which after simplifying it we can write $x<d$. This is satisfied in the region $x<-d$, so the expression (\ref{eq:eq333}) decays.

Let us proceed to the region $-d<x<d$ where we have
{\medmuskip=0mu
\thinmuskip=0mu
\thickmuskip=0mu
\begin{align}
\psi_\omega(x)&\frac{\psi_{\omega}(d^+)}{\omega-\omega'\mp i\xi}\approx-\frac{\chi^2 }{\omega-\omega'\mp i\xi}\left(\frac{1}{\pi(\kappa R)^\frac{1}{2}}\right.\nonumber\\
&e^{-i\sigma\varrho_r R^\frac{1}{2}}e^{(2\mu)^\frac{1}{2}\left|\sin\left(\frac{1}{2}\theta\right)\right| R^{\frac{1}{2}}\left[\varepsilon 2\mu\left(x+\frac{V_0}{\varepsilon}-d\right)-\sigma\right]}e^{i\varpi_i R^{\frac{1}{2}}\left(x+\frac{V_0}{\varepsilon}-d\right)}\nonumber\\
&-\frac{\sigma}{8\pi(\kappa R)^\frac{3}{2}}\nonumber\\
&e^{i\varrho_r R^\frac{1}{2}(\sigma-2\beta)}e^{(2\mu)^\frac{1}{2}\left|\sin\left(\frac{1}{2}\theta\right)\right| R^{\frac{1}{2}}\left[-\varepsilon2\mu\left(x+\frac{V_0}{\varepsilon}+d\right)+(\sigma-2\beta)\right]}e^{-i\varpi_i R^{\frac{1}{2}}\left(x+\frac{V_0}{\varepsilon}+d\right)}\nonumber\\
&-\frac{4(\kappa R)^\frac{1}{2}}{\pi\sigma}\nonumber\\
&e^{-i(2\beta+\sigma)\varrho_r R^\frac{1}{2}}e^{(2\mu)^\frac{1}{2}\left|\sin\left(\frac{1}{2}\theta\right)\right| R^{\frac{1}{2}}\left[\varepsilon2\mu\left(x+\frac{V_0}{\varepsilon}+d\right)-(2\beta+\sigma)\right]}e^{i\varpi_i R^{\frac{1}{2}}\left(x+\frac{V_0}{\varepsilon}+d\right)}\nonumber\\
&+\left.\frac{1}{\pi(\kappa R)^\frac{1}{2}}\right.\nonumber\\
&\left.e^{i\varrho_r R^\frac{1}{2}(\sigma-4\beta)}e^{(2\mu)^\frac{1}{2}\left|\sin\left(\frac{1}{2}\theta\right)\right| R^{\frac{1}{2}}\left[-\varepsilon2\mu\left(x+\frac{V_0}{\varepsilon}-d\right)+(\sigma-4\beta)\right]}e^{-i\varpi_i R^{\frac{1}{2}}\left(x+\frac{V_0}{\varepsilon}-d\right)}\right)\label{eq:eq334}
\end{align}}
In (\ref{eq:eq334}) we have four different terms that needs to be checked separately. The first two exponentials in (\ref{eq:eq334}) decay if
\begin{align}
\varepsilon 2\mu\left(x+\frac{V_0}{\varepsilon}-d\right)-\sigma&<0 & -\varepsilon2\mu\left(x+\frac{V_0}{\varepsilon}+d\right)+(\sigma-2\beta)&<0\nonumber\\
&\Downarrow & &\Downarrow\nonumber\\
d&>x & -3d&<x\label{eq:eq334.1}
\end{align}
These conditions are satisfied in the current region $-d<x<d$.

The second two parts of (\ref{eq:eq334}) converge to zero if
\begin{align}
\varepsilon2\mu\left(x+\frac{V_0}{\varepsilon}+d\right)-(2\beta+\sigma)&<0 & -\varepsilon2\mu\left(x+\frac{V_0}{\varepsilon}-d\right)+(\sigma-4\beta)<&0\nonumber\\
&\Downarrow & &\Downarrow\nonumber\\
d&>x & -3d&<x\label{eq:eq334.2}
\end{align}
which are obeyed as well for the same reasons. This leads to the integral in (\ref{eq:eq331}) vanishing for this region.

Let us explore the last region $d<x$. Here we get using (\ref{eq:eq332}) the following.
{\medmuskip=0mu
\thinmuskip=0mu
\thickmuskip=0mu
\begin{align}
\psi_\omega(x)\frac{\psi_{\omega}(d^+)}{\omega-\omega'\mp i\xi}&\approx\frac{\chi^2}{\omega-\omega'\mp i\xi}\left(\frac{1}{\pi(\kappa R)^\frac{1}{2}}e^{\varepsilon (2\mu)^\frac{3}{2}\left|\sin\left(\frac{1}{2}\theta\right)\right| R^{\frac{1}{2}}(x-d)}e^{i\varpi_i R^{\frac{1}{2}}(x-d)}\right.\nonumber\\
&-\frac{\sigma }{4(\kappa R)^\frac{3}{2}}e^{i2\beta\varrho_r R^\frac{1}{2}}e^{(2\mu)^\frac{1}{2}\left|\sin\left(\frac{1}{2}\theta\right)\right| R^{\frac{1}{2}}\left[-\varepsilon2\mu(x+d)+2\beta\right]}e^{-i\varpi_i R^{\frac{1}{2}}(x+d)}\nonumber\\
&-\frac{4(\kappa R)^\frac{1}{2}}{\pi\sigma}e^{-i2\beta\varrho_r R^\frac{1}{2}}e^{(2\mu)^\frac{1}{2}\left|\sin\left(\frac{1}{2}\theta\right)\right| R^{\frac{1}{2}}\left[\varepsilon2\mu(x+d)-2\beta\right]}e^{i\varpi_i R^{\frac{1}{2}}(x+d)}\nonumber\\
&\left.+\frac{1}{\pi(\kappa R)^\frac{1}{2}}e^{\varepsilon (2\mu)^\frac{3}{2}\left|\sin\left(\frac{1}{2}\theta\right)\right| R^{\frac{1}{2}}(d-x)}e^{-i\varpi_i R^{\frac{1}{2}}(x-d)}\right)\label{eq:eq335}
\end{align}}
This is a similar situation as in the previous case. We proceed therefore in an almost identical way. The first two terms in this expression decay exponentially if $x<d$ and $\varepsilon2\mu(x+d)-2\beta<0$ which can be also written as $x>d$ respectively. For the last two terms in (\ref{eq:eq335}) we have $\varepsilon2\mu(x+d)-2\beta<0$ or $x<d$ and $d<x$. We got a variety of answers from which we can deduce that with respect to the current region $d<x$ the first and the third exponential in (\ref{eq:eq335}) grow while the second and fourth term decay for $R\to\infty$. The sum of these four terms gives us exponential growth, so the integrand in (\ref{eq:eq331}) diverges for $d<x$.

After analysing the integrand on the right hand side in (\ref{eq:eq331}) we can conclude, that it decays exponentially only when $x<d$ in the limit $R\to\infty$. Going back to (\ref{eq:eq331}) we see, that for this region we have in the limit $R\to\infty$
\begin{align}
\frac{2\pi}{i}\sum_j\mathrm{Res}\left(\psi_\omega(x)\frac{\psi_{\omega}(d^+)}{\omega-\omega'\mp i\xi},\omega_j\right)=P_{\xi}(\omega')\label{eq:eq339}
\end{align}
and $f^\pm(x)$ in (\ref{eq:eq328}) becomes
{\medmuskip=0mu
\thinmuskip=0mu
\thickmuskip=0mu
\begin{align}
f^\pm(x)&=\frac{\chi\varepsilon}{\rho^2}\int_{-\infty}^\infty a(\omega')\left[\mp i\frac{|\det\mathbf{M}(\omega')|}{p(\omega')}\right]\frac{2\pi}{i}\sum_j\mathrm{Res}\left(\psi_\omega(x)\frac{\psi_{\omega}(d^+)}{\omega-\omega'\mp i\xi},\omega_j\right)\mathrm{d}\omega'\label{eq:eq340}
\end{align}}
which is the sought expansion, but one needs to evaluate the residua. The expression in the residue function should first be expressed as
{\medmuskip=-1mu
\thinmuskip=0mu
\thickmuskip=0mu
\begin{align}
&\psi_\omega(x)\frac{\psi_{\omega}(d^+)}{\omega-\omega'\mp i\xi}=\frac{\chi^2}{\omega-\omega'\mp i\xi}\nonumber\\
&\left\{
\begin{array}{l l}
\frac{2}{\pi^2}\left(\frac{1}{\overline{\det \mathbf{M}}(\omega)}\mathrm{Ci}^+\left(\mu\left(d+\frac{\omega}{\varepsilon}\right)\right)-\frac{1}{\det\mathbf{M}(\omega)}\mathrm{Ci}^-\left(\mu\left(d+\frac{\omega}{\varepsilon}\right)\right)\right)\mathrm{Ai}\left(\mu\left(x+\frac{\omega}{\varepsilon}\right)\right) & x<-d\\
\frac{2}{\pi}\left[(B_1'A_0-B_1A_0')\mathrm{Ai}\left(\mu\left(x+\frac{\omega+V_0}{\varepsilon}\right)\right)\right. & \\
\left.+(A_1A_0'-A_1'A_0)\mathrm{Bi}\left(\mu\left(x+\frac{\omega+V_0}{\varepsilon}\right)\right)\right] & \\
\left(\frac{1}{\overline{\det \mathbf{M}}(\omega)}\mathrm{Ci}^+\left(\mu\left(d+\frac{\omega}{\varepsilon}\right)\right)-\frac{1}{\det\mathbf{M}(\omega)}\mathrm{Ci}^-\left(\mu\left(d+\frac{\omega}{\varepsilon}\right)\right)\right) & -d<x<d\\
\left[\mathrm{Ci}^+\left(\mu\left(x+\frac{\omega}{\varepsilon}\right)\right)\mathrm{Ci}^-\left(\mu\left(d+\frac{\omega}{\varepsilon}\right)\right)+\mathrm{Ci}^-\left(\mu\left(x+\frac{\omega}{\varepsilon}\right)\right)\mathrm{Ci}^+\left(\mu\left(d+\frac{\omega}{\varepsilon}\right)\right)\right] & \\
-\frac{\overline{\det\mathbf{M}}(\omega)}{\det \mathbf{M}(\omega)}\mathrm{Ci}^+\left(\mu\left(x+\frac{\omega}{\varepsilon}\right)\right)\mathrm{Ci}^+\left(\mu\left(d+\frac{\omega}{\varepsilon}\right)\right) & \\
-\frac{\det\mathbf{M}(\omega)}{\overline{\det \mathbf{M}}(\omega)}\mathrm{Ci}^-\left(\mu\left(x+\frac{\omega}{\varepsilon}\right)\right)\mathrm{Ci}^-\left(\mu\left(d+\frac{\omega}{\varepsilon}\right)\right) & d<x
\end{array}\right.\label{eq:eq342}
\end{align}}
The resonant eigenstates $\omega_j$ in (\ref{eq:eq339}) are the zero points of $\det\mathbf{M}(\omega)$ located in the lower complex frequency half-plane and similarly the zeros of $\overline{\det\mathbf{M}}(\omega)$ are in the upper half. It is clear, that the poles of (\ref{eq:eq342}) are determined by zeros of $\det\mathbf{M}(\omega)$, since $C_R$ is in the lower half-plane, see (\ref{eq:eq331}). Remembering this we can ignore those terms that do not have $\omega_j$ as poles in the lower complex half-plane. Then the residues become
\begin{align}
&\mathrm{Res}\left(\psi_\omega(x)\frac{\psi_{\omega}(d^+)}{\omega-\omega'\mp i\xi},\omega_j\right)=\frac{\chi}{\omega_j-\omega'\mp i\xi}\psi_j(x)\label{eq:eq344}
\end{align}
with
{\medmuskip=0mu
\thinmuskip=0mu
\thickmuskip=0mu
\begin{align}
\psi_j(x)&=\chi\lim_{\omega\to\omega_j}\left\{\frac{\omega-\omega_j}{\det\mathbf{M}(\omega)}\right\}\nonumber\\
&\left\{
\begin{array}{l l}
-\frac{2}{\pi^2}\mathrm{Ci}^-_j\left(\mu\left(d+\frac{\omega}{\varepsilon}\right)\right)\mathrm{Ai}_j\left(\mu\left(x+\frac{\omega}{\varepsilon}\right)\right) & x<-d\\
-\frac{2}{\pi}\left[(B_1'A_0-B_1A_0')_j\mathrm{Ai}_j\left(\mu\left(x+\frac{\omega+V_0}{\varepsilon}\right)\right)\right. & -d<x<d\\
\left.+(A_1A_0'-A_1'A_0)_j\mathrm{Bi}_j\left(\mu\left(x+\frac{\omega+V_0}{\varepsilon}\right)\right)\right]\mathrm{Ci}^-_j\left(\mu\left(d+\frac{\omega}{\varepsilon}\right)\right) & \\
-\overline{\det\mathbf{M}}(\omega_j)\mathrm{Ci}^+_j\left(\mu\left(d+\frac{\omega}{\varepsilon}\right)\right)\mathrm{Ci}^+_j\left(\mu\left(x+\frac{\omega}{\varepsilon}\right)\right) & d<x
\end{array}\right.\label{eq:eq345}
\end{align}}
where we denoted $\mathrm{Ai}_j\left(\mu\left(x+\frac{\omega}{\varepsilon}\right)\right)=\mathrm{Ai}\left(\mu\left(x+\frac{\omega}{\varepsilon}\right)\right)\Big|_{\omega=\omega_j}$,\\ $(B_1'A_0-B_1A_0')_j=(B_1'A_0-B_1A_0')\Big|_{\omega=\omega_j}$ etc. One can see that the functions $\psi_j(x)$ are proportional to the resonant states (\ref{eq:eq132}). Substituting (\ref{eq:eq344}) back into (\ref{eq:eq340}) and taking the limit $\xi\to 0$ we get
\begin{align}
\lim_{\xi\to 0} f^\pm(x)&=\frac{\chi\varepsilon}{\rho^2}\int_{-\infty}^\infty a(\omega')\left[\mp i\frac{|\det\mathbf{M}(\omega')|}{p(\omega')}\right]\frac{2\pi}{i}\sum_j\frac{\chi}{\omega_j-\omega'\mp i\xi}\psi_j(x)\mathrm{d}\omega'\nonumber\\
&=\sum_j\frac{\chi2\pi \varepsilon}{\rho^2}\int_{-\infty}^\infty a(\omega')\left[\mp \frac{|\det\mathbf{M}(\omega')|}{p(\omega')}\right]\frac{\chi}{\omega_j-\omega'\mp i\xi}\mathrm{d}\omega'\psi_j(x)\nonumber\\
&=\sum_jc^\pm_j\psi_j(x)\label{eq:eq346}
\end{align}
where the coefficients are
\begin{align}
c^\pm_j&=\frac{\chi2\pi \varepsilon}{\rho^2}\int_{-\infty}^\infty a(\omega')\left[\mp \frac{|\det\mathbf{M}(\omega')|}{p(\omega')}\right]\frac{\chi}{\omega_j-\omega'\mp i\xi}\mathrm{d}\omega'\label{eq:eq347}
\end{align}
with $\rho=2\mu\varepsilon$.

It is worth mentioning that these results are surprising, because the convergence of the sum in (\ref{eq:eq286}) does not depend on the depth of the well $V_0$. The same result was obtained in the previous chapter with a different proof. As a consequence of this, one might ask what if we set $V_0=0$? Then the width of the well $d$ would lost its meaning, so would we get convergence everywhere? To answer this we must remember that for this case we need to consider $V_0=0$ all the way from the beginning. Let us see what do the scattering states look like in this special case. For convenience we choose $d>0$. The variable transformation or arguments to the Airy functions in the region $-d<x<d$ now becomes the same as in the outside of the well. So we have
\begin{align}
\psi_\omega(x)&=\chi\left\{
\begin{array}{l l}
2i\mathrm{Ai}\left(\mu\left(x+\frac{\omega}{\varepsilon}\right)\right) & x<d\\
\mathrm{Ci}^{+}\left(\mu\left(x+\frac{\omega}{\varepsilon}\right)\right)-\mathrm{Ci}^{-}\left(\mu\left(x+\frac{\omega}{\varepsilon}\right)\right) & d<x
\end{array}\right.\nonumber\\
&=\chi 2i\mathrm{Ai}(y_1(x))\quad\forall x\label{eq:eq347.8}
\end{align}
and the determinant in this case becomes
\begin{align}
\det\mathbf{M}_0(\omega)&=C_3A'_3-C'_3A_3=(B_3+iA_3)A'_3-(B'_3+iA'_3)A_3=B_3A'_3-B'_3A_3\nonumber\\
&=-\frac{1}{\pi}\label{eq:eq347.3}
\end{align}
with some normalization constant $\chi$. We find out that this constant in this case is $\chi=2^{-\frac{2}{3}}\varepsilon^{-\frac{1}{6}}$. In both of the proofs we used the complex contour $\Gamma_R$ to prove that the function $f(x)$ can be represented as a linear combination of the resonant states and we got the equations (\ref{eq:eq193}) and (\ref{eq:eq331}). We needed to compute the integral along this contour using Cauchy's residue theorem which gives us the sum over the poles of the function that we are interested in. In both cases, the poles were determined by the zero points $\omega_j$ of the determinant $\det\mathbf{M}(\omega)$. In our case $V_0=0$, the determinant is $\det\mathbf{M}_0(\omega)=-\frac{1}{\pi}$ which has no zeros, hence the residues become zero. The equation we are left with is now from (\ref{eq:eq205})
\begin{align}
\delta(x-x')&=-\lim_{R\to\infty}\int_{C_R}\psi_\omega(x)\psi_\omega(x')\mathrm{d}\omega\label{eq:eq347.9}
\end{align}
which is the completness relation we started with.

As another consequence of the fact that the determinant has no zeros is, that there are no energy eigenvalues, so resonant states can not exist. However, scattering states can exist. Scattering states are states where an in-going wave with amplitude $a$ generates a transmitted and reflected wave. A resonant state is a state where there is no in-going wave $a=0$. Such solutions are non-trivial only if the determinant is equal to zero which in the case when $V_0=0$ is not possible, because $\det\mathbf{M}_0(\omega)=-\frac{1}{\pi}$.

Let us discuss the difference between these two proofs for the square well. The first version of the proof showed us that the integrand in (\ref{eq:eq286}) vanishes if the function $f(x)$ has its compact support confined in the region $x<d$. However, the second version provides a more exact condition. The integrand in (\ref{eq:eq286}) decays pointwise for all $x<d$. This means, that if the compact support of the function $f(x)$ is focused on $x=d$, then the part where $x<d$ will converge, whereas the part $x>d$ will diverge.

\begin{figure}[H]
\centering
\begin{subfigure}{.5\textwidth}
  \centering
  \includegraphics[scale=0.16]{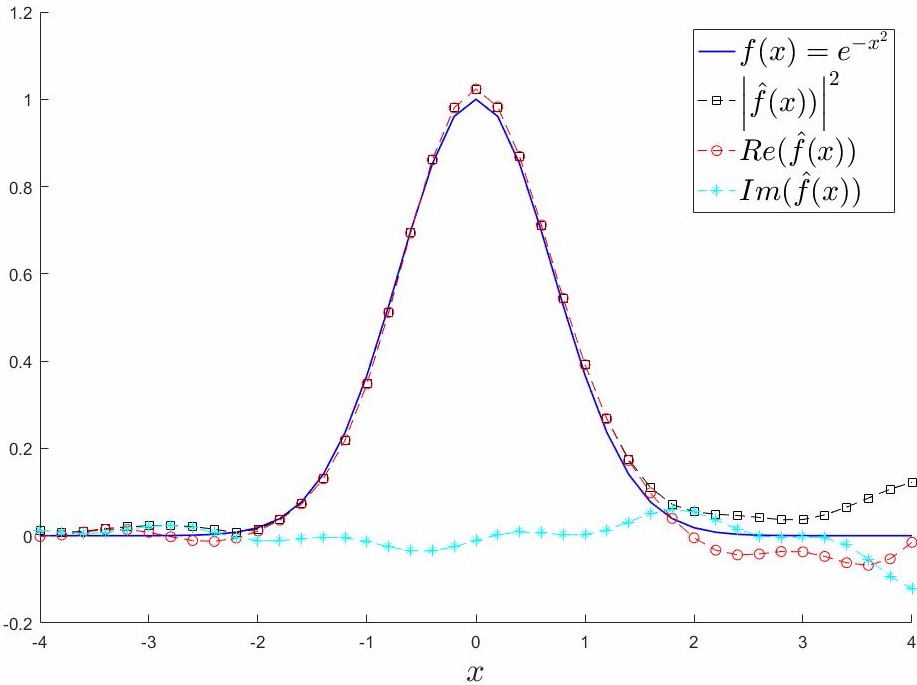}
  \caption{200 terms used in the expansion}
  \label{fig:fig0a}
\end{subfigure}%
\begin{subfigure}{.5\textwidth}
  \centering
  \includegraphics[scale=0.16]{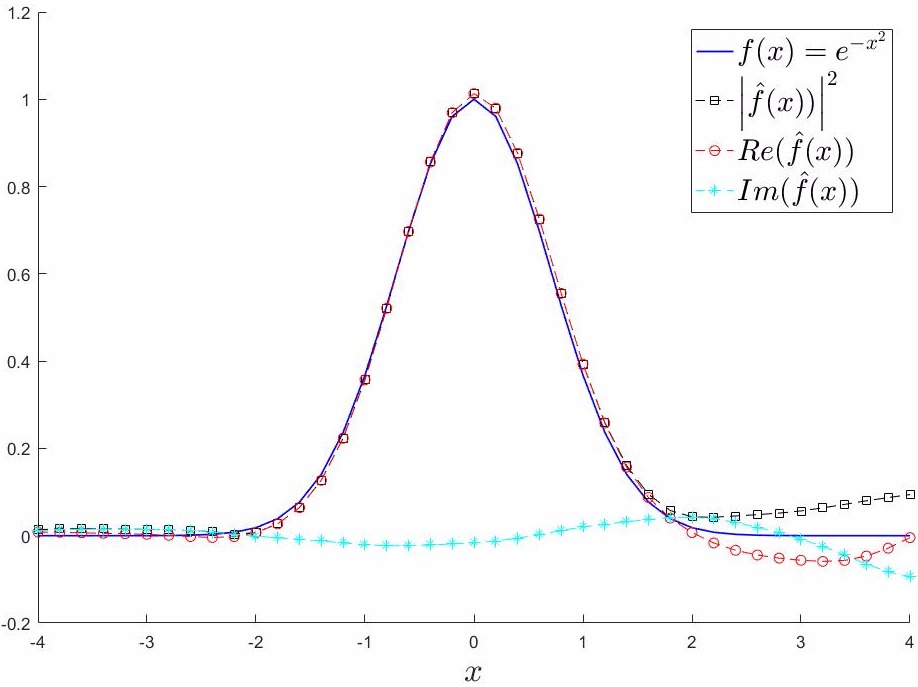}
  \caption{500 terms used in the expansion}
  \label{fig:fig0b}
\end{subfigure}
\caption{The test function $f(x)=e^{-x^2}$ and the corresponding expansion $\hat{f}(x)$ using resonant states from A-series. There have been 200 and 500 resonant states used in this picture. The parameters are $V_0=1,d=4,\varepsilon=0.03$.}
\label{fig:fig0}
\end{figure}

We have done some numerical simulations to test and visualize the obtained results. A particular test-function $f(x)$ was chosen and the expansion coefficients were calculated using resonant states from the A-series. The figure (\ref{fig:fig0}) shows these numerical results where we used a test-function $f(x)=e^{-x^2}$ and computed the first 500 terms from the A-series in the expansion $\hat{f}(x)=\sum_nc_n\psi_{\omega_n}(x)$. The parameters we used were $V_0=1,d=4,\varepsilon=0.03$. The coefficients that are ensuring the continuity of the resonant states at the points $x=-d,d$ used in this example were calculated numerically. To compute the corresponding eigenvalues, we used a Newton method-based algorithm to numerically evaluate them using the determinant of a 4x4 matrix that encodes the continuity conditions. These acquired numerical eigenvalues were fed into the matrix and their eigensystems were obtained. For each energy eigenvalue, we obtained 4 matrix-eigenvalues where one of them was always the smallest, but not 0 since we are dealing only with numerical values. The eigenvector belonging to this eigenvalue then contained the desired coefficients.

The subfigures in figure (\ref{fig:fig0}) show us, that the real part of the approximation is dominating over the imaginary part. The fact, that the number of terms in the expansion increased more than twice, affected the approximation $\hat{f}(x)$ very little. This could be caused by the convergence rate that should be studied in more detail. However, taking a function of a temporal evolution in the system is a test-function would provide a better approximated than a ``common target function'' as we used here. To gain more knowledge about how fast and for what type of functions this expansion converges in the well, we continue with a chapter dedicated to this subject.

\section{Asymptotic series}\label{s5}
In this section we directly investigate the behaviour of the sum
\begin{align}
f(x)=\sum_{n}c_n\psi_{\omega_n}(x)\label{eq348}
\end{align}
for a specific $f(x)$ at the point $x=0$. In particular, the convergence rate is what we are interested in. As we could see earlier, the energy eigenvalues can be categorized into 2 groups. The first is the A-series, which are located in the fourth quadrant in the complex plane and the second is the C-series located in the third quadrant. First, we look at the part of the sum corresponding to the eigenvalues from A-series and after that a brief summing up of the C-series.

We are going to consider resonant states that can be normalized. In order to achieve the normalization, we rely on the technique introduced in \cite{Gyarmati}. This technique uses a complex contour $\mathcal{L}$ on which the resonant states are evaluated. One can see from the form of the resonant states that they decay on the negative axis while on the positive part, where the outgoing part is situated, they show exponential growth. It turns out that on this complex contour the outgoing part decays (see \cite{Kolesik2}) and hence, can be used to normalize the states. This ability to normalize the resonant states is achieved by replacing the space of integration by this contour. The contour $\mathcal{L}$ also supports the orthogonality relation of the states and gives an opportunity to numerically handle the computation of time-dependent Schr\"odinger equation since this domain realizes transparent boundary conditions for an open system. The formula of this contour has the form form
\begin{equation}
\mathcal{L}=z(x)=\left\{
\begin{array}{l l}
x & x<x_c\\
x_c+e^{i\theta}(x-x_c) & x> x_c
\end{array}\right.\label{Aeq0.0}
\end{equation}
which serves its purpose for any chosen contour parameters $x_c>0$ and $0<\theta\leq\frac{\pi}{2}$. In the following, we will consider $\theta=\frac{\pi}{2}$.

To reach the terms in the expansion (\ref{eq348}) with high index number $n$, one must calculate the asymptotic formulas for the energy eigenvalues. The key is to realize there is a correspondence between the index of the eigenvalue and its asymptotic value \cite{Stegun}. This allows us to look deep into the convergence of the expansion and disclose some features of this expansion. Throughout the rest of this paper we use this correspondence which will give rise to new results.

The basis we use in (\ref{eq348}) are the resonant states
\begin{align}
\psi_\omega(x)=\chi\left\{
\begin{array}{l l}
a_1\mathrm{Ai}(y_1(x)) & x<-d\\
a_2\mathrm{Ai}(y_2(x))+a_3\mathrm{Bi}(y_2(x)) & -d< x< d\\
a_4\mathrm{Ci}^+(y_1(x)) & d<x<x_c\\
a_4\mathrm{Ci}^+(\tilde{y}(x)) & x_c<x
\end{array}\right.\label{Aeq1}
\end{align}
where
\begin{align}
y_1(x)&=-2(2\varepsilon)^{-\frac{2}{3}}(\varepsilon x+\omega)\label{Aeq1.1}\\
y_2(x)&=-2(2\varepsilon)^{-\frac{2}{3}}(\varepsilon x+V_0+\omega)\label{Aeq2}\\
\tilde{y}(x)&=-2(2\varepsilon)^{-\frac{2}{3}}\left(i\varepsilon x+\varepsilon x_c(1-i)+\omega\right)\label{Aeq2.0}
\end{align}
and the constants $a_1,\ldots,a_4$ are ensuring the continuity of the states. Let our function we want to expand $f(x)$ be
\begin{align}
f(x)=\left\{\begin{array}{c c}
1 & -a\leq x \leq a\\
0 & \mathrm{otherwise}
\end{array}\right.\label{Aeq2.1}
\end{align}
for any $0<a<d$. We rewrite the expansion into the form
\begin{align}
f(x)=\sum_p\frac{\left(f,\psi_{\omega_p}\right)}{\left(\psi_{\omega_p},\psi_{\omega_p}\right)}\psi_{\omega_p}(x)=\sum_p\frac{b_p(x)}{N_p}\label{Aeq2.2}
\end{align}
where we denoted
\begin{align}
b_p(x)&=\left(f,\psi_{\omega_p}\right)\psi_{\omega_p}(x)\label{Aeq2.2.1}\\
N_p&=\left(\psi_{\omega_p},\psi_{\omega_p}\right)\label{Aeq2.2.2}\\
c_p&=\frac{b_p(0)}{N_p}\label{Aeq2.2.3}
\end{align}
and we are interested in the coefficients $b_p$ of this expansion at the point $x=0$. Using the actual formulas of the resonant state we have
{\medmuskip=0mu
\thinmuskip=0mu
\thickmuskip=0mu
\begin{align}
b_p(0)&=\psi_{\omega_p}(0)\int_{-a}^a\psi_{\omega_p}(x)\mathrm{d}x=\left(a_2\mathrm{Ai}\left( \left.y_2(0)\right|_{\omega=\omega_p}\right)+a_3\mathrm{Bi}\left( \left.y_2(0)\right|_{\omega=\omega_p}\right)\right)\nonumber\\
&\int_{-a}^a\left(a_2\mathrm{Ai}\left( \left.y_2(x)\right|_{\omega=\omega_p}\right)+a_3\mathrm{Bi}\left( \left.y_2(x)\right|_{\omega=\omega_p}\right)\right)\mathrm{d}x\label{Aeq2.3}
\end{align}}
As we mentioned earlier, the correspondence between the index of an eigenvalue and its asymptotic value can be represented through the following formula
\begin{align}
\omega_p\sim\left(\frac{3\pi p}{2}\right)^\frac{2}{3}\label{Aeq2.4}
\end{align}
provided that the index $p$ is reasonably large. This formula is true for A-series eigenvalues. But since we know that the A-series eiganvalues lie in the fourth quadrant and are ``close'' to the real axis, the transformation we use to compute the asymptotic formula for the eigenvalues is
\begin{align}
\omega_p(\xi)\sim\left(\frac{3\pi \xi_p}{2}\right)^\frac{2}{3}\label{Aeq2.5}
\end{align}
where $\xi_p=p+s$ is a complex number consisting of a dominating real part $p$ and a smaller complex number $s$, so the assumption is $s\ll p$. The computations of asymptotic eigenvalues and the coefficients $c_p$ are presented in Appendix \ref{App:AppendixB}.

We found at (\ref{Aeq34}) that $b_p(0)$ is of order 1 and that $N_p\sim p^{\frac{1}{3}}$ which means, that the series (\ref{Aeq2.2}) is diverging at the point $x=0$ at the rate $\sim p^{-\frac{1}{3}}$. The idea how to use these results is to realize that by integrating in (\ref{Aeq2.3}) we got a factor $\frac{1}{g_2(p)}$ which is of order $\sim p^{-\frac{1}{3}}$ (see \ref{Aeq20})). This tells us that each time we integrate, we get this factor in addition. Let us say, we want to compute the expansion coefficients of a ``smoother'' function
\begin{align}
f_1(x)=\left\{
\begin{array}{c c}
1+x & -a<x<0\\
1-x & 0<x<a\\
0 & \mathrm{otherwise}
\end{array}\right.\label{Aeq80.1}
\end{align}
The coefficients $b_p(0)$ in this case will be
\begin{align}
b_p(0)=\psi_{\omega_p}(0)\int_{-a}^af_1(x)\psi_{\omega_p}(x)\mathrm{d}x\label{Aeq81}
\end{align}
We know that the leading order of $\psi_{\omega_p}(0)$ is $p^{-\frac{1}{6}}p^\frac{1}{3}$, where the term $p^{-\frac{1}{6}}$ comes from $z^{-\frac{1}{4}}$ and $p^\frac{1}{3}$ from the sine or cosine (see \ref{Aeq16}, \ref{Aeq30}, \ref{Aeq31})). So (\ref{Aeq81}) can be approximately written as
\begin{align}
b_p(0)\sim p^{-\frac{1}{6}}p^\frac{1}{3}\int_{-a}^af_1(x)p^{-\frac{1}{6}}\left[\sin(g_1(p)+xg_2(p))+\cos(g_1(p)+xg_2(p))\right]\mathrm{d}x\label{Aeq84}
\end{align}
where the terms are explained in (\ref{Aeq19}), (\ref{Aeq20}). Using integration by parts we get
{\medmuskip=-1mu
\thinmuskip=0mu
\thickmuskip=0mu
\begin{align}
b_j(0)&\sim p^{-\frac{1}{6}}p^\frac{1}{3}\nonumber\\
&\int_{-a}^af_1(x)p^{-\frac{1}{6}}\left[a_2\sin\left( g_1(p)+xg_2(p)+\frac{\pi}{4}\right)+a_3\cos\left( g_1(p)+xg_2(p)+\frac{\pi}{4}\right)\right]\mathrm{d}x\nonumber\\
&=f_1(x)\frac{1}{g_2(p)}\left[-a_2\cos\left( g_1(p)+xg_2(p)+\frac{\pi}{4}\right)+a_3\sin\left( g_1(p)+xg_2(p)+\frac{\pi}{4}\right)\right]\Big|_{x=-a}^{x=a}\nonumber\\
&-\frac{1}{g_2(p)}\int_{-a}^af_1'(x)\left[-a_2\cos\left( g_1(p)+xg_2(p)+\frac{\pi}{4}\right)+a_3\sin\left( g_1(p)+xg_2(p)+\frac{\pi}{4}\right)\right]\mathrm{d}x\nonumber\\
&=-\frac{1}{g_2(p)}\nonumber\\
&\int_{-a}^af_1'(x)\left[-a_2\cos\left( g_1(p)+xg_2(p)+\frac{\pi}{4}\right)+a_3\sin\left( g_1(p)+xg_2(p)+\frac{\pi}{4}\right)\right]\mathrm{d}x\label{Aeq85}
\end{align}}
The first half disappeared because the function $f_1(x)$ is 0 at the boundary. Its derivative happens to be
\begin{align}
f_1'(x)=f(x)=\left\{
\begin{array}{c c}
1 & -a<x<0\\
-1 & 0<x<a\\
0 & \mathrm{otherwise}
\end{array}\right.\label{Aeq86}
\end{align}
With this function, the integral we are left with in (\ref{Aeq85}) is similar to (\ref{Aeq84}) up to some constants. Observe that the leading order will not change. So with the variable transformation $y=g_1(p)+xg_2(p)+\frac{\pi}{4}$ in this case we get
{\medmuskip=0mu
\thinmuskip=0mu
\thickmuskip=0mu
\begin{align}
b_p(0)&=-\frac{1}{g_2^2(p)}\left(\int_{g_1(p)-ag_2(p)+\frac{\pi}{4}}^{g_1(p)+\frac{\pi}{4}}\left[a_2\sin(y)+a_3\cos(y)\right]\mathrm{d}y\right.\nonumber\\
&\left.-\int_{g_1(p)+\frac{\pi}{4}}^{g_1(p)+ag_2(p)+\frac{\pi}{4}}\left[a_2\sin(y)+a_3\cos(y)\right]\mathrm{d}y\right)\nonumber\\
&=-\frac{2(\cos(ag_2(p))-1)}{g_2^2(p)}\left[a_2\cos\left(g_1(p)+\frac{\pi}{4}\right)-a_3\sin\left(g_1(p)+\frac{\pi}{4}\right)\right]\label{Aeq86.1}
\end{align}}
and we see that the structure is similar to what we got in (\ref{Aeq23}) with the original function $f(x)$. Since $g_2(p)$ does not contain any leading term of a complex nature, $\cos(ag_2(p))$ is of order 1. With this in mind observe that we gained another term of the form $\frac{1}{g_2(p)}$ after integrating (\ref{Aeq85}). We would like to emphasize that this step is of the most importance. Converting the above expression into simplified version including only the eigenvalue index, the remaining sine and cosine give us $p^\frac{1}{3}$. So in total we get
\begin{align}
b_p(0)&=-\frac{2(\cos(ag_2(p))-1)}{g_2^2(p)}\left[a_2\cos\left(g_1(p)+\frac{\pi}{4}\right)-a_3\sin\left(g_1(p)+\frac{\pi}{4}\right)\right]\nonumber\\
&\sim -\frac{1}{p^\frac{2}{3}}p^\frac{1}{3}\sim p^{-\frac{1}{3}}\label{Aeq87}
\end{align}
By normalizing the coefficients $b_p(0)$ with $N_p$ the final answer is $\sim p^{-\frac{2}{3}}$. This is an algebraic decay, which is not converging. There is however a pattern here. Each time, we would integrate by parts, we would get an extra term of the form $\frac{1}{g_2(p)}\sim p^{-\frac{1}{3}}$. Hence, we can deduce, that if we started with a function whose second derivative is our step function $f(x)$, we would get an additional term of order $\sim p^{-\frac{1}{3}}$. The total order of the coefficients would become $p^{-1}$. This rate is still not sufficient to converge, but the next step would result into $p^{-\frac{4}{3}}$ that already converges. To find analytically the series of such functions $f_i(x)$ such that by gradual derivation we end up with a step function similar to (\ref{Aeq86}) and all the "between" functions satisfy the desired conditions is not easy. One example of such a sequence of functions are
{\medmuskip=0mu
\thinmuskip=0mu
\thickmuskip=0mu
\begin{align}
f_0(x)&=\left\{
\begin{array}{c c}
2 & -a<x<0\\
-2 & 0<x<a\\
0 & \mathrm{otherwise}
\end{array}\right.
& &f_1(x)=\left\{
\begin{array}{c c}
a+2x & -a<x<0\\
a-2x & 0<x<a\\
0 & \mathrm{otherwise}
\end{array}\right.\label{Aeq88}\\
f_2(x)&=\left\{
\begin{array}{c c}
ax+x^2 & -a<x<0\\
ax-x^2 & 0<x<a\\
0 & \mathrm{otherwise}
\end{array}\right.
& &f_3(x)=\left\{
\begin{array}{c c}
-\frac{a^3}{6}+\frac{ax^2}{2}+\frac{x^3}{3} & -a<x<0\\
-\frac{a^3}{6}+\frac{ax^2}{2}-\frac{x^3}{3} & 0<x<a\\
0 & \mathrm{otherwise}
\end{array}\right.\label{Aeq90}
\end{align}}
The next step in this series would be a function $f_4(x)$ that can not exist such that it satisfies all the conditions $\lim_{x\to 0^-}f_4(x)=\lim_{x\to 0^+}f_4(x), f_4(-a)=f_4(a)=0$. On the other hand we see that by the argumentation above we have that the expasion (\ref{Aeq2.2}) diverges at $x=0$ for $f_{0,1,2}(x)$, but it converges algebraically for $f_3(x)$ at the rate $\sim p^{-\frac{4}{3}}$. Let us confirm the validity of the computation of asymptotic expression in Appendix \ref{App:AppendixB}.

In figure \ref{fig:fig2} we see how accurate the asymptotic expression are. They are showing two set of points, namely the exact and approximated ratios $\left|\frac{b_p(0)}{N_p}\right|=\left|\frac{b_p}{N_p}\right|$ for different ranges of $p$. For smaller values of $p$, the approximation seems to fluctuate around the exact values and does not seem to catch up the swinging or the breaks. As we move forward to higher values, we get a reasonably satisfying approximations. In the last subfigure, the swinging of the values (both exact and approximations) can appear to be lost, but note, that the range of the variable $p$ is the largest here $\left(\right.$around $\left.10^{11}\right)$ so the values look more scattered. However, our model is following these irregularities very satisfactorily.

\begin{figure}[t]
\centering
\begin{subfigure}{.5\textwidth}
  \centering
  \includegraphics[scale=0.16]{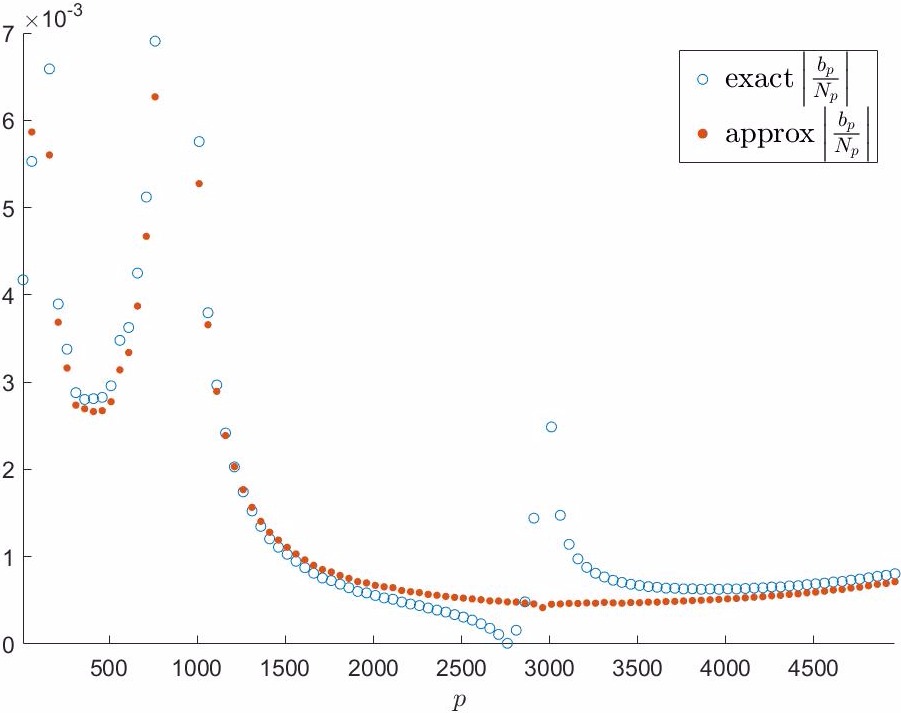}
  \caption{$ $}
  \label{fig:fig2a}
\end{subfigure}%
\begin{subfigure}{.5\textwidth}
  \centering
  \includegraphics[scale=0.16]{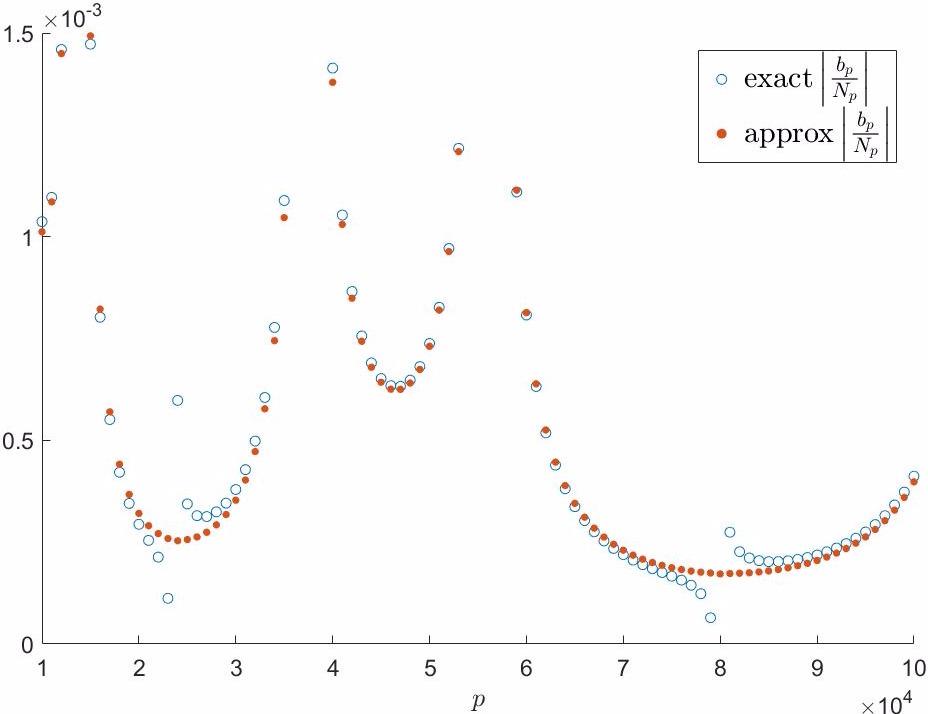}
  \caption{$ $}
  \label{fig:fig2b}
\end{subfigure}%

\centering
\begin{subfigure}{.5\textwidth}
  \centering
  \includegraphics[scale=0.16]{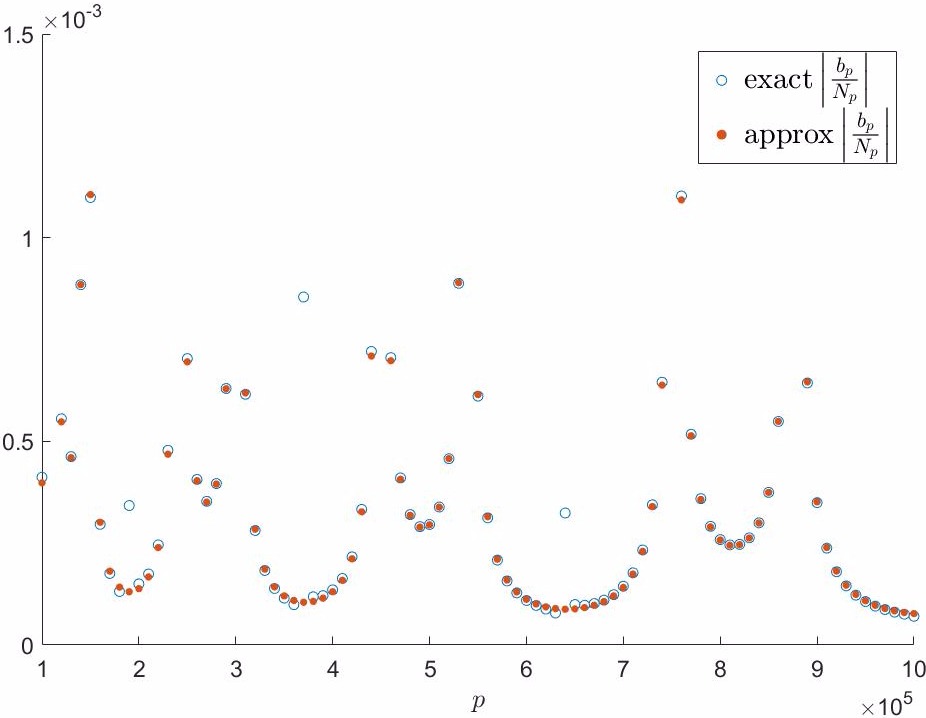}
  \caption{$ $}
  \label{fig:fig2c}
\end{subfigure}%
\begin{subfigure}{.5\textwidth}
  \centering
  \includegraphics[scale=0.16]{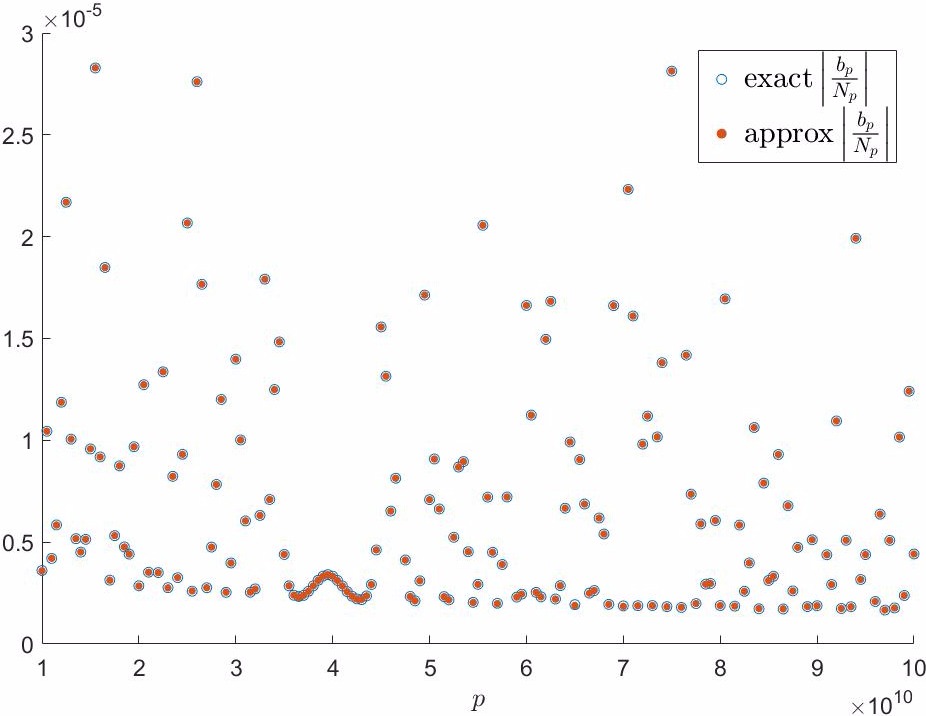}
  \caption{$ $}
  \label{fig:fig2d}
\end{subfigure}
\caption{The absolute value of the exact and asymptotic ratios $\left|\frac{b_p}{N_p}\right|$ for different ranges of eigenvalue index $p$. The precision of the approximations is improving with higher values of $p$ as expected. The anticipated order of the ratios is $\sim p^{-\frac{1}{3}}$ up to 1 order of magnitude since there are some factors too. This can be also be seen that is obeyed.}
\label{fig:fig2}
\end{figure}

When it comes to the continuity coefficients $a_1,\ldots,a_4$ (\ref{Aeq1}), one might ask whether they also depend on $p$ or not. But since this vector $(a_1,\ldots,a_4)$ is an eigenvector belonging to the eigenvalue 0, it can be multiplied by any constant or normalized such that its length will be less than 1. In this way, these coefficients  will not have any effect on the result. However, in the figures above, they were not normalized but despite of that we got a very satisfying match with the computations. A detailed analysis is done in the appendix chapter \ref{Aai}.

Now, how can this be used to show whether the series is convergent for a more general function $f(x)$ and what are the assumptions we are requesting from $f(x)$ to apply these results? Let us first of all assume we have a continuous function $f(x)$ with a compact support in the region $(-d,d)$. We perform the integration by parts three times with $b_p$ as before and we get
\begin{align}
\frac{b_p(0)}{N_p}\sim p^{-\frac{1}{3}}\frac{1}{g_2^3(p)}\int_{-d}^d f^{(3)}(x)\psi_p^{(-1)}(x)\mathrm{d}x\sim p^{-\frac{4}{3}}\int_{-d}^d f^{(3)}(x)\psi_p^{(-1)}(x)\mathrm{d}x\label{Aeq92}
\end{align}
where $\psi_p^{(-1)}(x)$ is the primitive of the function $\psi_p(x)$. For big $p$ the function $\psi_p(x)$ is a sum of a sine and cosine, so integrating twice we get the same back. If we assume in addition, that the third derivative of $f(x)$ has a compact support, it is continuous and that $\left|f^{(3)}\right|\leq M$, for some real number $M<\infty$, then we can estimate
\begin{align}
\left|\frac{b_p(0)}{N_p}\right|\lesssim p^{-\frac{4}{3}}M\int_{-d}^d \psi_p^{(-1)}(x)\mathrm{d}x\lesssim p^{-\frac{4}{3}}M\frac{1}{g_2(p)}p^\frac{1}{3}\lesssim p^{-\frac{4}{3}}M\label{Aeq93}
\end{align}
This means, we can extend the space of such functions, for which we get convergence by demanding that $\left|f^{(n)}\right|\leq M$ in general. Then the coefficients in the series can be estimated as $\left|\frac{b_p(0)}{N_p}\right|\lesssim p^{-\frac{n+1}{3}}M$.

Until now, we talked only about that part of the expansion (\ref{eq348}) that belongs to the A-series eigenvalues. As we know, besides those there also are resonant states belonging to C-series. However, we found out in Appendix \ref{App:AppendixC} that these resonant states decay exponentially so they do not have any considerable effect on the expansion (\ref{eq348}).

\section[Conclusion]{Conclusion}
\begin{figure}[t]
\captionsetup{width=0.85\textwidth}
\begin{center}
\includegraphics[scale=0.3]{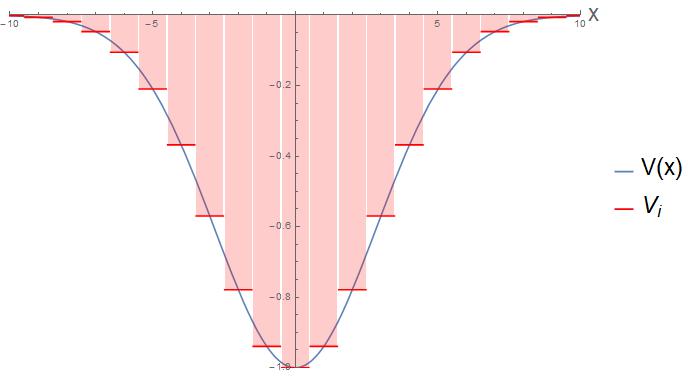}
\end{center}
\caption{A continuous potential $V(x)$ with a suitable disretization $V_i$.\label{fig:fig21}}
\end{figure}

We have shown that a convergence of the resonant state expansion is present for the region $x<d$ and it does not depend on the depth of the well $V_0$. We also discussed the option when $V_0=0$, where we concluded that it is not possible to get any resonant states in this case, so one must have $V_0>0$. In chapter 5 we have confirmed this statement and moreover we have shown how smooth must a function be so that its expansion becomes convergent. This conclusion gives rise to an interesting thought. 
Let us have two such square wells next to each other with different depths. Using the same methods, similar results could be shown. If we would proceed further, we could consider a continuous potential well $V(x)$ over the whole space which decays at $\pm\infty$ approximated with narrow square wells with suitable depths $V_i$ and widths $\Delta x$. The outermost square wells would be then more and more shallow such that $0<V_i\ll 1$. Without our result one would think, that we would loose the convergence in those places. Our result however tells us, it is not necessarily the case. On the other side we do not know if this result can be generalized to this case when one has many square wells one after each other, therefore it is a conjecture as we stated in the introduction. This statement could be supported only by explicit computations of more examples. If our result turns out to be true for these more complicated examples, one could think of constructing a formal proof of this statement. It would be interesting to see the properties of those resonant states obtained by eventually taking the limit in that potential discretization $\lim_{i\to\infty,\Delta x\to0}V_i=V(x)$. We can see an example of this approximation on Figure \ref{fig:fig21}.
\noindent
As we mentioned in the introduction, this idea of a continuous potential without a compact support is considerably an important conjecture. It can be formulated for example as a statement which says that with a potential we just described, we would get a convergence for all $x$. It is a very important result of this work since the solution to that discovery would reveal other interesting facts as well.

\newpage

\renewcommand{\thesection}{\Alph{section}}
\numberwithin{equation}{section}
\section{}\label{App:Appendix}
This Appendix gives details on computing the integral $\Upsilon_{\xi}^\pm(\omega,\omega')$ in (\ref{eq:eq326.1}). Observe that we can write the resonant states as
\begin{align}
\psi_\omega(s)&=\Psi\left(y_{1,2}(s,\omega)\right)\label{eq:eq316}\\
\psi_{\omega'\pm i\xi}^\pm(s)&=\Psi^\pm\left(y_{1,2}(s,\omega'\pm i\xi)\right)\label{eq:eq317}
\end{align}
where the functions $\Psi$, $\Psi^\pm$ are linear combinations of Airy functions in general on intervals $(-\infty,-d)$ and $(d,\infty)$ and $y_{1,2}(x,\omega)$ are the variable transformations written with $\omega$ in the argument which one can write as
\begin{align}
y_{1,2}(s,\omega)&=-2\mu\varepsilon\left(s_{1,2}(s)+\frac{\omega}{\varepsilon}\right)\label{eq:eq318}
\end{align}
where we defined
\begin{align}
s_i(s)=\left\{
\begin{array}{l l}
s & i=1\\
s+\frac{V_0}{\varepsilon} & i=2
\end{array}\right.\label{eq:eq319}
\end{align}
For the next step we can use the formula (3.53) in \cite{Valee} which says that if $A(s)$ and $B(s)$ are any linear combinations of Airy functions then
{\medmuskip=0mu
\thinmuskip=0mu
\thickmuskip=0mu
\begin{align}
\int A(\rho(s+\beta_1))&B(\rho(s+\beta_2))\mathrm{d}s\nonumber\\
&=\frac{1}{\rho^2(\beta_1-\beta_2)}\left[A'(\rho(s+\beta_1))B(\rho(s+\beta_2))-A(\rho(s+\beta_1))B'(\rho(s+\beta_2))\right]\label{eq:eq320}
\end{align}}
Using this in (\ref{eq:eq313}) and the fact that the resonant states $\psi_\omega(s)$ decay at $\pm\infty$, letting $A(s)=\Psi\left(y_{1,2}(s,\omega)\right),B(s)=\Psi^\pm\left(y_{1,2}(s,\omega'\pm i\xi)\right)$ and defying $\rho=2\mu\varepsilon$ we get
{\medmuskip=0mu
\thinmuskip=0mu
\thickmuskip=0mu
\begin{align}
\Upsilon_{\xi}^\pm(\omega,\omega')&=\int_{-\infty}^{-d}\Psi\left(-\rho\left(s_1(s)+\frac{\omega}{\varepsilon}\right)\right)\Psi^\pm\left(-\rho\left(s_1(s)+\frac{\omega'\pm i\xi}{\varepsilon}\right)\right)\mathrm{d}s\nonumber\\
&+\int_{d}^{\infty}\Psi\left(-\rho\left(s_1(s)+\frac{\omega}{\varepsilon}\right)\right)\Psi^\pm\left(-\rho\left( s_1(s)+\frac{\omega'\pm i\xi}{\varepsilon}\right)\right)\mathrm{d}s\nonumber\\
&+\int_{-d}^{d}\Psi\left(-\rho\left(s_2(s)+\frac{\omega}{\varepsilon}\right)\right)\Psi^\pm\left(-\rho\left( s_2(s)+\frac{\omega'\pm i\xi}{\varepsilon}\right)\right)\mathrm{d}s\nonumber\\
&=\frac{\varepsilon}{\rho^2(\omega-\omega'\mp i\xi)}\nonumber\\
&\left.\left[\Psi'\left(y_{1}(s,\omega)\right)\Psi^\pm\left(y_{1}(s,\omega'\pm i\xi)\right)-\Psi\left(y_{1}(s,\omega)\right)\Psi'^\pm\left(y_{1}(s,\omega'\pm i\xi)\right)\right]\right|_{s=d^+}^{s=-d^-}\nonumber\\
&+\frac{\varepsilon}{\rho^2(\omega-\omega'\mp i\xi)}\nonumber\\
&\left.\left[\Psi'\left(y_{2}(s,\omega)\right)\Psi^\pm\left(y_{2}(s,\omega'\pm i\xi)\right)-\Psi\left(y_{2}(s,\omega)\right)\Psi'^\pm\left(y_{2}(s,\omega'\pm i\xi)\right)\right]\right|_{s=-d^+}^{s=d^-}\label{eq:eq321}
\end{align}}
We use the continuity of the derivative at $x=-d,d$ for $\Psi(s)$, the continuity of the derivative at $x=-d$ for $\Psi^\pm(s)$ (\ref{eq:eq316}), (\ref{eq:eq317}) and from (\ref{eq:eq321}) we get
{\medmuskip=0mu
\thinmuskip=0mu
\thickmuskip=0mu
\begin{align}
\Upsilon_{\xi}^\pm(\omega,\omega')&=\frac{\varepsilon}{\rho^2(\omega-\omega'\mp i\xi)}\nonumber\\
&\left.\left[\Psi'\left(y_{1}(s,\omega)\right)\Psi^\pm\left(y_{1}(s,\omega'\pm i\xi)\right)-\Psi\left(y_{1}(s,\omega)\right)\Psi'^\pm\left(y_{1}(s,\omega'\pm i\xi)\right)\right]\right|_{s-d^+}^{s=-d^-}\nonumber\\
&+\frac{\varepsilon}{\rho^2(\omega-\omega'\mp i\xi)}\nonumber\\
&\left.\left[\Psi'\left(y_{2}(s,\omega)\right)\Psi^\pm\left(y_{2}(s,\omega'\pm i\xi)\right)-\Psi\left(y_{2}(s,\omega)\right)\Psi'^\pm\left(y_{2}(s,\omega'\pm i\xi)\right)\right]\right|_{s=-d^+}^{s=d^-}\nonumber\\
&=\frac{\varepsilon}{\rho^2(\omega-\omega'\mp i\xi)}\left[\Psi'\left(y_{1}(-d^-,\omega)\right)\Psi^\pm\left(y_{1}(-d^-,\omega'\pm i\xi)\right)\right.\nonumber\\
&-\Psi\left(y_{1}(-d^-,\omega)\right)\Psi'^\pm\left(y_{1}(-d^-,\omega'\pm i\xi)\right)\nonumber\\
&-\Psi'\left(y_{1}(d^+,\omega)\right)\Psi^\pm\left(y_{1}(d^+,\omega'\pm i\xi)\right)\nonumber\\
&\left.+\Psi\left(y_{1}(d^+,\omega)\right)\Psi'^\pm\left(y_{1}(d^+,\omega'\pm i\xi)\right)+\Psi'\left(y_{2}(d^-,\omega)\right)\Psi^\pm\left(y_{2}(d^-,\omega'\pm i\xi)\right)\right.\nonumber\\
&\left.-\Psi\left(y_{2}(d^-,\omega)\right)\Psi'^\pm\left(y_{2}(d^-,\omega'\pm i\xi)\right)\right.\nonumber\\
&-\Psi'\left(y_{2}(-d^+,\omega)\right)\Psi^\pm\left(y_{2}(-d^+,\omega'\pm i\xi)\right)\nonumber\\
&\left.+\Psi\left(y_{2}(-d^+,\omega)\right)\Psi'^\pm\left(y_{2}(-d^+,\omega'\pm i\xi)\right)\right]\nonumber\\
&=\frac{\varepsilon}{\rho^2(\omega-\omega'\mp i\xi)}\nonumber\\
&\left[\Psi\left(y_{1}(d^+,\omega)\right)\Psi'^\pm\left(y_{1}(d^+,\omega'\pm i\xi)\right)-\Psi\left(y_{2}(d^-,\omega)\right)\Psi'^\pm\left(y_{2}(d^-,\omega'\pm i\xi)\right)\right]\nonumber\\
&=\frac{\varepsilon\psi_\omega(d^+)}{\rho^2(\omega-\omega'\mp i\xi)}\left[\Psi'^\pm\left(y_{1}(d^+,\omega')\right)-\Psi'^\pm\left(y_{2}(d^-,\omega')\right)\right]\label{eq:eq322}
\end{align}}
where in the last line we have put $\xi=0$ in all terms without any loss of generality. From our explicit expressions for the states $\psi_\omega^\pm(x)$ in (\ref{eq:eq303}), (\ref{eq:eq304}) and its derivative at $x=d$ we obtain
\begin{align}
&\Psi'^\pm\left(y_{1}(d^+,\omega')\right)-\Psi'^\pm\left(y_{2}(d^-,\omega')\right)\nonumber\\
&=\pm\chi i\left(\frac{\det\mathbf{M}^\mp(\omega')}{\det\mathbf{M}^\pm(\omega')}\right)^{\frac{1}{2}}\mathrm{Ci'}^\pm(y_1(d^+))\mp \chi i\left(\frac{\det\mathbf{M}^\mp(\omega')}{\det\mathbf{M}^\pm(\omega')}\right)^{\frac{1}{2}}\mathrm{Ci}^\pm(y_1(d^-,\omega')))\nonumber\\
&\frac{\left[(B_1'A_0-B_1A_0')\mathrm{Ai'}(y_2(d^-,\omega'))+(A_1A_0'-A_1'A_0)\mathrm{Bi'}(y_2(d^-,\omega'))\right]}{p(\omega')}\nonumber\\
&=\pm \chi i\left(\frac{\det\mathbf{M}^\mp(\omega')}{\det\mathbf{M}^\pm(\omega')}\right)^{\frac{1}{2}}\mathrm{Ci'}^\pm(y_1(d,\omega'))\mp\chi i\left(\frac{\det\mathbf{M}^\mp(\omega')}{\det\mathbf{M}^\pm(\omega')}\right)^{\frac{1}{2}}\mathrm{Ci}^\pm(y_1(d,\omega')))\nonumber\\
&\frac{(B_1'A_0-B_1A_0')A_2'+(A_1A_0'-A_1'A_0)B_2'}{(B_1'A_0-B_1A_0')A_2+(A_1A_0'-A_1'A_0)B_2}\label{eq:eq323}
\end{align}
where we denoted $\det\mathbf{M}^+(\omega')=\det\mathbf{M}(\omega')$ and $\det\mathbf{M}^-(\omega')=\overline{\det\mathbf{M}}(\omega')$ and\\ every quantity $A,B,C$ contains $\omega'$. Let us look at the determinant $\det\mathbf{M}(\omega)$:
\begin{align}
\det\mathbf{M}(\omega)&=(A_0A'_1-A'_0A_1)(B_2C'_3-B'_2C_3)-(A_0B'_1-A'_0B_1)(A_2C'_3-A'_2C_3)\nonumber\\
&=(A_0A'_1-A'_0A_1)B_2C'_3-(A_0A'_1-A'_0A_1)B'_2C_3\nonumber\\
&-(A_0B'_1-A'_0B_1)A_2C'_3+(A_0B'_1-A'_0B_1)A'_2C_3\nonumber\\
&=C_3\left[(A_0B'_1-A'_0B_1)A'_2+(A'_0A_1-A_0A'_1)B'_2\right]\nonumber\\
&-C_3'\left[(A_0B'_1-A'_0B_1)A_2+(A'_0A_1-A_0A'_1)B_2\right]\label{eq:eq324}
\end{align}
Using this observation and $\mathrm{Ci}^\pm(y_1(d,\omega'))=C^\pm_3$, where $C^+_3=C_3$ and $C^-_3=D_3$, (\ref{eq:eq323}) becomes
{\medmuskip=-1mu
\thinmuskip=0mu
\thickmuskip=0mu
\begin{align}
\Psi'^\pm&\left(y_{1}(d^+,\omega')\right)-\Psi'^\pm\left(y_{2}(d^-,\omega')\right)=\pm \chi i\left(\frac{\det\mathbf{M}^\mp(\omega')}{\det\mathbf{M}^\pm(\omega')}\right)^{\frac{1}{2}}C'^\pm_3\mp \chi i\left(\frac{\det\mathbf{M}^\mp(\omega')}{\det\mathbf{M}^\pm(\omega')}\right)^{\frac{1}{2}}\nonumber\\
&\frac{\det\mathbf{M}^\pm(\omega')+C'^\pm_3\left[(B_1'A_0-B_1A_0')A_2+(A_1A_0'-A_1'A_0)B_2\right]}{(B_1'A_0-B_1A_0')A_2+(A_1A_0'-A_1'A_0)B_2}\nonumber\\
&=\pm \chi i\left(\frac{\det\mathbf{M}^\mp(\omega')}{\det\mathbf{M}^\pm(\omega')}\right)^{\frac{1}{2}}C'^\pm_3\mp \chi i\left(\frac{\det\mathbf{M}^\mp(\omega')}{\det\mathbf{M}^\pm(\omega')}\right)^{\frac{1}{2}}C'^\pm_3\mp \chi i\frac{|\det\mathbf{M}(\omega')|}{p(\omega')}\nonumber\\
&=\mp \chi i\frac{|\det\mathbf{M}(\omega')|}{p(\omega')}\label{eq:eq325}
\end{align}}
where $p(\omega')=(B_1'A_0-B_1A_0')A_2+(A_1A_0'-A_1'A_0)B_2$. Hence we can write (\ref{eq:eq322}) using (\ref{eq:eq325}) as
\begin{align}
\Upsilon_{\xi}^\pm(\omega,\omega')&=\frac{\chi\varepsilon}{\rho^2}\frac{\psi_\omega(d^+)}{\omega-\omega'\mp i\xi}\left(\mp i\frac{|\det\mathbf{M}(\omega')|}{p(\omega')}\right)\label{eq:eq326}
\end{align}

\section{}\label{App:AppendixB}
In this Appendix we find the asymptotic expression for the A-series coefficients $c_p$ in the sum(\ref{eq348}) at the point $x=0$, where
\begin{align}
c_p=\frac{b_p(0)}{N_p}\label{Aeq0}
\end{align}

We are going to transform the problem into a $\xi$-plane, where we find the asymptotic determinant and solve it to get the eigenvalues in this $\xi$-plane. From \cite{Stegun} we have the relation between the zero of $\mathrm{Ai}(x)$ and its index $p$. Realizing, that the position of A-series eigenvalues are approximately the same as the zeros of $\mathrm{Ai}(x)$, we can assume the relation (\ref{Aeq2.4}). Since the A-series energies are shifted downwards from the real axis, we assume the index $p$ to be complex $\xi_p=p+s$, where $s$ is a complex correction to the zero and $|s|\ll p$. Then we choose the transformation from the $\omega$-plane into $\xi$-plane to be
\begin{align}
\omega(\xi)=\gamma^{-1}\left(\frac{3\pi}{2}\xi\right)^\frac{2}{3}\label{Aeq3}
\end{align}
where we denoted
\begin{align}
\gamma=2(2\varepsilon)^{-\frac{2}{3}}\label{Aeq4}
\end{align}
The role of the constant $\gamma^{-1}$ will be clear later. In the next subsection in this appendix we compute the eigenvalues from the determinant.

\subsection{Asymptotic A-series determinant}
The matrix $\mathbf{M}$ that contains the continuity conditions for the resonant state has its determinant of the form
\begin{align}
\det\mathbf{M}(\omega)=(A_0A'_1-A'_0A_1)(B_2C'_3-B'_2C_3)-(A_0B'_1-A'_0B_1)(A_2C'_3-A'_2C_3)\label{Aeq136}
\end{align}
together with
\begin{align}
\begin{array}{l l}
A_0=\mathrm{Ai}\left(-2\mu\left(-d\varepsilon+\omega\right)\right) & A_1=\mathrm{Ai}\left(-2\mu\left(-d\varepsilon+\omega+V_0\right)\right)\\
B_1=\mathrm{Bi}\left(-2\mu\left(-d\varepsilon+\omega+V_0\right)\right) & C_3=\mathrm{Ci}^{+}\left(-2\mu\left(d\varepsilon+\omega\right)\right)\\
A_2=\mathrm{Ai}\left(-2\mu\left(d\varepsilon+\omega+V_0\right)\right) & B_2=\mathrm{Bi}\left(-2\mu\left(d\varepsilon+\omega+V_0\right)\right) 
\end{array}\label{Aeq137}
\end{align}
where $\mu=(2\varepsilon)^{-\frac{2}{3}}$. We denote the arguments as
\begin{align}
y_1(x)&=-2\mu\left(x\varepsilon+\omega\right)\label{Aeq138}\\
y_2(x)&=-2\mu\left(x\varepsilon+\omega+V_0\right)\label{Aeq139}
\end{align}
The asymptotic behaviours of Airy functions according to \cite{Stegun} are
{\medmuskip=-1mu
\thinmuskip=-1mu
\thickmuskip=-1mu
\begin{align}
\mathrm{Ai}(-z)&\approx\frac{1}{2i\sqrt{\pi}}z^{-\frac{1}{4}}\left(e^{i\left(\zeta+\frac{\pi}{4}\right)}-e^{-i\left(\zeta+\frac{\pi}{4}\right)}\right)\nonumber\\
\mathrm{Ai'}(-z)&\approx-\frac{1}{2\sqrt{\pi}}z^{\frac{1}{4}}\left(e^{i\left(\zeta+\frac{\pi}{4}\right)}+e^{-i\left(\zeta+\frac{\pi}{4}\right)}\right)\label{Aeq140}\\
\mathrm{Bi}(-z)&\approx\frac{1}{2\sqrt{\pi}}z^{-\frac{1}{4}}\left(e^{i\left(\zeta+\frac{\pi}{4}\right)}+e^{-i\left(\zeta+\frac{\pi}{4}\right)}\right)\nonumber\\
\mathrm{Bi'}(-z)&\approx\frac{1}{2i\sqrt{\pi}}z^{\frac{1}{4}}\left(e^{i\left(\zeta+\frac{\pi}{4}\right)}-e^{-i\left(\zeta+\frac{\pi}{4}\right)}\right)\label{Aeq141}\\
\mathrm{Ci}^+(-z)&\approx\pi^{-\frac{1}{2}}z^{-\frac{1}{4}}e^{i\left(\zeta+\frac{\pi}{4}\right)}\nonumber\\
\mathrm{Ci'}^+(-z)&\approx-i\pi^{-\frac{1}{2}}z^{\frac{1}{4}}e^{i\left(\zeta+\frac{\pi}{4}\right)}\label{Aeq142}
\end{align}}
where $\zeta=\frac{2}{3}z^\frac{3}{2}$. The choice of these functions depends of the complex argument of the input, that is in the region $|\arg(z)|<\frac{2\pi}{3}$. Next, we define new quantities and make simplifications with respect to the asymptotic behaviour of the functions. Define
\begin{align}
z_{1,2}(x)=-y_{1,2}(x)\label{Aeq143}
\end{align}
We evaluate these at the points $x=-d,d$ and consequently raise to powers of $\frac{1}{4}$ and $\frac{3}{2}$ so we compute these up front using Taylor expansion. Defying $\beta=\gamma\varepsilon d$ we have
\begin{align}
\zeta^{\pm}_1&=\frac{2}{3}(z_1(\pm d))^\frac{3}{2}=\frac{2}{3}(\pm \beta+\gamma\omega)^\frac{3}{2}=\frac{2}{3}(\gamma\omega)^\frac{3}{2}\left(\pm\frac{\beta}{\gamma\omega}+1\right)^\frac{3}{2}\nonumber\\
&\approx\frac{2}{3}(\gamma\omega)^\frac{3}{2}\left(1\pm\frac{3\beta}{2\gamma\omega}\right)=\frac{2}{3}(\gamma\omega)^\frac{3}{2}\pm\beta(\gamma\omega)^\frac{1}{2}\label{Aeq144}\\
\zeta^\pm_2&=\frac{2}{3}(z_2(\pm d))^\frac{3}{2}=\frac{2}{3}(\pm \beta+\gamma V_0+\gamma\omega)^\frac{3}{2}=\frac{2}{3}(\gamma\omega)^\frac{3}{2}\left(\frac{\gamma V_0\pm\beta}{\gamma\omega}+1\right)^\frac{3}{2}\nonumber\\
&\approx\frac{2}{3}(\gamma\omega)^\frac{3}{2}\left(1+\frac{3(\gamma V_0\pm\beta)}{2\gamma\omega}\right)=\frac{2}{3}(\gamma\omega)^\frac{3}{2}+(\gamma V_0\pm\beta)(\gamma\omega)^\frac{1}{2}\label{Aeq145}
\end{align}
Given the structure of the determinant (\ref{Aeq136}) we see, that it contains products of Airy functions and their derivatives. This tells us that we should expect quantities of the form $z_{1,2}(\pm d)^{\pm\frac{1}{4}}z_{1,2}(\pm d)^{\mp\frac{1}{4}}$. We therefore compute
\begin{align}
z_1(\pm d)^{-\frac{1}{4}}z_2(\pm d)^{\frac{1}{4}}&=(\pm\beta+\gamma\omega)^{-\frac{1}{4}}\left(\gamma V_0\pm\beta+\gamma\omega\right)^{\frac{1}{4}}\nonumber\\
&=(\gamma\omega)^{-\frac{1}{4}}\left(1\pm\frac{\beta}{\gamma\omega}\right)^{-\frac{1}{4}}(\gamma\omega)^{\frac{1}{4}}\left(1+\frac{\gamma V_0\pm\beta}{\gamma\omega}\right)^{\frac{1}{4}}\nonumber\\
&=\left(1\mp\frac{\beta}{4\gamma\omega}\right)\left(1+\frac{\gamma V_0\pm\beta}{4\gamma\omega}\right)\nonumber\\
&=1+\frac{\gamma V_0\pm\beta}{4\gamma\omega}\mp\frac{\beta}{4\gamma\omega}\mp\frac{\beta(\gamma V_0\pm\beta)}{16(\gamma\omega)^2}\approx1+\frac{\gamma V_0}{4\gamma\omega}=1+\frac{V_0}{4\omega}\label{Aeq146}\\
z_1(\pm d)^{\frac{1}{4}}z_2(\pm d)^{-\frac{1}{4}}&=(\pm\beta+\gamma\omega)^{\frac{1}{4}}(\gamma V_0\pm\beta+\gamma\omega)^{-\frac{1}{4}}\nonumber\\
&=(\gamma\omega)^{\frac{1}{4}}\left(1\pm\frac{\beta}{\gamma\omega}\right)^{\frac{1}{4}}(\gamma\omega)^{-\frac{1}{4}}\left(1+\frac{\gamma V_0\pm\beta}{\gamma\omega}\right)^{-\frac{1}{4}}\nonumber\\
&=\left(1\pm\frac{\beta}{4\gamma\omega}\right)\left(1-\frac{\gamma V_0\pm\beta}{4\gamma\omega}\right)\nonumber\\
&=1-\frac{\gamma V_0\pm\beta}{4\gamma\omega}\pm\frac{\beta}{4\gamma\omega}\mp\frac{\beta(\gamma V_0\pm\beta)}{16(\gamma\omega)^2}\approx1-\frac{\gamma V_0}{4\gamma\omega}=1-\frac{V_0}{4\omega}\label{Aeq147}
\end{align}
The next step is to put all these pieces together according to (\ref{Aeq136}) and obtain the asymptotic form of the determinant in the $\omega$-plane. This step is straightforward and the final result is
\begin{align}
\det\mathbf{M}(\omega)&\approx-\frac{1}{\pi^2}+\frac{V_0}{4i\pi^2\omega}e^{i\left(\frac{4}{3}(\gamma\omega)^\frac{3}{2}+2\beta(\gamma\omega)^\frac{1}{2}\right)}-\frac{V_0}{4i\pi^2\omega}e^{i\left(\frac{4}{3}(\gamma\omega)^\frac{3}{2}-2\beta(\gamma\omega)^\frac{1}{2}\right)}\nonumber\\
&=-\frac{1}{\pi^2}+\frac{V_0}{2\pi^2\omega}e^{i\frac{4}{3}(\gamma\omega)^\frac{3}{2}}\sin\left(2\beta(\gamma\omega)^\frac{1}{2}\right)\label{Aeq148}
\end{align}
We turn now our attention back to the transformation (\ref{Aeq3}) and use it in (\ref{Aeq148}). Determinant of the continuity matrix in $\xi$-plane is then
\begin{align}
\det\mathbf{M}\left(\xi\right)=-\frac{1}{\pi^2}+c_1\xi^{-\frac{2}{3}}e^{i2\pi\xi}\sin\left(c_3\xi^\frac{1}{3}\right)\label{Aeq5}
\end{align}
where
\begin{align}
c_1&=\frac{V_0\gamma}{2^{\frac{1}{3}}3^\frac{2}{3}\pi^\frac{8}{3}}\label{Aeq6}\\
c_3&=2d(3\pi\varepsilon)^\frac{1}{3}\label{Aeq7}
\end{align}
The determinant is set to be zero and the equation solved for $\xi=\xi_p=p+s$, where $p\in\mathbb{N},s\in\mathbb{C}$ and by assumption $|s|\ll p$. This assumption allows us to write $\xi_p^{-\frac{2}{3}}\approx p^{-\frac{2}{3}}$ and $\xi_p^\frac{1}{3}\approx p^\frac{1}{3}$ and we get the equation
\begin{align}
-\frac{1}{\pi^2}+c_1p^{-\frac{2}{3}}e^{i2\pi(p+s)}\sin\left(c_3p^\frac{1}{3}\right)=0
\end{align}
which can be simply solved for $s$ and the asymptotic formula for the zeros of the determinant becomes
\begin{align}
\xi_p=p-i\rho_1\ln(p)+i\rho_2+\frac{i}{2\pi}\ln\left(\left|\sin\left(c_3p^\frac{1}{3}\right)\right|\right)+\frac{1}{2\pi}h(p)\label{Aeq8}
\end{align}
where $p$ is the index of the zero and is large by assumption. The constant involved in (\ref{Aeq8}) are
\begin{align}
\rho_1&=\frac{1}{3\pi}\label{Aeq9}\\
\rho_2&=\frac{\ln\left(\pi^2c_1\right)}{2\pi}\label{Aeq10}\\
h(p)&=\arg\left(\sin\left(c_3p^\frac{1}{3}\right)\right)\label{Aeq10.1}
\end{align}

\subsection{\texorpdfstring{Coefficients $b_p(0)$}{Abp}}
We move to the numerator in (\ref{Aeq0}). According to (\ref{Aeq2.3}), we need the asymptotic formulas to Airy functions $\mathrm{Ai}(z)$, $\mathrm{Bi}(z)$ using the newly computed points $\xi_p$. Even in the new $\xi$-plane, the complex angle of the argument stays in the region $\arg|z|<\frac{2\pi}{3}$. We have the general asymptotic formula for $\arg|z|<\frac{2\pi}{3}$ in (\ref{Aeq140}), (\ref{Aeq141}) and we define our $z$ as
\begin{align}
z(\xi_p)=&-y_2\left.(x)\right|_{\omega=\xi_p}=\gamma\left(\varepsilon x+V_0+\omega(\xi_p)\right)=\gamma(x\varepsilon+V_0)+\gamma\gamma^{-1}\left(\frac{3\pi}{2}\xi_p\right)^\frac{2}{3}\nonumber\\
=&\gamma(x\varepsilon+V_0)+\mu\xi_p^\frac{2}{3}\label{Aeq13}
\end{align}
where we defined
\begin{align}
\mu=\left(\frac{3\pi}{2}\right)^\frac{2}{3}\label{Aeq14}
\end{align}
The asymptotic function $\mathrm{Ai}(-z)$ in (\ref{Aeq140}) then becomes
\begin{align}
\mathrm{Ai}(-z(\xi_p))&\approx\pi^{-\frac{1}{2}}z^{-\frac{1}{4}}\sin\left(\frac{2}{3}z^\frac{3}{2}+\frac{\pi}{4}\right)\nonumber\\
&=\pi^{-\frac{1}{2}}\left(\gamma(x\varepsilon+V_0)+\mu\xi_p^\frac{2}{3}\right)^{-\frac{1}{4}}\sin\left(\frac{2}{3}\left(\gamma(x\varepsilon+V_0)+\mu\xi_p^\frac{2}{3}\right)^\frac{3}{2}+\frac{\pi}{4}\right)\label{Aeq15}
\end{align}
We use Taylor expansions to simplify the power terms in (\ref{Aeq15}).
{\medmuskip=0mu
\thinmuskip=0mu
\thickmuskip=0mu
\begin{align}
\left(\gamma(x\varepsilon+V_0)+\mu\xi_p^\frac{2}{3}\right)^{-\frac{1}{4}}&=\mu^{-\frac{1}{4}}\xi_p^{-\frac{1}{6}}\left(1+\frac{\gamma(x\varepsilon+V_0)}{\mu \xi_p^\frac{2}{3}}\right)^{-\frac{1}{4}}\approx \mu^{-\frac{1}{4}}\xi_p^{-\frac{1}{6}}\left(1-\frac{\gamma(x\varepsilon+V_0)}{4\mu \xi_p^\frac{2}{3}}\right)\nonumber\\
&=\left(\frac{3\pi}{2}\right)^{-\frac{1}{6}}\xi_p^{-\frac{1}{6}}-\frac{\gamma(x\varepsilon+V_0)}{4\mu^\frac{5}{4} \xi_p^\frac{5}{6}}\approx\left(\frac{3\pi}{2}\right)^{-\frac{1}{6}}\xi_p^{-\frac{1}{6}}\approx \left(\frac{3\pi}{2}\right)^{-\frac{1}{6}}p^{-\frac{1}{6}}\label{Aeq16}
\end{align}}
where we also simplified the term $\xi^{-\frac{1}{6}}$ to $p^{-\frac{1}{6}}$ in the second line. The input of the sine term in (\ref{Aeq14}) is
{\medmuskip=0mu
\thinmuskip=0mu
\thickmuskip=0mu
\begin{align}
\frac{2}{3}\left(\gamma(x\varepsilon+V_0)+\mu\xi_p^\frac{2}{3}\right)^\frac{3}{2}&\approx \frac{2}{3}\mu^\frac{3}{2}\xi_p\left(1+\frac{3\gamma(x\varepsilon+V_0)}{2\mu\xi_p^\frac{2}{3}}\right)=\frac{2}{3}\mu^\frac{3}{2}\xi_p+\mu^\frac{1}{2}\xi_p^\frac{1}{3}\gamma(x\varepsilon+V_0)\nonumber\\
&\approx \frac{2}{3}\mu^\frac{3}{2}\xi_p+\mu^\frac{1}{2}p^\frac{1}{3}\gamma(x\varepsilon+V_0)\nonumber\\
&=\frac{2}{3}\mu^\frac{3}{2}\xi_p+\mu^\frac{1}{2}p^\frac{1}{3}\gamma V_0+x\mu^\frac{1}{2}p^\frac{1}{3}\gamma\varepsilon=g_1(p)+xg_2(p)\label{Aeq18}
\end{align}}
where we used a similar approach as in(\ref{Aeq16}) and wrote $\xi_p^\frac{1}{3}$ in the second term as $p^\frac{1}{3}$ and we defined
\begin{align}
g_1(p)=&\frac{2}{3}\mu^\frac{3}{2}\xi_p+\mu^\frac{1}{2}p^\frac{1}{3}\gamma V_0\label{Aeq19}\\
g_2(p)=&\mu^\frac{1}{2}p^\frac{1}{3}\gamma\varepsilon\label{Aeq20}
\end{align}
With these simplifications we can write (\ref{Aeq15}) and similarly $\mathrm{Bi}(-z)$ as
\begin{align}
\mathrm{Ai}(-z(\xi_p))&\approx\pi^{-\frac{1}{2}}\left(\frac{3\pi}{2}\right)^{-\frac{1}{6}}p^{-\frac{1}{6}}\sin\left(g_1(p)+xg_2(p)+\frac{\pi}{4}\right)\label{Aeq21}\\
\mathrm{Bi}(-z(\xi_p))&\approx\pi^{-\frac{1}{2}}\left(\frac{3\pi}{2}\right)^{-\frac{1}{6}}p^{-\frac{1}{6}}\cos\left(g_1(p)+xg_2(p)+\frac{\pi}{4}\right)\label{Aeq22}
\end{align}
As a reminder of our coefficients let us recall their form.
\begin{align}
b_p(0)&=\left(a_2\mathrm{Ai}\left( \left.y_2(0)\right|_{\omega=\omega_p}\right)+a_3\mathrm{Bi}\left( \left.y_2(0)\right|_{\omega=\omega_p}\right)\right)\nonumber\\
&\int_{-a}^a\left(a_2\mathrm{Ai}\left( \left.y_2(x)\right|_{\omega=\omega_p}\right)+a_3\mathrm{Bi}\left( \left.y_2(x)\right|_{\omega=\omega_p}\right)\right)\mathrm{d}x\label{Aeq22.1}
\end{align}
Now we are ready to integrate just as we see (\ref{Aeq22.1}). Under a transformation $y=g_1(p)+xg_2(p)+\frac{\pi}{4}$ we have $\mathrm{d}y=g_2(p)\mathrm{d}x$.
{\medmuskip=0mu
\thinmuskip=0mu
\thickmuskip=0mu
\begin{align}
&\int_{-a}^a\left(a_2\mathrm{Ai}\left( \left.y_2(x)\right|_{\omega=\omega_p}\right)+a_3\mathrm{Bi}\left( \left.y_2(x)\right|_{\omega=\omega_p}\right)\right)\mathrm{d}x\nonumber\\
&=\pi^{-\frac{1}{2}}\left(\frac{3\pi}{2}\right)^{-\frac{1}{6}}p^{-\frac{1}{6}}\frac{1}{g_2(p)}\int_{g_1(p)-ag_2(p)+\frac{\pi}{4}}^{g_1(p)+ag_2(p)+\frac{\pi}{4}}\left[a_2\sin(y)+a_3\cos(y)\right]\mathrm{d}y\nonumber\\
&=\pi^{-\frac{1}{2}}\left(\frac{3\pi}{2}\right)^{-\frac{1}{6}}p^{-\frac{1}{6}}\frac{1}{g_2(p)}\left[a_3\sin(y)-a_2\cos(y)\right]\Big|_{y=g_1(p)-ag_2(p)+\frac{\pi}{4}}^{y=g_1(p)+ag_2(p)+\frac{\pi}{4}}\nonumber\\
&=\pi^{-\frac{1}{2}}\left(\frac{3\pi}{2}\right)^{-\frac{1}{6}}p^{-\frac{1}{6}}\frac{1}{g_2(p)}\left[a_3\sin\left(g_1(p)+ag_2(p)+\frac{\pi}{4}\right)-a_2\cos\left(g_1(p)+ag_2(p)+\frac{\pi}{4}\right)\right.\nonumber\\
&\left.-a_3\sin\left(g_1(p)-ag_2(p)+\frac{\pi}{4}\right)+a_2\cos\left(g_1(p)-ag_2(p)+\frac{\pi}{4}\right)\right]\nonumber\\
&=\pi^{-\frac{1}{2}}\left(\frac{3\pi}{2}\right)^{-\frac{1}{6}}p^{-\frac{1}{6}}\frac{2\sin\left(ag_2(p)\right)}{g_2(p)}\left[a_3\cos\left(g_1(p)+\frac{\pi}{4}\right)+a_2\sin\left(g_1(p)+\frac{\pi}{4}\right)\right]\label{Aeq23}
\end{align}}
We need to find the asymptotic expressions for $\mathrm{Ai}(y_2(0))$ and $\mathrm{Bi}(y_2(0))$ as they appear in (\ref{Aeq2.3}).  To find them, we can use the formulas (\ref{Aeq21}), (\ref{Aeq22}).
\begin{align}
\mathrm{Ai}\left(-z(\xi_p)|_{x=0}\right)&\approx\pi^{-\frac{1}{2}}\left(\frac{3\pi}{2}\right)^{-\frac{1}{6}}p^{-\frac{1}{6}}\sin\left(g_1(p)+\frac{\pi}{4}\right)\label{Aeq24}\\
\mathrm{Bi}(-z(\xi_p)|_{x=0})&\approx\pi^{-\frac{1}{2}}\left(\frac{3\pi}{2}\right)^{-\frac{1}{6}}p^{-\frac{1}{6}}\cos\left(g_1(p)+\frac{\pi}{4}\right)\label{Aeq25}
\end{align}
Using (\ref{Aeq23}), (\ref{Aeq24}) and (\ref{Aeq25}) the coefficients $b_p$ in (\ref{Aeq2.3}) become
{\medmuskip=0mu
\thinmuskip=0mu
\thickmuskip=0mu
\begin{align}
b_p&\approx\pi^{-\frac{1}{2}}\left(\frac{3\pi}{2}\right)^{-\frac{1}{6}}p^{-\frac{1}{6}}\left[a_2\sin\left(g_1(p)+\frac{\pi}{4}\right)+a_3\cos\left(g_1(p)+\frac{\pi}{4}\right)\right]\nonumber\\
&\quad\pi^{-\frac{1}{2}}\left(\frac{3\pi}{2}\right)^{-\frac{1}{6}}p^{-\frac{1}{6}}\frac{2\sin\left(ag_2(p)\right)}{g_2(p)}\left[a_3\cos\left(g_1(p)+\frac{\pi}{4}\right)+a_2\sin\left(g_1(p)+\frac{\pi}{4}\right)\right]\nonumber\\
&=\pi^{-1}\left(\frac{3\pi}{2}\right)^{-\frac{1}{3}}p^{-\frac{1}{3}}\frac{2\sin\left(ag_2(p)\right)}{g_2(p)}\left[a_2\sin\left(g_1(p)+\frac{\pi}{4}\right)+a_3\cos\left(g_1(p)+\frac{\pi}{4}\right)\right]^2\label{Aeq26}
\end{align}}
We take a further look at the trigonometric terms. After writing them in exponential forms, we find that one of the two exponentials can be neglected. Therefore we get
\begin{align}
\sin\left(g_1(p)+\frac{\pi}{4}\right)&\approx\frac{1}{2i}(-1)^p p^\frac{1}{3}\left(\pi^2c_1\right)^{-\frac{1}{2}} \left|\sin\left(c_3p^\frac{1}{3}\right)\right|^{-\frac{1}{2}} e^{i\left(\mu^\frac{1}{2}p^\frac{1}{3}\gamma V_0+\frac{1}{2}h(p)+\frac{\pi}{4}\right)}\label{Aeq30}\\
\cos\left(g_1(p)+\frac{\pi}{4}\right)&\approx\frac{1}{2}(-1)^p p^\frac{1}{3}\left(\pi^2c_1\right)^{-\frac{1}{2}}\left|\sin\left(c_3p^\frac{1}{3}\right)\right|^{-\frac{1}{2}} e^{i\left(\mu^\frac{1}{2}p^\frac{1}{3}\gamma V_0+\frac{1}{2}h(p)+\frac{\pi}{4}\right)}\label{Aeq31}\\
\sin\left(ag_2(p)\right)&=\sin\left(a(3\pi\varepsilon p)^\frac{1}{3}\right)\label{Aeq32}
\end{align}
Using (\ref{Aeq30}) - (\ref{Aeq32}) we write (\ref{Aeq26}) as
{\medmuskip=0mu
\thinmuskip=0mu
\thickmuskip=0mu
\begin{align}
b_p\approx&\pi^{-1}\left(\frac{3\pi}{2}\right)^{-\frac{1}{3}}p^{-\frac{1}{3}}\frac{2\sin\left(a(3\pi\varepsilon p)^\frac{1}{3}\right)}{g_2(p)}\left[a_2\frac{1}{2i}(-1)^p p^\frac{1}{3}\left(\pi^2c_1\right)^{-\frac{1}{2}} \left|\sin\left(c_3p^\frac{1}{3}\right)\right|^{-\frac{1}{2}} \right.\nonumber\\
&e^{i\left(\mu^\frac{1}{2}p^\frac{1}{3}\gamma V_0+\frac{1}{2}h(p)+\frac{\pi}{4}\right)}\nonumber\\
&\left.+a_3\frac{1}{2}(-1)^p p^\frac{1}{3}\left(\pi^2c_1\right)^{-\frac{1}{2}}\left|\sin\left(c_3p^\frac{1}{3}\right)\right|^{-\frac{1}{2}} e^{i\left(\mu^\frac{1}{2}p^\frac{1}{3}\gamma V_0+\frac{1}{2}h(p)+\frac{\pi}{4}\right)}\right]^2\nonumber\\
&=\pi^{-1}\left(\frac{3\pi}{2}\right)^{-\frac{1}{3}}p^{-\frac{1}{3}}\frac{2\sin\left(a(3\pi\varepsilon p)^\frac{1}{3}\right)}{\mu^\frac{1}{2}p^\frac{1}{3}\gamma\varepsilon}\nonumber\\
&\left(\frac{1}{2}(-1)^p p^\frac{1}{3}\left(\pi^2c_1\right)^{-\frac{1}{2}}\left|\sin\left(c_3p^\frac{1}{3}\right)\right|^{-\frac{1}{2}} e^{i\left(\mu^\frac{1}{2}p^\frac{1}{3}\gamma V_0+\frac{1}{2}h(p)+\frac{\pi}{4}\right)}\right)^2\left(\frac{a_2}{i}+a_3\right)^2\nonumber\\
&=\left(\frac{3\pi}{2}\right)^{-\frac{1}{3}}p^{-\frac{1}{3}}\frac{\sin\left(a(3\pi\varepsilon p)^\frac{1}{3}\right)}{2(3\pi\varepsilon p)^\frac{1}{3}}p^\frac{2}{3}\left(\pi^3c_1\right)^{-1}\left|\sin\left(2d(3\pi\varepsilon p)^\frac{1}{3}\right)\right|^{-1}\nonumber\\
& e^{i2\left(\mu^\frac{1}{2}p^\frac{1}{3}\gamma V_0+\frac{1}{2}h(p)+\frac{\pi}{4}\right)}\left(\frac{a_2}{i}+a_3\right)^2\nonumber\\
&=\left(\frac{3\pi}{2}\right)^{-\frac{1}{3}}\frac{\sin\left(a(3\pi\varepsilon p)^\frac{1}{3}\right)}{2(3\pi\varepsilon)^\frac{1}{3}}\left(\pi^3c_1\right)^{-1}\left|\sin\left(2d(3\pi\varepsilon p)^\frac{1}{3}\right)\right|^{-1}\nonumber\\
&e^{i2\left(\mu^\frac{1}{2}p^\frac{1}{3}\gamma V_0+\frac{1}{2}h(p)+\frac{\pi}{4}\right)}\left(\frac{a_2}{i}+a_3\right)^2\label{Aeq34}
\end{align}}

\subsection{Normalization}
The goal is to find the normalization coefficients of the resonant states that emerge in the coefficients $c_p$ in (\ref{Aeq0}). Essentially it comes down to computing the integral
\begin{align}
N_p=\int_\mathcal{L}\psi_{\omega_p}^2(z)\mathrm{d}z\label{Aeq35}
\end{align}
where $\mathcal{L}$ is the curve
\begin{align}
\mathcal{L}=u(x)=\left\{\begin{array}{c c}
x & x<x_c\\
x_c+i(x-x_c) & x_c<x
\end{array}\right.\label{Aeq36}
\end{align}
on which we know the resonant states decay on both ends. We gave more attention to this subject in section \ref{s5} We split the integral using the corresponding regions of $\psi_{\omega_p}(x)$ as
\begin{align}
N_p=\int_{-\infty}^{-d}a_1^2\mathrm{Ai}^2&(y_1(x))\mathrm{d}x+\int_{-d}^d\left(a_2\mathrm{Ai}(y_2(x))+a_3\mathrm{Bi}(y_2(x))\right)^2\mathrm{d}x\nonumber\\
&+\int_{d}^{x_c}a_4^2\left(\mathrm{Ci}^+(y_1(x))\right)^2\mathrm{d}x+i\int_{x_c}^\infty a_4^2\left(\mathrm{Ci}^+(\tilde{y}(x))\right)^2\mathrm{d}x\label{Aeq37}
\end{align}
where
\begin{align}
y_1(x)&=-2(2\varepsilon)^{-\frac{2}{3}}(\varepsilon x+\omega_p)\label{Aeq38}\\
y_2(x)&=-2(2\varepsilon)^{-\frac{2}{3}}(\varepsilon x+V_0+\omega_p)\label{Aeq39}\\
\tilde{y}(x)&=-2(2\varepsilon)^{-\frac{2}{3}}\left(i\varepsilon x+\varepsilon x_c(1-i)+\omega_p\right)\label{Aeq39.1}
\end{align}
To compute the integrals we use a known formula. Let $\mathrm{A}(x)$ and $\mathrm{B}(x)$ be any linear combination of Airy functions and $\mu,\nu$ be any real numbers, then
\begin{align}
\int\mathrm{A}(\mu(x+\nu))&\mathrm{B}(\mu(x+\nu))\mathrm{d}x\nonumber\\
&=(x+\nu)\mathrm{A}(\mu(x+\nu))\mathrm{B}(\mu(x+\nu))-\frac{1}{\mu}\mathrm{A}'(\mu(x+\nu))\mathrm{B}'(\mu(x+\nu))\label{Aeq40}
\end{align}
Let us denote the integrals in (\ref{Aeq37}) in order as they appear $I_1,\ldots,I_4$. We can use many formulas computed in the previous section, but we need to set up a different notation. So let
\begin{align}
\omega(\xi)&=\gamma^{-1}\left(\frac{3\pi\xi}{2}\right)^\frac{2}{3}\label{Aeq41}\\
\gamma&=2(2\varepsilon)^{-\frac{2}{3}}\label{Aeq42}\\
\mu&=\left(\frac{3\pi}{2}\right)^\frac{2}{3}\label{Aeq43}\\
z_1(\xi_p)&=-y_1(\xi_p)=\gamma x\varepsilon+\mu\xi_p^\frac{2}{3}\label{Aeq44}\\
z_2(\xi_p)&=-y_2(\xi_p)=\gamma(x\varepsilon+V_0)+\mu\xi_p^\frac{2}{3}\label{Aeq45}\\
\tilde{z}(\xi_p)&=-\tilde{y}(\xi_p)=\gamma(ix\varepsilon+\varepsilon x_c(1-i))+\mu\xi_p^\frac{2}{3}\label{Aeq45.1}
\end{align}
and $\xi_p$ defined as (\ref{Aeq8}). The asymptotic expressions for Airy functions and their derivatives are defined as in (\ref{Aeq140}), (\ref{Aeq141}) and (\ref{Aeq142}). We will use one additional term from the Taylor expansion in the asymptotic Airy functions. Using (\ref{Aeq16}) - (\ref{Aeq22}) we can write the asymptotic formulas as
\begin{align}
\mathrm{Ai}(-z_{1,2}(\xi_p))&\approx\pi^{-\frac{1}{2}}z_{1,2}\left(\xi_p\right)^{-\frac{1}{4}}\sin\left(s_{1,2}(p)+xt(p)+\frac{\pi}{4}\right)\label{Aeq52}\\
\mathrm{Ai}'(-z_{1,2}(\xi_p))&\approx-\pi^{-\frac{1}{2}}z_{1,2}\left(\xi_p\right)^{\frac{1}{4}}\cos\left(s_{1,2}(p)+xt(p)+\frac{\pi}{4}\right)\label{Aeq53}\\
\mathrm{Bi}(-z_{1,2}(\xi_p))&\approx\pi^{-\frac{1}{2}}z_{1,2}\left(\xi_p\right)^{-\frac{1}{4}}\cos\left(s_{1,2}(p)+xt(p)+\frac{\pi}{4}\right)\label{Aeq54}\\
\mathrm{Bi}'(-z_{1,2}(\xi_p))&\approx\pi^{-\frac{1}{2}}z_{1,2}\left(\xi_p\right)^{\frac{1}{4}}\sin\left(s_{1,2}(p)+xt(p)+\frac{\pi}{4}\right)\label{Aeq55}\\
\mathrm{Ci}^+(-z_{1,2}(\xi_p))&=\pi^{-\frac{1}{2}}z_{1,2}\left(\xi_p\right)^{-\frac{1}{4}}e^{i\left(s_{1,2}(p)+xt(p)+\frac{\pi}{4}\right)}\label{Aeq56}\\
\mathrm{Ci}^+(-\tilde{z}(\xi_p))&=\pi^{-\frac{1}{2}}\tilde{z}\left(\xi_p\right)^{-\frac{1}{4}}e^{i\left(\tilde{s}(p)+x\tilde{t} (p)+\frac{\pi}{4}\right)}\label{Aeq56.1}\\
\mathrm{Ci'}^+(-\tilde{z}(\xi_p))&=-i\pi^{-\frac{1}{2}}\tilde{z}\left(\xi_p\right)^{\frac{1}{4}}e^{i\left(\tilde{s}(p)+x\tilde{t} (p)+\frac{\pi}{4}\right)}\label{Aeq56.2}\\
\mathrm{Ci'}^+(-z_{1,2}(\xi_p))&=-i\pi^{-\frac{1}{2}}z_{1,2}\left(\xi_p\right)^{\frac{1}{4}}e^{i\left(s_{1,2}(p)+xt(p)+\frac{\pi}{4}\right)}\label{Aeq57}
\end{align}
Observe, that the Airy functions appear in quadratic form under the integrals, which means we need to simplify the following terms
{\medmuskip=0mu
\thinmuskip=0mu
\thickmuskip=0mu
\begin{align}
z_1(\xi_p)^{-\frac{1}{2}}&\approx\left(\frac{3\pi p}{2}\right)^{-\frac{1}{3}}-\frac{\gamma x\varepsilon}{2}\left(\frac{3\pi p}{2}\right)^{-1}\nonumber\\
 z_1(\xi_p)^{\frac{1}{2}}&\approx\left(\frac{3\pi p}{2}\right)^{\frac{1}{3}}+\frac{\gamma x\varepsilon}{2}\left(\frac{3\pi p}{2}\right)^{-\frac{1}{3}}\label{Aeq62.3}\\
z_2(\xi_p)^{-\frac{1}{2}}&\approx\left(\frac{3\pi p}{2}\right)^{-\frac{1}{3}}-\frac{\gamma(x\varepsilon+V_0)}{2}\left(\frac{3\pi p}{2}\right)^{-1}\nonumber\\
z_2(\xi_p)^{\frac{1}{2}}&\approx\left(\frac{3\pi p}{2}\right)^{\frac{1}{3}}+\frac{\gamma(x\varepsilon+V_0)}{2}\left(\frac{3\pi p}{2}\right)^{-\frac{1}{3}}\label{Aeq69.1}
\end{align}}
The arguments of sines and cosines are defined as
\begin{align}
s_1(p)&=\frac{2}{3}\mu^\frac{3}{2}\xi_p=\pi\xi_p & s_2(p)&=\frac{2}{3}\mu^\frac{3}{2}\xi_p+\mu^\frac{1}{2}p^\frac{1}{3}\gamma V_0=\pi\xi_p+\mu^\frac{1}{2}p^\frac{1}{3}\gamma V_0\label{Aeq59}\\
\tilde{s}(p)&=\pi\xi_p+\mu^\frac{1}{2}p^\frac{1}{3}\gamma\varepsilon x_c(1-i) & \tilde{t}(p)&=i\mu^\frac{1}{2}p^\frac{1}{3}\gamma\varepsilon\label{Aeq59.1}\\
t(p)&=\mu^\frac{1}{2}p^\frac{1}{3}\gamma\varepsilon & &\label{Aeq60}
\end{align}
At this point we have the set up to compute the integrals $I_1,\ldots,I_4$. Each integral has its own subsection.

\subsubsection{\texorpdfstring{$I_1$}{AI1}}
Let us take the first integral in (\ref{Aeq37}) $I_1$ and the formula according to (\ref{Aeq40}).
\begin{align}
I_1&=\int_{-\infty}^{-d}a_1^2\mathrm{Ai}^2\left(-\gamma\varepsilon\left(x+\frac{\mu\xi_p^\frac{2}{3}}{\gamma\varepsilon}\right)\right)\mathrm{d}x\nonumber\\
&=a_1^2\left(x+\frac{\mu\xi_p^\frac{2}{3}}{\gamma\varepsilon}\right)\mathrm{Ai}^2\left(y_1(x)\right)+\frac{a_1^2}{\gamma\varepsilon}\left(\mathrm{Ai}'\left(y_1(x)\right)\right)^2\Big|_{x=-\infty}^{x=-d}\nonumber\\
&=a_1^2\left(-d+\frac{\mu\xi_p^\frac{2}{3}}{\gamma\varepsilon}\right)\mathrm{Ai}^2\left(y_1(-d)\right)+\frac{a_1^2}{\gamma\varepsilon}\left(\mathrm{Ai}'\left(y_1(-d)\right)\right)^2\label{Aeq61}
\end{align}
Before we proceed, we express the term $\xi_p^\frac{2}{3}$ by taking one extra term in its Taylor expansion. The reason is because it allows us to reach the leading order.
\begin{align}
\xi_p^{\frac{2}{3}}&\approx p^\frac{2}{3}+p^{-\frac{1}{3}}f(p)\label{Aeq62}
\end{align}
where
\begin{align}
f(p)=-i\frac{2\rho_1\ln(p)}{3}+i\frac{2\rho_2}{3}+\frac{i}{3\pi}\ln\left(\left|\sin\left(c_3p^\frac{1}{3}\right)\right|\right)+\frac{1}{3\pi}h(p)\label{Aeq62.1}
\end{align}
The integral $I_1$ becomes
{\medmuskip=0mu
\thinmuskip=0mu
\thickmuskip=0mu
\begin{align}
I_1\approx& a_1^2\pi^{-1}\left(-d+\frac{\mu\left(p^\frac{2}{3}+p^{-\frac{1}{3}}f(p)\right)}{\gamma\varepsilon}\right)\left(\left(\frac{3\pi p}{2}\right)^{-\frac{1}{3}}-\frac{\gamma(-d)\varepsilon}{2}\left(\frac{3\pi p}{2}\right)^{-1}\right)\nonumber\\
&\left[\sin\left(s_1(p)-dt(p)+\frac{\pi}{4}\right)\right]^2\nonumber\\
&+\frac{a_1^2}{\gamma\varepsilon}\pi^{-1}\left(\left(\frac{3\pi p}{2}\right)^{\frac{1}{3}}+\frac{\gamma(-d)\varepsilon}{2}\left(\frac{3\pi p}{2}\right)^{-\frac{1}{3}}\right)\left[\cos\left(s_1(p)-dt(p)+\frac{\pi}{4}\right)\right]^2\nonumber\\
=&a_1^2\pi^{-1}\left[-d\left(\frac{\gamma d\varepsilon}{2}\left(\frac{3\pi p}{2}\right)^{-1}\right)+\frac{\mu p^{-\frac{1}{3}}f(p)}{\gamma\varepsilon}\left(\left(\frac{3\pi p}{2}\right)^{-\frac{1}{3}}+\frac{\gamma d\varepsilon}{2}\left(\frac{3\pi p}{2}\right)^{-1}\right)\right]\nonumber\\
&\left[\sin\left(s_1(p)-dt(p)+\frac{\pi}{4}\right)\right]^2+\frac{a_1^2}{\gamma\varepsilon}\pi^{-1}\left(\left(\frac{3\pi p}{2}\right)^{\frac{1}{3}}-\frac{\gamma d\varepsilon}{2}\left(\frac{3\pi p}{2}\right)^{-\frac{1}{3}}\right)\nonumber\\
\approx& a_1^2\pi^{-1}\left[-d\left(\frac{\gamma d\varepsilon}{2}\left(\frac{3\pi p}{2}\right)^{-1}\right)+\frac{\mu p^{-\frac{1}{3}}f(p)}{\gamma\varepsilon}\left(\left(\frac{3\pi p}{2}\right)^{-\frac{1}{3}}+\frac{\gamma d\varepsilon}{2}\left(\frac{3\pi p}{2}\right)^{-1}\right)\right]\nonumber\\
&\left[\sin\left(s_1(p)-dt(p)+\frac{\pi}{4}\right)\right]^2+\frac{a_1^2}{\gamma\varepsilon}\pi^{-1}\left(\frac{3\pi p}{2}\right)^{\frac{1}{3}}\label{Aeq63}
\end{align}}
Let us expand the term $\sin\left(s_1(p)-dt(p)+\frac{\pi}{4}\right)$.
{\medmuskip=0mu
\thinmuskip=0mu
\thickmuskip=0mu
\begin{align}
&\sin\left(s_1(p)-dt(p)+\frac{\pi}{4}\right)\approx\frac{1}{2i}(-1)^pp^\frac{1}{3}\left(\pi^2c_1\right)^{-\frac{1}{2}}\left|\sin\left(c_3p^\frac{1}{3}\right)\right|^{-\frac{1}{2}} e^{i\left(\frac{1}{2}h(p)-d\mu^\frac{1}{2}p^\frac{1}{3}\gamma\varepsilon+\frac{\pi}{4}\right)}\label{Aeq64}
\end{align}}
where we neglected the second term of the sinus, because it contains the term\newline $p^{-\frac{1}{3}}$ which is much smaller than the term left standing. The expression (\ref{Aeq63}) then becomes
\begin{align}
I_1\approx&-a_1^2\pi^{-1}\frac{f(p)}{\gamma\varepsilon}\left(\frac{3\pi}{2}\right)^{\frac{1}{3}}\frac{1}{4}\left(\pi^2c_1\right)^{-1}\left|\sin\left(c_3p^\frac{1}{3}\right)\right|^{-1}e^{i2\left(\frac{1}{2}h(p)-d\mu^\frac{1}{2}p^\frac{1}{3}\gamma\varepsilon+\frac{\pi}{4}\right)}\nonumber\\
&+\frac{a_1^2}{\gamma\varepsilon}\pi^{-1}\left(\frac{3\pi p}{2}\right)^{\frac{1}{3}}\approx\frac{a_1^2}{\gamma\varepsilon}\pi^{-1}\left(\frac{3\pi p}{2}\right)^{\frac{1}{3}}\label{Aeq65}
\end{align}

\subsubsection{\texorpdfstring{$I_2$}{AI2}}
The second integral $I_2$ has the form
{\medmuskip=0mu
\thinmuskip=0mu
\thickmuskip=0mu
\begin{align}
I_2=&\int_{-d}^d\left(a_2\mathrm{Ai}\left(-\gamma\varepsilon\left(x+\frac{\mu\xi_p^\frac{2}{3}+\gamma V_0}{\gamma\varepsilon}\right)\right)+a_3\mathrm{Bi}\left(-\gamma\varepsilon\left(x+\frac{\mu\xi_p^\frac{2}{3}+\gamma V_0}{\gamma\varepsilon}\right)\right)\right)^2\mathrm{d}x\nonumber\\
=&\left(x+\frac{\mu\xi_p^\frac{2}{3}+\gamma V_0}{\gamma\varepsilon}\right)\left(a_2\mathrm{Ai}\left(y_2(x)\right)+a_3\mathrm{Bi}\left(y_2(x)\right)\right)^2\nonumber\\
&+\frac{1}{\gamma\varepsilon}\left(a_2\mathrm{Ai}'\left(y_2(x)\right)+a_3\mathrm{Bi}'\left(y_2(x)\right)\right)^2\Big|_{x=-d}^{x=d}\nonumber\\
=&\left(d+\frac{\mu\xi_p^\frac{2}{3}+\gamma V_0}{\gamma\varepsilon}\right)\left(a_2\mathrm{Ai}\left(y_2(d)\right)+a_3\mathrm{Bi}\left(y_2(d)\right)\right)^2\nonumber\\
&+\frac{1}{\gamma\varepsilon}\left(a_2\mathrm{Ai}'\left(y_2(d)\right)+a_3\mathrm{Bi}'\left(y_2(d)\right)\right)^2\nonumber\\
&-\left[\left(-d+\frac{\mu\xi_p^\frac{2}{3}+\gamma V_0}{\gamma\varepsilon}\right)\left(a_2\mathrm{Ai}\left(y_2(-d)\right)+a_3\mathrm{Bi}\left(y_2(-d)\right)\right)^2\right.\nonumber\\
&\left.+\frac{1}{\gamma\varepsilon}\left(a_2\mathrm{Ai}'\left(y_2(-d)\right)+a_3\mathrm{Bi}'\left(y_2(-d)\right)\right)^2\right]\label{Aeq66}
\end{align}}
Let us first express the term $\left(a_2\mathrm{Ai}\left(y_2(d)\right)+a_3\mathrm{Bi}\left(y_2(d)\right)\right)^2$. We can directly use (\ref{Aeq21}), (\ref{Aeq22}) and that $g_1(p)=s_2(p)$ and $g_2(p)=t(p)$.
{\medmuskip=0mu
\thinmuskip=0mu
\thickmuskip=0mu
\begin{align}
\left(a_2\mathrm{Ai}\left(y_2(d)\right)+a_3\mathrm{Bi}\left(y_2(d)\right)\right)&^2\approx\pi^{-1}\left(\left(\frac{3\pi p}{2}\right)^{-\frac{1}{3}}-\frac{\gamma(d\varepsilon+V_0)}{2}\left(\frac{3\pi p}{2}\right)^{-1}\right)\nonumber\\
&\left[a_2\sin\left(s_2(p)+dt(p)+\frac{\pi}{4}\right)+a_3\cos\left(s_2(p)+dt(p)+\frac{\pi}{4}\right)\right]^2\label{Aeq67}
\end{align}}
And similarly its derivative $\left(a_2\mathrm{Ai}'\left(y_2(d)\right)+a_3\mathrm{Bi}'\left(y_2(d)\right)\right)^2$:
{\medmuskip=-1mu
\thinmuskip=0mu
\thickmuskip=0mu
\begin{align}
\left(a_2\mathrm{Ai}'\left(y_2(d)\right)+a_3\mathrm{Bi}'\left(y_2(d)\right)\right)&^2\approx\pi^{-1}\left(\left(\frac{3\pi p}{2}\right)^{\frac{1}{3}}+\frac{\gamma(d\varepsilon+V_0)}{2}\left(\frac{3\pi p}{2}\right)^{-\frac{1}{3}}\right)\nonumber\\
&\left[a_3\sin\left(s_2(p)+dt(p)+\frac{\pi}{4}\right)-a_2\cos\left(s_2(p)+dt(p)+\frac{\pi}{4}\right)\right]^2\label{Aeq68}
\end{align}}
Taking (\ref{Aeq62}), (\ref{Aeq67}) and (\ref{Aeq68}) we have
{\medmuskip=-2mu
\thinmuskip=0mu
\thickmuskip=0mu
\begin{align}
I_2\approx&\left(d+\frac{\mu \left(p^\frac{2}{3}+p^{-\frac{1}{3}}f(p)\right)+\gamma V_0}{\gamma\varepsilon}\right)\pi^{-1}\left[\left(\frac{3\pi p}{2}\right)^{-\frac{1}{3}}-\frac{\gamma(d\varepsilon+V_0)}{2}\left(\frac{3\pi p}{2}\right)^{-1}\right]\nonumber\\
&\left[a_2\sin\left(s_2(p)+dt(p)+\frac{\pi}{4}\right)+a_3\cos\left(s_2(p)+dt(p)+\frac{\pi}{4}\right)\right]^2\nonumber\\
&+\frac{1}{\gamma\varepsilon}\pi^{-1}\left[\left(\frac{3\pi p}{2}\right)^{\frac{1}{3}}+\frac{\gamma(d\varepsilon+V_0)}{2}\left(\frac{3\pi p}{2}\right)^{-\frac{1}{3}}\right]\nonumber\\
&\left[a_3\sin\left(s_2(p)+dt(p)+\frac{\pi}{4}\right)-a_2\cos\left(s_2(p)+dt(p)+\frac{\pi}{4}\right)\right]^2\nonumber\\
&-\left[\left(-d+\frac{\mu\left(p^\frac{2}{3}+p^{-\frac{1}{3}}f(p)\right)+\gamma V_0}{\gamma\varepsilon}\right)\pi^{-1}\left[\left(\frac{3\pi p}{2}\right)^{-\frac{1}{3}}-\frac{\gamma(-d\varepsilon+V_0)}{2}\left(\frac{3\pi p}{2}\right)^{-1}\right]\right.\nonumber\\
&\left.\left[a_2\sin\left(s_2(p)-dt(p)+\frac{\pi}{4}\right)+a_3\cos\left(s_2(p)-dt(p)+\frac{\pi}{4}\right)\right]^2\right.\nonumber\\
&\left.+\frac{1}{\gamma\varepsilon}\pi^{-1}\left[\left(\frac{3\pi p}{2}\right)^{\frac{1}{3}}+\frac{\gamma(-d\varepsilon+V_0)}{2}\left(\frac{3\pi p}{2}\right)^{-\frac{1}{3}}\right]\right.\nonumber\\
&\left.\left[a_3\sin\left(s_2(p)-dt(p)+\frac{\pi}{4}\right)-a_2\cos\left(s_2(p)-dt(p)+\frac{\pi}{4}\right)\right]^2\right]\label{Aeq69.3}
\end{align}}
which can by simplified as
{\medmuskip=0mu
\thinmuskip=0mu
\thickmuskip=0mu
\begin{align}
I_2=&\pi^{-1}\left(d\left[\left(\frac{3\pi p}{2}\right)^{-\frac{1}{3}}-\frac{\gamma(d\varepsilon+V_0)}{2}\left(\frac{3\pi p}{2}\right)^{-1}\right]+\frac{\mu\left(p^\frac{2}{3}+p^{-\frac{1}{3}}f(p)\right)+\gamma V_0}{\gamma\varepsilon}\right.\nonumber\\
&\left.\left[\left(\frac{3\pi p}{2}\right)^{-\frac{1}{3}}-\frac{\gamma(d\varepsilon+V_0)}{2}\left(\frac{3\pi p}{2}\right)^{-1}\right]\right)\nonumber\\
&\left[a_2\sin\left(s_2(p)+dt(p)+\frac{\pi}{4}\right)+a_3\cos\left(s_2(p)+dt(p)+\frac{\pi}{4}\right)\right]^2\nonumber\\
&+\frac{1}{\gamma\varepsilon}\pi^{-1}\left[\left(\frac{3\pi p}{2}\right)^{\frac{1}{3}}+\frac{\gamma(d\varepsilon+V_0)}{2}\left(\frac{3\pi p}{2}\right)^{-\frac{1}{3}}\right]\nonumber\\
&\left[a_3\sin\left(s_2(p)+dt(p)+\frac{\pi}{4}\right)-a_2\cos\left(s_2(p)+dt(p)+\frac{\pi}{4}\right)\right]^2-\nonumber\\
&\left[\left(-d\left[\left(\frac{3\pi p}{2}\right)^{-\frac{1}{3}}-\frac{\gamma(-d\varepsilon+V_0)}{2}\left(\frac{3\pi p}{2}\right)^{-1}\right]+\frac{\mu\left(p^\frac{2}{3}+p^{-\frac{1}{3}}f(p)\right)+\gamma V_0}{\gamma\varepsilon}\right.\right.\nonumber\\
&\left.\left.\left[\left(\frac{3\pi p}{2}\right)^{-\frac{1}{3}}-\frac{\gamma(-d\varepsilon+V_0)}{2}\left(\frac{3\pi p}{2}\right)^{-1}\right]\right)\pi^{-1}\right.\nonumber\\
&\left.\left[a_2\sin\left(s_2(p)-dt(p)+\frac{\pi}{4}\right)+a_3\cos\left(s_2(p)-dt(p)+\frac{\pi}{4}\right)\right]^2\right.\nonumber\\
&\left.+\frac{1}{\gamma\varepsilon}\pi^{-1}\left[\left(\frac{3\pi p}{2}\right)^{\frac{1}{3}}+\frac{\gamma(-d\varepsilon+V_0)}{2}\left(\frac{3\pi p}{2}\right)^{-\frac{1}{3}}\right]\right.\nonumber\\
&\left.\left[a_3\sin\left(s_2(p)-dt(p)+\frac{\pi}{4}\right)-a_2\cos\left(s_2(p)-dt(p)+\frac{\pi}{4}\right)\right]^2\right]\label{Aeq69.3.1}
\end{align}
}
We use the following identities to simplify the expression above.
\begin{align}
\mathrm{1.}\quad&\left(a_2\sin(x+y)+a_3\cos(x+y)\right)^2+\left(a_3\sin(x+y)-a_2\cos(x+y)\right)^2\nonumber\\
&-\left(a_2\sin(x-y)+a_3\cos(x-y)\right)^2-\left(a_3\sin(x-y)-a_2\cos(x-y)\right)^2=0\label{Aeq69.4}\\
\mathrm{2.}\quad&\left(a_2\sin(x+y)+a_3\cos(x+y)\right)^2+\left(a_3\sin(x+y)-a_2\cos(x+y)\right)^2\nonumber\\
&+\left(a_2\sin(x-y)+a_3\cos(x-y)\right)^2+\left(a_3\sin(x-y)-a_2\cos(x-y)\right)^2\nonumber\\
&=2\left(a_2^2+a_3^2\right)\label{Aeq69.4.1}\\
\mathrm{3.}\quad&-\left(a_2\sin(x+y)+a_3\cos(x+y)\right)^2+\left(a_3\sin(x+y)-a_2\cos(x+y)\right)^2\nonumber\\
&+\left(a_2\sin(x-y)+a_3\cos(x-y)\right)^2-\left(a_3\sin(x-y)-a_2\cos(x-y)\right)^2\nonumber\\
&=-2\left(2a_2a_3\cos(2x)+\left(a_2^2-a_3^2\right)\sin(2x)\right)\sin(2y)\label{Aeq69.4.2}
\end{align}
for some values $x,y$. We see that this is also true if the expression is multiplied by any number. Let us focus on the terms in front of sines and cosines in (\ref{Aeq69.3.1}). The first is in order as they appear is
{\medmuskip=0mu
\thinmuskip=0mu
\thickmuskip=0mu
\begin{align}
&d\left[\left(\frac{3\pi p}{2}\right)^{-\frac{1}{3}}-\frac{\gamma(d\varepsilon+V_0)}{2}\left(\frac{3\pi p}{2}\right)^{-1}\right]\nonumber\\
&+\frac{\mu\left(p^\frac{2}{3}+p^{-\frac{1}{3}}f(p)\right)+\gamma V_0}{\gamma\varepsilon}\left[\left(\frac{3\pi p}{2}\right)^{-\frac{1}{3}}-\frac{\gamma(d\varepsilon+V_0)}{2}\left(\frac{3\pi p}{2}\right)^{-1}\right]\nonumber\\
&=d\left(\frac{3\pi p}{2}\right)^{-\frac{1}{3}}-\left(\frac{\gamma d^2\varepsilon}{2}+\frac{dV_0}{2}\right)\left(\frac{3\pi p}{2}\right)^{-1}+\frac{\mu p^\frac{2}{3}+\mu p^{-\frac{1}{3}}f(p)+\gamma V_0}{\gamma\varepsilon}\left(\frac{3\pi p}{2}\right)^{-\frac{1}{3}}\nonumber\\
&-\left(\mu p^\frac{2}{3}+\mu p^{-\frac{1}{3}}f(p)+\gamma V_0\right)\frac{d}{2}\left(\frac{3\pi p}{2}\right)^{-1}-\left(\mu p^\frac{2}{3}+\mu p^{-\frac{1}{3}}f(p)+\gamma V_0\right)\frac{V_0}{2\varepsilon}\left(\frac{3\pi p}{2}\right)^{-1}\nonumber\\
&=d\left(\frac{3\pi p}{2}\right)^{-\frac{1}{3}}-\left(\frac{\gamma d^2\varepsilon}{2}+\frac{d\gamma V_0}{2}\right)\left(\frac{3\pi p}{2}\right)^{-1}+\frac{1}{\gamma\varepsilon}\left(\frac{3\pi p}{2}\right)^{\frac{1}{3}}\nonumber\\
&+\frac{\mu p^{-\frac{1}{3}}f(p)+\gamma V_0}{\gamma\varepsilon}\left(\frac{3\pi p}{2}\right)^{-\frac{1}{3}}-\frac{d}{2}\left(\frac{3\pi p}{2}\right)^{-\frac{1}{3}}-\left(\mu p^{-\frac{1}{3}}f(p)+\gamma V_0\right)\frac{d}{2}\left(\frac{3\pi p}{2}\right)^{-1}\nonumber\\
&-\frac{V_0}{2\varepsilon}\left(\frac{3\pi p}{2}\right)^{-\frac{1}{3}}-\left(\mu p^{-\frac{1}{3}}f(p)+\gamma V_0\right)\frac{V_0}{2\varepsilon}\left(\frac{3\pi p}{2}\right)^{-1}\label{Aeq69.5}
\end{align}
}
The third coefficient in (\ref{Aeq69.3.1}) is similar
{\medmuskip=0mu
\thinmuskip=0mu
\thickmuskip=0mu
\begin{align}
&d\left(\frac{3\pi p}{2}\right)^{-\frac{1}{3}}+\left(\frac{\gamma d^2\varepsilon}{2}-\frac{d\gamma V_0}{2}\right)\left(\frac{3\pi p}{2}\right)^{-1}-\frac{1}{\gamma\varepsilon}\left(\frac{3\pi p}{2}\right)^{\frac{1}{3}}\nonumber\\
&-\frac{\mu p^{-\frac{1}{3}}f(p)+\gamma V_0}{\gamma\varepsilon}\left(\frac{3\pi p}{2}\right)^{-\frac{1}{3}}-\frac{d}{2}\left(\frac{3\pi p}{2}\right)^{-\frac{1}{3}}-\left(\mu p^{-\frac{1}{3}}f(p)+\gamma V_0\right)\frac{d}{2}\left(\frac{3\pi p}{2}\right)^{-1}\nonumber\\
&+\frac{V_0}{2\varepsilon}\left(\frac{3\pi p}{2}\right)^{-\frac{1}{3}}+\left(\mu p^{-\frac{1}{3}}f(p)+\gamma V_0\right)\frac{V_0}{2\varepsilon}\left(\frac{3\pi p}{2}\right)^{-1}\label{Aeq69.6}
\end{align}
}
and the terms in front of the second and fourth are also similar. We ignore the $\pi^{-1}$ because it appears in each term, but we keep it in our minds. The second and fourth term in (\ref{Aeq69.3.1}) is
{\medmuskip=0mu
\thinmuskip=0mu
\thickmuskip=0mu
\begin{align}
&\frac{1}{\gamma\varepsilon}\left[\left(\frac{3\pi p}{2}\right)^{\frac{1}{3}}+\frac{\gamma(d\varepsilon+V_0)}{2}\left(\frac{3\pi p}{2}\right)^{-\frac{1}{3}}\right]=\frac{1}{\gamma\varepsilon}\left(\frac{3\pi p}{2}\right)^{\frac{1}{3}}+\frac{d}{2}\left(\frac{3\pi p}{2}\right)^{-\frac{1}{3}}\label{Aeq69.7}\\
&+\frac{V_0}{2\varepsilon}\left(\frac{3\pi p}{2}\right)^{-\frac{1}{3}}-\frac{1}{\gamma\varepsilon}\left[\left(\frac{3\pi p}{2}\right)^{\frac{1}{3}}+\frac{\gamma(-d\varepsilon+V_0)}{2}\left(\frac{3\pi p}{2}\right)^{-\frac{1}{3}}\right]=-\frac{1}{\gamma\varepsilon}\left(\frac{3\pi p}{2}\right)^{\frac{1}{3}}\nonumber\\
&+\frac{d}{2}\left(\frac{3\pi p}{2}\right)^{-\frac{1}{3}}-\frac{V_0}{2\varepsilon}\left(\frac{3\pi p}{2}\right)^{-\frac{1}{3}}\label{Aeq69.8}
\end{align}}
Now we can spot those terms in these four expression which are the same and have the signs $+,+,-,-$ or $-,-,+,+$ in the respective order as they appear and remove them from (\ref{Aeq69.3.1}) because of (\ref{Aeq69.4}). Also, for every only positive term we can use the identity (\ref{Aeq69.4.1}). It is easy to spot, that the term $\frac{1}{\gamma\varepsilon}\left(\frac{3\pi p}{2}\right)^{\frac{1}{3}}$ has this property, namely that it has signs $+,+,-,-$, so we can use (\ref{Aeq69.4}) in (\ref{Aeq69.3.1}). The term $\frac{d}{2}\left(\frac{3\pi p}{2}\right)^{-\frac{1}{3}}$ appears with signs $+,+,+,+$, so we use the identity (\ref{Aeq69.4.1}) on this one. The term $\frac{V_0}{2\varepsilon}\left(\frac{3\pi p}{2}\right)^{-\frac{1}{3}}$ appears in the configuration satisfying the identity (\ref{Aeq69.4.2}). Writing the expression (\ref{Aeq69.3.1}) in the expanded form we have
{\medmuskip=0mu
\thinmuskip=0mu
\thickmuskip=0mu
\begin{align}
I_2=&\pi^{-1}\left[d\left(\frac{3\pi p}{2}\right)^{-\frac{1}{3}}-\left(\frac{\gamma d^2\varepsilon}{2}+\frac{d\gamma V_0}{2}\right)\left(\frac{3\pi p}{2}\right)^{-1}+\frac{1}{\gamma\varepsilon}\left(\frac{3\pi p}{2}\right)^{\frac{1}{3}}\right.\nonumber\\
&\left.+\frac{\mu p^{-\frac{1}{3}}f(p)+\gamma V_0}{\gamma\varepsilon}\left(\frac{3\pi p}{2}\right)^{-\frac{1}{3}}-\frac{d}{2}\left(\frac{3\pi p}{2}\right)^{-\frac{1}{3}}-\left(\mu p^{-\frac{1}{3}}f(p)+\gamma V_0\right)\frac{d}{2}\left(\frac{3\pi p}{2}\right)^{-1}\right.\nonumber\\
&\left.-\frac{V_0}{2\varepsilon}\left(\frac{3\pi p}{2}\right)^{-\frac{1}{3}}-\left(\mu p^{-\frac{1}{3}}f(p)+\gamma V_0\right)\frac{V_0}{2\varepsilon}\left(\frac{3\pi p}{2}\right)^{-1}\right]\nonumber\\
&\left[a_2\sin\left(s_2(p)+dt(p)+\frac{\pi}{4}\right)+a_3\cos\left(s_2(p)+dt(p)+\frac{\pi}{4}\right)\right]^2\nonumber\\
&+\pi^{-1}\left[\frac{1}{\gamma\varepsilon}\left(\frac{3\pi p}{2}\right)^{\frac{1}{3}}+\frac{d}{2}\left(\frac{3\pi p}{2}\right)^{-\frac{1}{3}}+\frac{V_0}{2\varepsilon}\left(\frac{3\pi p}{2}\right)^{-\frac{1}{3}}\right]\nonumber\\
&\left[a_3\sin\left(s_2(p)+dt(p)+\frac{\pi}{4}\right)-a_2\cos\left(s_2(p)+dt(p)+\frac{\pi}{4}\right)\right]^2\nonumber\\
&+\pi^{-1}\left[d\left(\frac{3\pi p}{2}\right)^{-\frac{1}{3}}+\left(\frac{\gamma d^2\varepsilon}{2}-\frac{d\gamma V_0}{2}\right)\left(\frac{3\pi p}{2}\right)^{-1}-\frac{1}{\gamma\varepsilon}\left(\frac{3\pi p}{2}\right)^{\frac{1}{3}}\right.\nonumber\\
&\left.-\frac{\mu p^{-\frac{1}{3}}f(p)+\gamma V_0}{\gamma\varepsilon}\left(\frac{3\pi p}{2}\right)^{-\frac{1}{3}}-\frac{d}{2}\left(\frac{3\pi p}{2}\right)^{-\frac{1}{3}}-\left(\mu p^{-\frac{1}{3}}f(p)+\gamma V_0\right)\frac{d}{2}\left(\frac{3\pi p}{2}\right)^{-1}\right.\nonumber\\
&\left.+\frac{V_0}{2\varepsilon}\left(\frac{3\pi p}{2}\right)^{-\frac{1}{3}}+\left(\mu p^{-\frac{1}{3}}f(p)+\gamma V_0\right)\frac{V_0}{2\varepsilon}\left(\frac{3\pi p}{2}\right)^{-1}\right]\nonumber\\
&\left[a_2\sin\left(s_2(p)-dt(p)+\frac{\pi}{4}\right)+a_3\cos\left(s_2(p)-dt(p)+\frac{\pi}{4}\right)\right]^2\nonumber\\
&+\pi^{-1}\left[-\frac{1}{\gamma\varepsilon}\left(\frac{3\pi p}{2}\right)^{\frac{1}{3}}+\frac{d}{2}\left(\frac{3\pi p}{2}\right)^{-\frac{1}{3}}-\frac{V_0}{2\varepsilon}\left(\frac{3\pi p}{2}\right)^{-\frac{1}{3}}\right]\nonumber\\
&\left[a_3\sin\left(s_2(p)-dt(p)+\frac{\pi}{4}\right)-a_2\cos\left(s_2(p)-dt(p)+\frac{\pi}{4}\right)\right]^2\label{Aeq69.9}
\end{align}}
Using the simplifications, (\ref{Aeq69.3.1}) becomes
{\medmuskip=0mu
\thinmuskip=0mu
\thickmuskip=0mu
\begin{align}
I_2&=\pi^{-1}\left[-\left(\frac{\gamma d^2\varepsilon}{2}+\frac{d\gamma V_0}{2}\right)\left(\frac{3\pi p}{2}\right)^{-1}+\frac{\mu p^{-\frac{1}{3}}f(p)+\gamma V_0}{\gamma\varepsilon}\left(\frac{3\pi p}{2}\right)^{-\frac{1}{3}}\right.\nonumber\\
&\left.-\left(\mu p^{-\frac{1}{3}}f(p)+\gamma V_0\right)\frac{d}{2}\left(\frac{3\pi p}{2}\right)^{-1}-\left(\mu p^{-\frac{1}{3}}f(p)+\gamma V_0\right)\frac{V_0}{2\varepsilon}\left(\frac{3\pi p}{2}\right)^{-1}\right]\nonumber\\
&\left[a_2\sin\left(s_2(p)+dt(p)+\frac{\pi}{4}\right)+a_3\cos\left(s_2(p)+dt(p)+\frac{\pi}{4}\right)\right]^2\nonumber\\
&+\pi^{-1}\left[\left(\frac{\gamma d^2\varepsilon}{2}-\frac{d\gamma V_0}{2}\right)\left(\frac{3\pi p}{2}\right)^{-1}-\frac{\mu p^{-\frac{1}{3}}f(p)+\gamma V_0}{\gamma\varepsilon}\left(\frac{3\pi p}{2}\right)^{-\frac{1}{3}}\right.\nonumber\\
&\left.-\left(\mu p^{-\frac{1}{3}}f(p)+\gamma V_0\right)\frac{d}{2}\left(\frac{3\pi p}{2}\right)^{-1}+\left(\mu p^{-\frac{1}{3}}f(p)+\gamma V_0\right)\frac{V_0}{2\varepsilon}\left(\frac{3\pi p}{2}\right)^{-1}\right]\nonumber\\
&\left[a_2\sin\left(s_2(p)-dt(p)+\frac{\pi}{4}\right)+a_3\cos\left(s_2(p)-dt(p)+\frac{\pi}{4}\right)\right]^2\nonumber\\
&+\frac{d}{\pi}\left(\frac{3\pi p}{2}\right)^{-\frac{1}{3}}\left(a^2+a_3^2\right)-\pi^{-1}\frac{V_0}{\varepsilon}\left(\frac{3\pi p}{2}\right)^{-\frac{1}{3}}\left(2a_2a_3\cos\left(2s_2(p)+\frac{\pi}{2}\right)\right.\nonumber\\
&\left.+\left(a_2^2-a_3^2\right)\sin\left(2s_2(p)+\frac{\pi}{2}\right)\right)\sin\left(2dt(p)\right)\label{Aeq69.9.1}
\end{align}}
Now we are ready to expand all the trigonometric terms in the expression.
\begin{align}
\sin\left(s_2(p)\pm dt(p)+\frac{\pi}{4}\right)\approx&\frac{1}{2i}(-1)^pp^\frac{1}{3}\left(\pi^2c_1\right)^{-\frac{1}{2}}\nonumber\\
&\left|\sin\left(c_3p^\frac{1}{3}\right)\right|^{-\frac{1}{2}} e^{i\left(\frac{1}{2}h(p)+\mu^\frac{1}{2}p^\frac{1}{3}\gamma(V_0\pm d\varepsilon)+\frac{\pi}{4}\right)}\label{Aeq70}\\
\cos\left(s_2(p)\pm dt(p)+\frac{\pi}{4}\right)\approx& \frac{1}{2}(-1)^pp^\frac{1}{3}\left(\pi^2c_1\right)^{-\frac{1}{2}}\nonumber\\
&\left|\sin\left(c_3p^\frac{1}{3}\right)\right|^{-\frac{1}{2}} e^{i\left(\frac{1}{2}h(p)+\mu^\frac{1}{2}p^\frac{1}{3}\gamma(V_0\pm d\varepsilon)+\frac{\pi}{4}\right)}\label{Aeq71}\\
\sin\left(2s_2(p)+\frac{\pi}{2}\right)\approx&\frac{1}{2}p^\frac{2}{3}\left(\pi^2c_1\right)^{-1}\left|\sin\left(c_3p^\frac{1}{3}\right)\right|^{-1} e^{i\left(h(p)+2\mu^\frac{1}{2}p^\frac{1}{3}\gamma V_0\right)}\label{Aeq71.1}\\
\cos\left(2s_2(p)+\frac{\pi}{2}\right)\approx&\frac{i}{2}p^\frac{2}{3}\left(\pi^2c_1\right)^{-1}\left|\sin\left(c_3p^\frac{1}{3}\right)\right|^{-1} e^{i\left(h(p)+2\mu^\frac{1}{2}p^\frac{1}{3}\gamma V_0\right)}\label{Aeq71.2}\\
\sin\left(2dt(p)\right)=&\sin\left(2d\mu^\frac{1}{2}p^\frac{1}{3}\gamma\varepsilon\right)\label{Aeq71.3}
\end{align}
With (\ref{Aeq70}) - (\ref{Aeq71.3}) we write (\ref{Aeq69.9.1}) as
{\medmuskip=0mu
\thinmuskip=0mu
\thickmuskip=0mu
\begin{align}
I_2&\approx\pi^{-1}\left[-\left(\frac{\gamma d^2\varepsilon}{2}+\frac{d\gamma V_0}{2}\right)\left(\frac{3\pi p}{2}\right)^{-1}+\frac{\mu p^{-\frac{1}{3}}f(p)+\gamma V_0}{\gamma\varepsilon}\left(\frac{3\pi p}{2}\right)^{-\frac{1}{3}}\right.\nonumber\\
&\left.-\left(\mu p^{-\frac{1}{3}}f(p)+\gamma V_0\right)\frac{d}{2}\left(\frac{3\pi p}{2}\right)^{-1}-\left(\mu p^{-\frac{1}{3}}f(p)+\gamma V_0\right)\frac{V_0}{2\varepsilon}\left(\frac{3\pi p}{2}\right)^{-1}\right]\nonumber\\
&\frac{i}{4}p^\frac{2}{3}\left(\pi^2c_1\right)^{-1}\left|\sin\left(c_3p^\frac{1}{3}\right)\right|^{-1} e^{i2\left(\frac{1}{2}h(p)+\mu^\frac{1}{2}p^\frac{1}{3}\gamma(V_0+ d\varepsilon)\right)}\left(\frac{a_2}{i}+a_3\right)^2\nonumber\\
&+\pi^{-1}\left[\left(\frac{\gamma d^2\varepsilon}{2}-\frac{d\gamma V_0}{2}\right)\left(\frac{3\pi p}{2}\right)^{-1}-\frac{\mu p^{-\frac{1}{3}}f(p)+\gamma V_0}{\gamma\varepsilon}\left(\frac{3\pi p}{2}\right)^{-\frac{1}{3}}\right.\nonumber\\
&\left.-\left(\mu p^{-\frac{1}{3}}f(p)+\gamma V_0\right)\frac{d}{2}\left(\frac{3\pi p}{2}\right)^{-1}+\left(\mu p^{-\frac{1}{3}}f(p)+\gamma V_0\right)\frac{V_0}{2\varepsilon}\left(\frac{3\pi p}{2}\right)^{-1}\right]\nonumber\\
&\frac{i}{4}p^\frac{2}{3}\left(\pi^2c_1\right)^{-1}\left|\sin\left(c_3p^\frac{1}{3}\right)\right|^{-1} e^{i2\left(\frac{1}{2}h(p)+\mu^\frac{1}{2}p^\frac{1}{3}\gamma(V_0- d\varepsilon)\right)}\left(\frac{a_2}{i}+a_3\right)^2\nonumber\\
&+\frac{d}{\pi}\left(\frac{3\pi p}{2}\right)^{-\frac{1}{3}}\left(a^2+a_3^2\right)-\frac{V_0}{\varepsilon\pi}\left(\frac{3\pi p}{2}\right)^{-\frac{1}{3}}\frac{1}{2}p^\frac{2}{3}\left(\pi^2c_1\right)^{-1}\left|\sin\left(c_3p^\frac{1}{3}\right)\right|^{-1}\nonumber\\
& e^{i\left(h(p)+2\mu^\frac{1}{2}p^\frac{1}{3}\gamma V_0\right)}\left(2a_2a_3i+\left(a_2^2-a_3^2\right)\right)\sin\left(2d\mu^\frac{1}{2}p^\frac{1}{3}\gamma\varepsilon\right)\label{Aeq71.4}
\end{align}}
which can simplified as
{\medmuskip=0mu
\thinmuskip=0mu
\thickmuskip=0mu
\begin{align}
I_2&\approx\pi^{-1}\frac{\mu p^{-\frac{1}{3}}f(p)+\gamma V_0}{\gamma\varepsilon}\left(\frac{3\pi}{2}\right)^{-\frac{1}{3}}p^\frac{1}{3}\frac{i}{4}\left(\pi^2c_1\right)^{-1}\left|\sin\left(c_3p^\frac{1}{3}\right)\right|^{-1} \nonumber\\
&e^{i2\left(\frac{1}{2}h(p)+\mu^\frac{1}{2}p^\frac{1}{3}\gamma(V_0+ d\varepsilon)\right)}\left(\frac{a_2}{i}+a_3\right)^2-\pi^{-1}\frac{\mu p^{-\frac{1}{3}}f(p)+\gamma V_0}{\gamma\varepsilon}\left(\frac{3\pi}{2}\right)^{-\frac{1}{3}}p^\frac{1}{3}\nonumber\\
&\frac{i}{4}\left(\pi^2c_1\right)^{-1}\left|\sin\left(c_3p^\frac{1}{3}\right)\right|^{-1}e^{i2\left(\frac{1}{2}h(p)+\mu^\frac{1}{2}p^\frac{1}{3}\gamma(V_0- d\varepsilon)\right)}\left(\frac{a_2}{i}+a_3\right)^2\nonumber\\
&-\pi^{-1}\frac{V_0}{\varepsilon}\left(\frac{3\pi}{2}\right)^{-\frac{1}{3}}\frac{1}{2}p^\frac{1}{3}\left(\pi^2c_1\right)^{-1}\left|\sin\left(c_3p^\frac{1}{3}\right)\right|^{-1} e^{i\left(h(p)+2\mu^\frac{1}{2}p^\frac{1}{3}\gamma V_0\right)}\nonumber\\
&\left(2a_2a_3i+\left(a_2^2-a_3^2\right)\right)\sin\left(2d\mu^\frac{1}{2}p^\frac{1}{3}\gamma\varepsilon\right)\nonumber\\
&=-\frac{\mu p^{-\frac{1}{3}}f(p)+\gamma V_0}{\pi\gamma\varepsilon}\left(\frac{3\pi}{2}\right)^{-\frac{1}{3}}p^\frac{1}{3}\frac{1}{2}\left(\pi^2c_1\right)^{-1}\left|\sin\left(c_3p^\frac{1}{3}\right)\right|^{-1} e^{i\left(h(p)+2\mu^\frac{1}{2}p^\frac{1}{3}\gamma V_0\right)}\nonumber\\
&\left(\frac{a_2}{i}+a_3\right)^2\sin\left(2\mu^\frac{1}{2}p^\frac{1}{3}\gamma d\varepsilon\right)-\pi^{-1}\frac{V_0}{\varepsilon}\left(\frac{3\pi}{2}\right)^{-\frac{1}{3}}\frac{1}{2}p^\frac{1}{3}\left(\pi^2c_1\right)^{-1}\left|\sin\left(c_3p^\frac{1}{3}\right)\right|^{-1}\nonumber\\
& e^{i\left(h(p)+2\mu^\frac{1}{2}p^\frac{1}{3}\gamma V_0\right)}\left(2a_2a_3i+\left(a_2^2-a_3^2\right)\right)\sin\left(2d\mu^\frac{1}{2}p^\frac{1}{3}\gamma\varepsilon\right)\label{Aeq71.4.1}
\end{align}}
Observe, that $-\left(\frac{a_2}{i}+a_3\right)^2=2a_2a_3i+\left(a_2^2-a_3^2\right)$, so the expression (\ref{Aeq71.4.1}) turns to
{\medmuskip=0mu
\thinmuskip=0mu
\thickmuskip=0mu
\begin{align}
I_2&=-\frac{\mu p^{-\frac{1}{3}}f(p)}{\pi\gamma\varepsilon}\left(\frac{3\pi}{2}\right)^{-\frac{1}{3}}p^\frac{1}{3}\frac{1}{2}\left(\pi^2c_1\right)^{-1}\left|\sin\left(c_3p^\frac{1}{3}\right)\right|^{-1} e^{i\left(h(p)+2\mu^\frac{1}{2}p^\frac{1}{3}\gamma V_0\right)}\nonumber\\
&\left(\frac{a_2}{i}+a_3\right)^2\sin\left(2\mu^\frac{1}{2}p^\frac{1}{3}\gamma d\varepsilon\right)\nonumber\\
&=-\frac{f(p)}{\gamma\varepsilon}\left(\frac{3\pi}{2}\right)^{\frac{1}{3}}\frac{1}{2}\left(\pi^3c_1\right)^{-1}\left|\sin\left(c_3p^\frac{1}{3}\right)\right|^{-1} e^{i\left(h(p)+2\mu^\frac{1}{2}p^\frac{1}{3}\gamma V_0\right)}\nonumber\\
&\left(\frac{a_2}{i}+a_3\right)^2\sin\left(2\mu^\frac{1}{2}p^\frac{1}{3}\gamma d\varepsilon\right)\label{Aeq72}
\end{align}}

\subsubsection{\texorpdfstring{$I_3$}{AI3}}
The third integral $I_3$ in (\ref{Aeq37}) is of the form
{\medmuskip=0mu
\thinmuskip=0mu
\thickmuskip=0mu
\begin{align}
I_3&=\int_{d}^{x_c}a_4^2\left(\mathrm{Ci}^+\left(-\gamma\varepsilon\left(x+\frac{\mu\xi_p^\frac{2}{3}}{\gamma\varepsilon}\right)\right)\right)^2\mathrm{d}x\nonumber\\
&=a_4^2\left(x+\frac{\mu\xi_p^\frac{2}{3}}{\gamma\varepsilon}\right)\left(\mathrm{Ci}^+\left(y_1(x)\right)\right)^2+\frac{a_4^2}{\gamma\varepsilon}\left(\mathrm{Ci'}^+\left(y_1(x)\right)\right)^2\Big|_{x=d}^{x=x_c}\nonumber\\
&=a_4^2\left(x_c+\frac{\mu\xi_p^\frac{2}{3}}{\gamma\varepsilon}\right)\left(\mathrm{Ci}^+\left(y_1(x_c)\right)\right)^2+\frac{a_4^2}{\gamma\varepsilon}\left(\mathrm{Ci'}^+\left(y_1(x_c)\right)\right)^2\nonumber\\
&-a_4^2\left(d+\frac{\mu\xi_p^\frac{2}{3}}{\gamma\varepsilon}\right)\left(\mathrm{Ci}^+\left(y_1(d)\right)\right)^2-\frac{a_4^2}{\gamma\varepsilon}\left(\mathrm{Ci'}^+\left(y_1(d)\right)\right)^2\label{Aeq73}
\end{align}}
Since $\mathrm{Ci}^+=\mathrm{Bi}+i\mathrm{Ai}$, we can use the expressions (\ref{Aeq56}), (\ref{Aeq57}) computed for the integral $I_2$ and let $a_2=ia_4,a_3=a_4$. Changing the boundary values in (\ref{Aeq69.3.1}), we have
{\medmuskip=0mu
\thinmuskip=0mu
\thickmuskip=0mu
\begin{align}
I_3\approx& \pi^{-1}\left(x_c\left[\left(\frac{3\pi p}{2}\right)^{-\frac{1}{3}}-\frac{\gamma x_c\varepsilon}{2}\left(\frac{3\pi p}{2}\right)^{-1}\right]\right.\nonumber\\
&\left.+\frac{\mu\left(p^\frac{2}{3}+p^{-\frac{1}{3}}f(p)\right)}{\gamma\varepsilon}\left[\left(\frac{3\pi p}{2}\right)^{-\frac{1}{3}}-\frac{\gamma x_c\varepsilon}{2}\left(\frac{3\pi p}{2}\right)^{-1}\right]\right)a_4^2e^{i2\left(s_1(p)+x_ct(p)+\frac{\pi}{4}\right)}\nonumber\\
&-\frac{a_4^2}{\gamma\varepsilon}\pi^{-1}\left[\left(\frac{3\pi p}{2}\right)^{\frac{1}{3}}+\frac{\gamma x_c\varepsilon}{2}\left(\frac{3\pi p}{2}\right)^{-\frac{1}{3}}\right]e^{i2\left(s_1(p)+x_ct(p)+\frac{\pi}{4}\right)}\nonumber\\
&-\left[\left(d\left[\left(\frac{3\pi p}{2}\right)^{-\frac{1}{3}}-\frac{\gamma d\varepsilon}{2}\left(\frac{3\pi p}{2}\right)^{-1}\right]\right.\right.\nonumber\\
&\left.\left.+\frac{\mu\left(p^\frac{2}{3}+p^{-\frac{1}{3}}f(p)\right)}{\gamma\varepsilon}\left[\left(\frac{3\pi p}{2}\right)^{-\frac{1}{3}}-\frac{\gamma d\varepsilon}{2}\left(\frac{3\pi p}{2}\right)^{-1}\right]\right) a_4^2\pi^{-1}e^{i2\left(s_1(p)+dt(p)+\frac{\pi}{4}\right)}\right.\nonumber\\
&\left.-\frac{a_4^2}{\gamma\varepsilon}\pi^{-1}\left[\left(\frac{3\pi p}{2}\right)^{\frac{1}{3}}+\frac{\gamma d\varepsilon}{2}\left(\frac{3\pi p}{2}\right)^{-\frac{1}{3}}\right]e^{i2\left(s_1(p)+dt(p)+\frac{\pi}{4}\right)}\right]\nonumber\\
=&\pi^{-1}\left(x_c\left[-\frac{\gamma x_c\varepsilon}{2}\left(\frac{3\pi p}{2}\right)^{-1}\right]+\frac{\mu p^{-\frac{1}{3}}f(p)}{\gamma\varepsilon}\left[\left(\frac{3\pi p}{2}\right)^{-\frac{1}{3}}-\frac{\gamma x_c\varepsilon}{2}\left(\frac{3\pi p}{2}\right)^{-1}\right]\right)\nonumber\\
&a_4^2e^{i2\left(s_1(p)+x_ct(p)+\frac{\pi}{4}\right)}-a_4^2\pi^{-1}\left[\left(d\left[-\frac{\gamma d\varepsilon}{2}\left(\frac{3\pi p}{2}\right)^{-1}\right]\right.\right.\nonumber\\
&\left.\left.+\frac{\mu p^{-\frac{1}{3}}f(p)}{\gamma\varepsilon}\left[\left(\frac{3\pi p}{2}\right)^{-\frac{1}{3}}-\frac{\gamma d\varepsilon}{2}\left(\frac{3\pi p}{2}\right)^{-1}\right]\right) e^{i2\left(s_1(p)+dt(p)+\frac{\pi}{4}\right)}\right]\nonumber\\
&=\left(-\frac{\gamma x_c^2\varepsilon}{2}\left(\frac{3\pi p}{2}\right)^{-1}+\frac{\mu p^{-\frac{1}{3}}f(p)}{\gamma\varepsilon}\left[\left(\frac{3\pi p}{2}\right)^{-\frac{1}{3}}-\frac{\gamma x_c\varepsilon}{2}\left(\frac{3\pi p}{2}\right)^{-1}\right]\right)\nonumber\\
&a_4^2\pi^{-1}e^{i2\left(s_1(p)+x_ct(p)+\frac{\pi}{4}\right)}\nonumber\\
&+\left(\frac{\gamma d^2\varepsilon}{2}\left(\frac{3\pi p}{2}\right)^{-1}-\frac{\mu p^{-\frac{1}{3}}f(p)}{\gamma\varepsilon}\left[\left(\frac{3\pi p}{2}\right)^{-\frac{1}{3}}-\frac{\gamma d\varepsilon}{2}\left(\frac{3\pi p}{2}\right)^{-1}\right]\right)\nonumber\\
&a_4^2\pi^{-1}e^{i2\left(s_1(p)+dt(p)+\frac{\pi}{4}\right)}
\label{Aeq74}
\end{align}}
The exponential terms in this expression can be simplified as
\begin{align}
&e^{i2\left(s_1(p)+xt(p)+\frac{\pi}{4}\right)}\approx p^\frac{2}{3}\left|\pi^2c_1\sin\left(c_3p^\frac{1}{3}\right)\right|^{-1}e^{i\left(h(p)+\frac{\pi}{2}\right)}e^{ix2(3\pi\varepsilon p)^\frac{1}{3}}\label{Aeq75}
\end{align}
Using this simplification, the integral (\ref{Aeq74}) becomes
{\medmuskip=0mu
\thinmuskip=0mu
\thickmuskip=0mu
\begin{align}
I_3\approx& \left(-\frac{\gamma x_c^2\varepsilon}{2}\left(\frac{3\pi p}{2}\right)^{-1}+\frac{\mu p^{-\frac{1}{3}}f(p)}{\gamma\varepsilon}\left[\left(\frac{3\pi p}{2}\right)^{-\frac{1}{3}}-\frac{\gamma x_c\varepsilon}{2}\left(\frac{3\pi p}{2}\right)^{-1}\right]\right)\nonumber\\
&a_4^2\pi^{-1}p^\frac{2}{3}\left|\pi^2c_1\sin\left(c_3p^\frac{1}{3}\right)\right|^{-1}e^{i\left(h(p)+\frac{\pi}{2}\right)}e^{ix_c2(3\pi\varepsilon p)^\frac{1}{3}}\nonumber\\
&+\left(\frac{\gamma d^2\varepsilon}{2}\left(\frac{3\pi p}{2}\right)^{-1}-\frac{\mu p^{-\frac{1}{3}}f(p)}{\gamma\varepsilon}\left[\left(\frac{3\pi p}{2}\right)^{-\frac{1}{3}}-\frac{\gamma d\varepsilon}{2}\left(\frac{3\pi p}{2}\right)^{-1}\right]\right)a_4^2\pi^{-1}\nonumber\\
&p^\frac{2}{3}\left|\pi^2c_1\sin\left(c_3p^\frac{1}{3}\right)\right|^{-1}e^{i\left(h(p)+\frac{\pi}{2}\right)}e^{id2(3\pi\varepsilon p)^\frac{1}{3}}\nonumber\\
\approx&\frac{\mu p^{-\frac{1}{3}}f(p)}{\gamma\varepsilon}\left(\frac{3\pi p}{2}\right)^{-\frac{1}{3}}a_4^2\pi^{-1}p^\frac{2}{3}\left|\pi^2c_1\sin\left(c_3p^\frac{1}{3}\right)\right|^{-1}e^{i\left(h(p)+\frac{\pi}{2}\right)}e^{ix_c2(3\pi\varepsilon p)^\frac{1}{3}}\nonumber\\
&-\frac{\mu p^{-\frac{1}{3}}f(p)}{\gamma\varepsilon}\left(\frac{3\pi p}{2}\right)^{-\frac{1}{3}}a_4^2\pi^{-1}p^\frac{2}{3}\left|\pi^2c_1\sin\left(c_3p^\frac{1}{3}\right)\right|^{-1}e^{i\left(h(p)+\frac{\pi}{2}\right)}e^{id2(3\pi\varepsilon p)^\frac{1}{3}}\nonumber\\
&=\frac{f(p)}{\gamma\varepsilon}\left(\frac{3\pi}{2}\right)^{\frac{1}{3}}a_4^2\left|\pi^3c_1\sin\left(c_3p^\frac{1}{3}\right)\right|^{-1}e^{i\left(h(p)+\frac{\pi}{2}\right)}\left(e^{ix_c2(3\pi\varepsilon p)^\frac{1}{3}}-e^{id2(3\pi\varepsilon p)^\frac{1}{3}}\right)
\label{Aeq76}
\end{align}}

\subsubsection{\texorpdfstring{$I_4$}{AI4}}
The fourth integral $I_4$ from (\ref{Aeq37}) is
\begin{align}
I_4&=i\int_{x_c}^\infty a_4^2\left(\mathrm{Ci}^+\left(-\gamma\varepsilon i\left(x+\frac{\mu\xi_p^\frac{2}{3}+\gamma \varepsilon x_c(1-i)}{\gamma\varepsilon i}\right)\right)\right)^2\mathrm{d}x\nonumber\\
&=ia_4^2\left(x+\frac{\mu\xi_p^\frac{2}{3}+\gamma \varepsilon x_c(1-i)}{\gamma\varepsilon i}\right)\left(\mathrm{Ci}^+\left(\tilde{y}(x)\right)\right)^2+i\frac{a_4^2}{\gamma\varepsilon i}\left(\mathrm{Ci'}^+\left(\tilde{y}(x)\right)\right)^2\Big|_{x=x_c}^{x=\infty}\nonumber\\
&=-ia_4^2\left(x_c+\frac{\mu\xi_p^\frac{2}{3}+\gamma \varepsilon x_c(1-i)}{\gamma\varepsilon i}\right)\left(\mathrm{Ci}^+\left(\tilde{y}(x_c)\right)\right)^2-\frac{a_4^2}{\gamma\varepsilon}\left(\mathrm{Ci'}^+\left(\tilde{y}(x_c)\right)\right)^2\label{Aeq77}
\end{align}
We can see that this integral is just the first two terms in the beginning of (\ref{Aeq74}) with a different sign and with a corresponding changes according to (\ref{Aeq56.1}) and (\ref{Aeq56.2}).
{\medmuskip=0mu
\thinmuskip=0mu
\thickmuskip=0mu
\begin{align}
I_4&\approx -i\pi^{-1}\left(x_c\left[\left(\frac{3\pi p}{2}\right)^{-\frac{1}{3}}-\frac{\gamma(ix_c\varepsilon+\varepsilon x_c(1-i))}{2}\left(\frac{3\pi p}{2}\right)^{-1}\right]\right.\nonumber\\
&\left.+\frac{\mu\left(p^\frac{2}{3}+p^{-\frac{1}{3}}f(p)\right)+\gamma \varepsilon x_c(1-i)}{\gamma\varepsilon i}\left[\left(\frac{3\pi p}{2}\right)^{-\frac{1}{3}}-\frac{\gamma (ix_c\varepsilon+\varepsilon x_c(1-i))}{2}\left(\frac{3\pi p}{2}\right)^{-1}\right]\right)\nonumber\\
&a_4^2e^{i2\left(\tilde{s}(p)+x_c\tilde{t}(p)+\frac{\pi}{4}\right)}\nonumber\\
&+\frac{a_4^2}{\gamma\varepsilon}\pi^{-1}\left[\left(\frac{3\pi p}{2}\right)^{\frac{1}{3}}+\frac{\gamma (ix_c\varepsilon+\varepsilon x_c(1-i))}{2}\left(\frac{3\pi p}{2}\right)^{-\frac{1}{3}}\right]e^{i2\left(\tilde{s}(p)+x_c\tilde{t}(p)+\frac{\pi}{4}\right)}\nonumber\\
=&-i\pi^{-1}\left(x_c\left[\left(\frac{3\pi p}{2}\right)^{-\frac{1}{3}}-\frac{\gamma\varepsilon x_c}{2}\left(\frac{3\pi p}{2}\right)^{-1}\right]\right.\nonumber\\
&\left.+\frac{\mu p^{-\frac{1}{3}}f(p)+\gamma \varepsilon x_c(1-i)}{\gamma\varepsilon i}\left[\left(\frac{3\pi p}{2}\right)^{-\frac{1}{3}}-\frac{\gamma \varepsilon x_c}{2}\left(\frac{3\pi p}{2}\right)^{-1}\right]\right)a_4^2e^{i2\left(\tilde{s}(p)+x_c\tilde{t}(p)+\frac{\pi}{4}\right)}\nonumber\\
&+\frac{a_4^2}{\gamma\varepsilon}\pi^{-1}\gamma\varepsilon x_c\left(\frac{3\pi p}{2}\right)^{-\frac{1}{3}}e^{i2\left(\tilde{s}(p)+x_c\tilde{t}(p)+\frac{\pi}{4}\right)}\nonumber\\
=&-i\pi^{-1}\left(x_c\left[\left(\frac{3\pi p}{2}\right)^{-\frac{1}{3}}-\frac{\gamma\varepsilon x_c}{2}\left(\frac{3\pi p}{2}\right)^{-1}\right]\right.\nonumber\\
&\left.+\frac{\mu p^{-\frac{1}{3}}f(p)}{\gamma\varepsilon i}\left[\left(\frac{3\pi p}{2}\right)^{-\frac{1}{3}}-\frac{\gamma \varepsilon x_c}{2}\left(\frac{3\pi p}{2}\right)^{-1}\right]\right.\nonumber\\
&\left.+\frac{x_c(1-i)}{i}\left[\left(\frac{3\pi p}{2}\right)^{-\frac{1}{3}}-\frac{\gamma \varepsilon x_c}{2}\left(\frac{3\pi p}{2}\right)^{-1}\right]\right)a_4^2e^{i2\left(\tilde{s}(p)+x_c\tilde{t}(p)+\frac{\pi}{4}\right)}\nonumber\\
&+a_4^2\pi^{-1} x_c\left(\frac{3\pi p}{2}\right)^{-\frac{1}{3}}e^{i2\left(\tilde{s}(p)+x_c\tilde{t}(p)+\frac{\pi}{4}\right)}\nonumber\\
=&-i\pi^{-1}\left(x_c\left[\left(\frac{3\pi p}{2}\right)^{-\frac{1}{3}}-\frac{\gamma\varepsilon x_c}{2}\left(\frac{3\pi p}{2}\right)^{-1}\right]\right.\nonumber\\
&\left.+\frac{\mu p^{-\frac{1}{3}}f(p)}{\gamma\varepsilon i}\left[\left(\frac{3\pi p}{2}\right)^{-\frac{1}{3}}-\frac{\gamma \varepsilon x_c}{2}\left(\frac{3\pi p}{2}\right)^{-1}\right]\right.\nonumber\\
&\left.+\frac{x_c}{i}\left[\left(\frac{3\pi p}{2}\right)^{-\frac{1}{3}}-\frac{\gamma \varepsilon x_c}{2}\left(\frac{3\pi p}{2}\right)^{-1}\right]-x_c\left[\left(\frac{3\pi p}{2}\right)^{-\frac{1}{3}}-\frac{\gamma \varepsilon x_c}{2}\left(\frac{3\pi p}{2}\right)^{-1}\right]\right)\nonumber\\
&a_4^2e^{i2\left(\tilde{s}(p)+x_c\tilde{t}(p)+\frac{\pi}{4}\right)}+a_4^2\pi^{-1} x_c\left(\frac{3\pi p}{2}\right)^{-\frac{1}{3}}e^{i2\left(\tilde{s}(p)+x_c\tilde{t}(p)+\frac{\pi}{4}\right)}\nonumber\\
=&-i\pi^{-1}\left(x_c\left[\left(\frac{3\pi p}{2}\right)^{-\frac{1}{3}}-\frac{\gamma\varepsilon x_c}{2}\left(\frac{3\pi p}{2}\right)^{-1}\right]\right.\nonumber\\
&\left.+\frac{\mu p^{-\frac{1}{3}}f(p)}{\gamma\varepsilon i}\left[\left(\frac{3\pi p}{2}\right)^{-\frac{1}{3}}-\frac{\gamma \varepsilon x_c}{2}\left(\frac{3\pi p}{2}\right)^{-1}\right]-\frac{\gamma \varepsilon x_c^2}{2i}\left(\frac{3\pi p}{2}\right)^{-1}\right.\nonumber\\
&\left.-x_c\left[\left(\frac{3\pi p}{2}\right)^{-\frac{1}{3}}-\frac{\gamma \varepsilon x_c}{2}\left(\frac{3\pi p}{2}\right)^{-1}\right]\right)a_4^2e^{i2\left(\tilde{s}(p)+x_c\tilde{t}(p)+\frac{\pi}{4}\right)}
\label{Aeq78.1}
\end{align}}
As before, let us expand the exponential in (\ref{Aeq78.1}).
\begin{align}
&e^{i2\left(\tilde{s}(p)+x_c\tilde{t}(p)+\frac{\pi}{4}\right)}\approx p^\frac{2}{3}\left(\pi^2c1\right)^{-1}\ln\left(\left|\sin\left(c_3p^\frac{1}{3}\right)\right|\right)^{-1}e^{i\left(h(p)+2\mu^\frac{1}{2}p^\frac{1}{3}\gamma\varepsilon x_c+\frac{\pi}{2}\right)}\label{Aeq79}
\end{align}
Using this in (\ref{Aeq78.1}) we get
{\medmuskip=0mu
\thinmuskip=0mu
\thickmuskip=0mu
\begin{align}
I_4&=-i\pi^{-1}\left(x_c\left[\left(\frac{3\pi p}{2}\right)^{-\frac{1}{3}}-\frac{\gamma\varepsilon x_c}{2}\left(\frac{3\pi p}{2}\right)^{-1}\right]\right.\nonumber\\
&\left.+\frac{\mu p^{-\frac{1}{3}}f(p)}{\gamma\varepsilon i}\left[\left(\frac{3\pi p}{2}\right)^{-\frac{1}{3}}-\frac{\gamma \varepsilon x_c}{2}\left(\frac{3\pi p}{2}\right)^{-1}\right]-\frac{\gamma \varepsilon x_c^2}{2i}\left(\frac{3\pi p}{2}\right)^{-1}\right.\nonumber\\
&\left.-x_c\left[\left(\frac{3\pi p}{2}\right)^{-\frac{1}{3}}-\frac{\gamma \varepsilon x_c}{2}\left(\frac{3\pi p}{2}\right)^{-1}\right]\right)a_4^2e^{i2\left(\tilde{s}(p)+x_c\tilde{t}(p)+\frac{\pi}{4}\right)}\nonumber\\
\approx&-i\pi^{-1}\left(x_c\left[\left(\frac{3\pi p}{2}\right)^{-\frac{1}{3}}-\frac{\gamma\varepsilon x_c}{2}\left(\frac{3\pi p}{2}\right)^{-1}\right]\right.\nonumber\\
&\left.+\frac{\mu p^{-\frac{1}{3}}f(p)}{\gamma\varepsilon i}\left[\left(\frac{3\pi p}{2}\right)^{-\frac{1}{3}}-\frac{\gamma \varepsilon x_c}{2}\left(\frac{3\pi p}{2}\right)^{-1}\right]-\frac{\gamma \varepsilon x_c^2}{2i}\left(\frac{3\pi p}{2}\right)^{-1}\right.\nonumber\\
&\left.-x_c\left[\left(\frac{3\pi p}{2}\right)^{-\frac{1}{3}}-\frac{\gamma \varepsilon x_c}{2}\left(\frac{3\pi p}{2}\right)^{-1}\right]\right)a_4^2p^\frac{2}{3}\left(\pi^2c1\right)^{-1}\ln\left(\left|\sin\left(c_3p^\frac{1}{3}\right)\right|\right)^{-1}\nonumber\\
&e^{i\left(h(p)+2\mu^\frac{1}{2}p^\frac{1}{3}\gamma\varepsilon x_c+\frac{\pi}{2}\right)}\nonumber\\
\approx&-i\pi^{-1}\left(\frac{\mu p^{-\frac{1}{3}}f(p)}{\gamma\varepsilon i}\left(\frac{3\pi p}{2}\right)^{-\frac{1}{3}}\right)a_4^2p^\frac{2}{3}\left(\pi^2c1\right)^{-1}\ln\left(\left|\sin\left(c_3p^\frac{1}{3}\right)\right|\right)^{-1}\nonumber\\
&e^{i\left(h(p)+2\mu^\frac{1}{2}p^\frac{1}{3}\gamma\varepsilon x_c+\frac{\pi}{2}\right)}\nonumber\\
=&-\frac{f(p)}{\gamma\varepsilon}\left(\frac{3\pi}{2}\right)^{\frac{1}{3}}a_4^2\left(\pi^3c1\right)^{-1}\ln\left(\left|\sin\left(c_3p^\frac{1}{3}\right)\right|\right)^{-1}e^{i\left(h(p)+2\mu^\frac{1}{2}p^\frac{1}{3}\gamma\varepsilon x_c+\frac{\pi}{2}\right)}\label{Aeq79.1}
\end{align}}

\subsection{Normalization coefficients}
We can now put together all the parts (\ref{Aeq65}), (\ref{Aeq72}), (\ref{Aeq76}) and (\ref{Aeq79.1}) of $N_p$ in (\ref{Aeq37}). Observe, that all terms are of order $\sim 1$ except $I_1$ which is $\sim p^\frac{1}{3}$, so this will dominate and we get a normalization factor
\begin{align}
&N_p\approx \frac{a_1^2}{\gamma\varepsilon}\pi^{-1}\left(\frac{3\pi p}{2}\right)^{\frac{1}{3}}\label{Aeq80}
\end{align}

\subsection{\texorpdfstring{Continuity coefficients}{Aai}}\label{Aai}
\begin{figure}[H]
\captionsetup{width=0.85\textwidth}
\begin{center}
\includegraphics[scale=0.28]{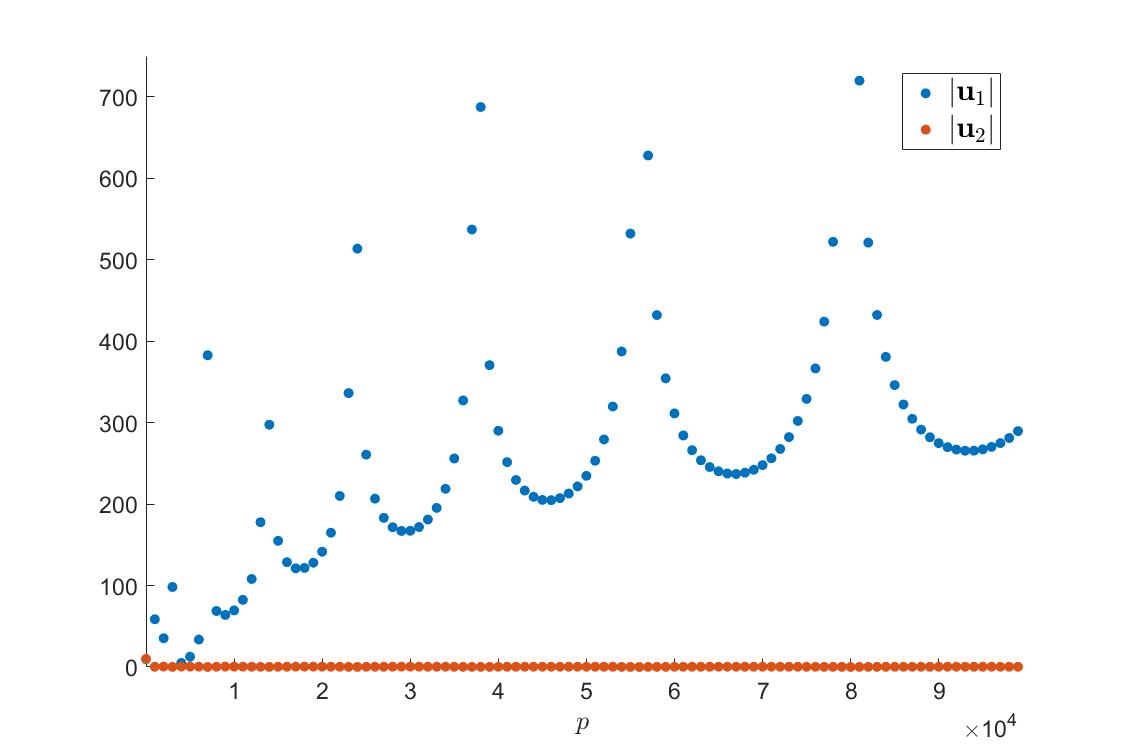}
\end{center}
\caption{The absolute value of the product $\mathbf{M}.\mathbf{n}_{1,2}=\mathbf{u}_{1,2}$ of the full continuity matrix $\mathbf{M}$ and both of the two vectors $\mathbf{n}_{1,2}$ computed from the leading order of its asymptotic form $\mathbf{M}_{\infty}$ as null-space. The expectation is that one of the vectors is staying low having small values in comparison to the second one. This helps to recognize which of them is belonging to the zero eigenvalue.\label{fig3}}
\end{figure}

When it comes to the continuity coefficients $a_1,\ldots,a_4$ used in the previous computations, we assumed, they were all independent of $p$ but we did not really show this. Let us do it in this chapter. The continuity matrix we use to compute the coefficients is
\begin{align}
\mathbf{M}=
\begin{pmatrix}
A_0 & -A_1 & -B_1 & 0\\
A'_0 & -A'_1 & -B'_1 & 0\\
0 & A_2 & B_2 & -C_3\\
0 & A'_2 & B'_2 & -C'_3
\end{pmatrix}\label{Aeq100}
\end{align}
where
\begin{align}
A_0&=\mathrm{Ai}(y_1(-d)) & A_1&=\mathrm{Ai}(y_2(-d)) & B_1&=\mathrm{Bi}(y_2(-d))\label{Aeq101}\\
A_2&=\mathrm{Ai}(y_2(d)) & B_2&=\mathrm{Bi}(y_2(d)) & C_3&=\mathrm{Ci}^+(y_1(d))\label{Aeq102}
\end{align}
We can gather all the relevant information to have an asymptotic formula to the above quantities up to the leading order.
{\medmuskip=-1mu
\thinmuskip=-1mu
\thickmuskip=-1mu
\begin{align}
A_0&\approx\pi^{-\frac{1}{2}}\left(\frac{3\pi}{2}\right)^{-\frac{1}{6}}p^{-\frac{1}{6}}\frac{1}{2i}(-1)^pp^\frac{1}{3}\left(\pi^2c_1\right)^{-\frac{1}{2}}\left|\sin\left(c_3p^\frac{1}{3}\right)\right|^{-\frac{1}{2}}e^{i\left(\frac{1}{2}h(p)- d\mu^\frac{1}{2}p^\frac{1}{3}\gamma\varepsilon+\frac{\pi}{4}\right)}\label{Aeq118}\\
A_1&\approx\pi^{-\frac{1}{2}}\left(\frac{3\pi}{2}\right)^{-\frac{1}{6}}p^{-\frac{1}{6}}\frac{1}{2i}(-1)^pp^\frac{1}{3}\left(\pi^2c_1\right)^{-\frac{1}{2}}\left|\sin\left(c_3p^\frac{1}{3}\right)\right|^{-\frac{1}{2}}\nonumber\\
& e^{i\left(\frac{1}{2}h(p)+\mu^\frac{1}{2}p^\frac{1}{3}\gamma(V_0- d\varepsilon)+\frac{\pi}{4}\right)}\label{Aeq119}\\
B_1&\approx\pi^{-\frac{1}{2}}\left(\frac{3\pi}{2}\right)^{-\frac{1}{6}}p^{-\frac{1}{6}}\frac{1}{2}(-1)^pp^\frac{1}{3}\left(\pi^2c_1\right)^{-\frac{1}{2}}\left|\sin\left(c_3p^\frac{1}{3}\right)\right|^{-\frac{1}{2}}\nonumber\\
&e^{i\left(\frac{1}{2}h(p)+\mu^\frac{1}{2}p^\frac{1}{3}\gamma(V_0- d\varepsilon)+\frac{\pi}{4}\right)}\label{Aeq120}\\
A_2&\approx\pi^{-\frac{1}{2}}\left(\frac{3\pi}{2}\right)^{-\frac{1}{6}}p^{-\frac{1}{6}}\frac{1}{2i}(-1)^pp^\frac{1}{3}\left(\pi^2c_1\right)^{-\frac{1}{2}}\left|\sin\left(c_3p^\frac{1}{3}\right)\right|^{-\frac{1}{2}}\nonumber\\
& e^{i\left(\frac{1}{2}h(p)+\mu^\frac{1}{2}p^\frac{1}{3}\gamma(V_0+ d\varepsilon)+\frac{\pi}{4}\right)}\label{Aeq121}\\
B_2&\approx\pi^{-\frac{1}{2}}\left(\frac{3\pi}{2}\right)^{-\frac{1}{6}}p^{-\frac{1}{6}}\frac{1}{2}(-1)^pp^\frac{1}{3}\left(\pi^2c_1\right)^{-\frac{1}{2}}\left|\sin\left(c_3p^\frac{1}{3}\right)\right|^{-\frac{1}{2}}\nonumber\\
& e^{i\left(\frac{1}{2}h(p)+\mu^\frac{1}{2}p^\frac{1}{3}\gamma(V_0+ d\varepsilon)+\frac{\pi}{4}\right)}\label{Aeq122}\\
C_3&\approx\pi^{-\frac{1}{2}}\left(\frac{3\pi}{2}\right)^{-\frac{1}{6}}p^{-\frac{1}{6}}(-1)^pp^\frac{1}{3}\left|\pi^2c_1\sin\left(c_3p^\frac{1}{3}\right)\right|^{-\frac{1}{2}}e^{i\frac{1}{2}\left(h(p)+\frac{\pi}{2}\right)}e^{id(3\pi\varepsilon p)^\frac{1}{3}}\label{Aeq123}
\end{align}}
And now their derivatives.
{\medmuskip=-1mu
\thinmuskip=-1mu
\thickmuskip=-1mu
\begin{align}
A'_0&\approx-\pi^{-\frac{1}{2}}\left(\frac{3\pi}{2}\right)^{\frac{1}{6}}p^{\frac{1}{6}}\frac{1}{2}(-1)^pp^\frac{1}{3}\left(\pi^2c_1\right)^{-\frac{1}{2}}\left|\sin\left(c_3p^\frac{1}{3}\right)\right|^{-\frac{1}{2}}e^{i\left(\frac{1}{2}h(p)- d\mu^\frac{1}{2}p^\frac{1}{3}\gamma\varepsilon+\frac{\pi}{4}\right)}\label{Aeq124}\\
A'_1&\approx-\pi^{-\frac{1}{2}}\left(\frac{3\pi}{2}\right)^{\frac{1}{6}}p^{\frac{1}{6}}\frac{1}{2}(-1)^pp^\frac{1}{3}\left(\pi^2c_1\right)^{-\frac{1}{2}}\left|\sin\left(c_3p^\frac{1}{3}\right)\right|^{-\frac{1}{2}}\nonumber\\
& e^{i\left(\frac{1}{2}h(p)+\mu^\frac{1}{2}p^\frac{1}{3}\gamma(V_0- d\varepsilon)+\frac{\pi}{4}\right)}\label{Aeq125}\\
B'_1&\approx\pi^{-\frac{1}{2}}\left(\frac{3\pi}{2}\right)^{\frac{1}{6}}p^{\frac{1}{6}}\frac{1}{2i}(-1)^pp^\frac{1}{3}\left(\pi^2c_1\right)^{-\frac{1}{2}}\left|\sin\left(c_3p^\frac{1}{3}\right)\right|^{-\frac{1}{2}}\nonumber\\
& e^{i\left(\frac{1}{2}h(p)+\mu^\frac{1}{2}p^\frac{1}{3}\gamma(V_0- d\varepsilon)+\frac{\pi}{4}\right)}\label{Aeq126}\\
A'_2&\approx-\pi^{-\frac{1}{2}}\left(\frac{3\pi}{2}\right)^{\frac{1}{6}}p^{\frac{1}{6}}\frac{1}{2}(-1)^pp^\frac{1}{3}\left(\pi^2c_1\right)^{-\frac{1}{2}}\left|\sin\left(c_3p^\frac{1}{3}\right)\right|^{-\frac{1}{2}}\nonumber\\
&e^{i\left(\frac{1}{2}h(p)+\mu^\frac{1}{2}p^\frac{1}{3}\gamma(V_0+ d\varepsilon)+\frac{\pi}{4}\right)}\label{Aeq127}\\
B'_2&\approx\pi^{-\frac{1}{2}}\left(\frac{3\pi}{2}\right)^{\frac{1}{6}}p^{\frac{1}{6}}\frac{1}{2i}(-1)^pp^\frac{1}{3}\left(\pi^2c_1\right)^{-\frac{1}{2}}\left|\sin\left(c_3p^\frac{1}{3}\right)\right|^{-\frac{1}{2}}\nonumber\\
&e^{i\left(\frac{1}{2}h(p)+\mu^\frac{1}{2}p^\frac{1}{3}\gamma(V_0+ d\varepsilon)+\frac{\pi}{4}\right)}\label{Aeq128}\\
C'_3&\approx-i\pi^{-\frac{1}{2}}\left(\frac{3\pi}{2}\right)^{\frac{1}{6}}p^{\frac{1}{6}}(-1)^pp^\frac{1}{3}\left|\pi^2c_1\sin\left(c_3p^\frac{1}{3}\right)\right|^{-\frac{1}{2}}e^{i\frac{1}{2}\left(h(p)+\frac{\pi}{2}\right)}e^{id(3\pi\varepsilon p)^\frac{1}{3}}\label{Aeq129}
\end{align}}
If we put these expression into the matrix (\ref{Aeq100}), we can remove the common factors in front of all the terms, since the equation we have is $\mathbf{M}\mathbf{a}=\mathbf{0}$. We see, that the common factors are\\
$\pi^{-\frac{1}{2}}(-1)^pp^\frac{1}{3}\left(\pi^2c_1\right)^{-\frac{1}{2}}\left|\sin\left(c_3p^\frac{1}{3}\right)\right|^{-\frac{1}{2}}e^{i\frac{1}{2}\left(h(p)+\frac{\pi}{2}\right)}$. So the matrix will be
\begin{align}
\left(\begin{matrix}
\left(\frac{3\pi}{2}\right)^{-\frac{1}{6}}p^{-\frac{1}{6}}\frac{1}{2i} e^{-i d\mu^\frac{1}{2}p^\frac{1}{3}\gamma\varepsilon} & -\left(\frac{3\pi}{2}\right)^{-\frac{1}{6}}p^{-\frac{1}{6}}\frac{1}{2i} e^{i\mu^\frac{1}{2}p^\frac{1}{3}\gamma(V_0- d\varepsilon)}\\
-\left(\frac{3\pi}{2}\right)^{\frac{1}{6}}p^{\frac{1}{6}}\frac{1}{2}e^{-i d\mu^\frac{1}{2}p^\frac{1}{3}\gamma\varepsilon} & \left(\frac{3\pi}{2}\right)^{\frac{1}{6}}p^{\frac{1}{6}}\frac{1}{2}e^{i\mu^\frac{1}{2}p^\frac{1}{3}\gamma(V_0- d\varepsilon)}\\
0 & \left(\frac{3\pi}{2}\right)^{-\frac{1}{6}}p^{-\frac{1}{6}}\frac{1}{2i} e^{i\mu^\frac{1}{2}p^\frac{1}{3}\gamma(V_0+ d\varepsilon)}\\
0 & -\left(\frac{3\pi}{2}\right)^{\frac{1}{6}}p^{\frac{1}{6}}\frac{1}{2} e^{i\mu^\frac{1}{2}p^\frac{1}{3}\gamma(V_0+ d\varepsilon)}
\end{matrix}\ldots\right.\nonumber\\
\left.
\quad\quad\quad\quad\ldots\begin{matrix}
-\left(\frac{3\pi}{2}\right)^{-\frac{1}{6}}p^{-\frac{1}{6}}\frac{1}{2}e^{i\mu^\frac{1}{2}p^\frac{1}{3}\gamma(V_0- d\varepsilon)} & 0\\
-\left(\frac{3\pi}{2}\right)^{\frac{1}{6}}p^{\frac{1}{6}}\frac{1}{2i}e^{i\mu^\frac{1}{2}p^\frac{1}{3}\gamma(V_0- d\varepsilon)} & 0\\
\left(\frac{3\pi}{2}\right)^{-\frac{1}{6}}p^{-\frac{1}{6}}\frac{1}{2} e^{i\mu^\frac{1}{2}p^\frac{1}{3}\gamma(V_0+ d\varepsilon)} & -\left(\frac{3\pi}{2}\right)^{-\frac{1}{6}}p^{-\frac{1}{6}}e^{id(3\pi\varepsilon p)^\frac{1}{3}}\\
\left(\frac{3\pi}{2}\right)^{\frac{1}{6}}p^{\frac{1}{6}}\frac{1}{2i}e^{i\mu^\frac{1}{2}p^\frac{1}{3}\gamma(V_0+ d\varepsilon)} & i\left(\frac{3\pi}{2}\right)^{\frac{1}{6}}p^{\frac{1}{6}}e^{id(3\pi\varepsilon p)^\frac{1}{3}}
\end{matrix}\right)
\label{Aeq130}
\end{align}
Note that $e^{i d\mu^\frac{1}{2}p^\frac{1}{3}\gamma\varepsilon}=e^{id(3\pi\varepsilon p)^\frac{1}{3}}$. If we denote
\begin{align}
A&=\left(\frac{3\pi}{2}\right)^{\frac{1}{6}}p^{\frac{1}{6}}\label{Aeq131}\\
B&=e^{i d\mu^\frac{1}{2}p^\frac{1}{3}\gamma\varepsilon}\label{Aeq132}\\
C&=e^{i V_0\mu^\frac{1}{2}p^\frac{1}{3}\gamma}\label{Aeq133}
\end{align}
we can write
{\medmuskip=1mu
\thinmuskip=1mu
\thickmuskip=1mu
\nulldelimiterspace=1pt
\scriptspace=1pt
\begin{align}
\mathbf{M}_{\infty}=
\begin{pmatrix}
A^{-1}\frac{1}{2i} B^{-1} & -A^{-1}\frac{1}{2i} B^{-1}C & -A^{-1}\frac{1}{2}B^{-1}C & 0\\
-A\frac{1}{2}B^{-1} & A\frac{1}{2}B^{-1}C & -A\frac{1}{2i}B^{-1}C & 0\\
0 & A^{-1}\frac{1}{2i}B C & A^{-1}\frac{1}{2}B C & -A^{-1}B\\
0 & -A\frac{1}{2}B C & A\frac{1}{2i}BC & iAB
\end{pmatrix}\label{Aeq134}
\end{align}}
We find, that its determinant is 0 and the null space are the vectors
\begin{align}
\mathbf{n}_1=\begin{pmatrix}
2i\\
\frac{2i}{C}\\
0\\
1
\end{pmatrix},\quad
\mathbf{n}_2=\begin{pmatrix}
0\\
-i\\
1\\
0
\end{pmatrix}\label{Aeq135}
\end{align}
There are two vectors in the null space of the matrix. This tells us, there is a degenerate eigenvalue 0, for which there exist these two vectors $\mathbf{n}_1,\mathbf{n}_2$. However, when we compute the vector of coefficients $\mathbf{a}$, we can not see any degeneration. One possible explanation to this is, that by taking the leading order of the matrix terms and removing all other smaller terms, we lose one eigenvalue of 4 in total and it gets merged with another one. We can write this as $\mathbf{M}\approx \mathbf{M}_{\infty}+\epsilon\mathbf{M}_1$, where $\mathbf{M}_{\infty}$ is the leading term matrix, $\epsilon$ is a small positive number and $\mathbf{M}_{1}$ is containing the next terms. So we have a degenerate eigenvalue for $\epsilon=0$. When $\epsilon$ becomes positive and non zero, this degenerate eigenvalue starts to split into two separate eigengvalues belonging to two different eigenvectors. In the very beginnning (for $\epsilon\ll 1$), they are close to each other, but as $\epsilon$ gets bigger, they tend to move from each other. We can ilustrate this by computing the product $\mathbf{M}.\mathbf{n}_{1,2}=\mathbf{u}_{1,2}$, taking their norm $\Vert\mathbf{u}_{1,2}\Vert$ and see which one of them is staying close to zero and which takes a different direction. The figure \ref{fig3} is showing the results. It is clear, that the correct asympototic vector of coefficients is $\mathbf{a}\approx\mathbf{n}_2$. Note, that the real vector of coefficients $\mathbf{a}$ is never going to be $\mathbf{n}_2$, because they are computed from the matrix $\mathbf{M}$ for which $\epsilon\neq 0$.

\section{}\label{App:AppendixC}
As we did in the previous Appendix for A-series, this one will be about finding the asymptotic expression for the C-series coefficients $c_p$ in the sum(\ref{eq348}) at the point $x=0$, where
\begin{align}
c_p=\frac{b_p(0)}{N_p}\label{Ceq0}
\end{align}
The process and the steps taken are basically the same. To keep track of things, we remind some formulas to help to follow. The basis we use are the resonant states
\begin{align}
\psi_\omega(x)=\chi\left\{
\begin{array}{l l}
a_1\mathrm{Ai}(y_1(x)) & x<-d\\
a_2\mathrm{Ai}(y_2(x))+a_3\mathrm{Bi}(y_2(x)) & -d< x< d\\
a_4\mathrm{Ci}^+(y_1(x)) & d<x<x_c\\
a_4\mathrm{Ci}^+(y_3(x)) & x_c<x
\end{array}\right.\label{Ceq1}
\end{align}
where
\begin{align}
y_1(x)&=-\gamma(\varepsilon x+\omega)\label{Ceq1.1}\\
y_2(x)&=-\gamma(\varepsilon x+V_0+\omega)\label{Ceq2}\\
y_3(x)&=-\gamma\left(i\varepsilon x+\varepsilon x_c(1-i)+\omega\right)\label{Ceq2.0}
\end{align}
where we denoted
\begin{align}
\gamma=2(2\varepsilon)^{-\frac{2}{3}}\label{Ceq5}
\end{align}
and the constants $a_1,\ldots,a_4$ are ensuring the continuity of the states. Let our function we want to expand $f(x)$ be
\begin{align}
f(x)=\left\{\begin{array}{c c}
1 & -a\leq x \leq a\\
0 & \mathrm{otherwise}
\end{array}\right.\label{Ceq2.1}
\end{align}
for $0<a<d$. In \cite{Stegun} we can choose between two types of asymptotic formulas for $\mathrm{Bi}$: 10.4.63 or 10.4.65, which we can then use to express the determinant and solve it in the $\xi$-plane. We found out, that the formula 10.4.63 gives us solutions to the determinant $\xi_p$ such that if we look at the angle of the argument $z$ to $\mathrm{Bi}$, we get $|\arg(z)|>\frac{\pi}{3}$. Which is violating the usage condition of this asymptotic formula. On the other hand, if we choose the formula 10.4.65 and compute $\xi_p$, we find that the angle of the argument is near 0, which is safe. That is why we stay with the formula 10.4.65 further on. This formula is
\begin{align}
\mathrm{Bi}\left(ze^{\pm i\frac{\pi}{3}}\right)&\approx\sqrt{\frac{2}{\pi}}e^{\pm i\frac{\pi}{6}}z^{-\frac{1}{4}}\frac{1}{2i}\left(e^{i\left(\frac{2}{3}z^\frac{3}{2}+\frac{\pi}{4}\mp\frac{i}{2}\ln2\right)}-e^{-i\left(\frac{2}{3}z^\frac{3}{2}+\frac{\pi}{4}\mp\frac{i}{2}\ln2\right)}\right)\label{Ceq3.1}
\end{align}
The argument is therefore
\begin{align}
y_{1,2,3}(x)&=ze^{\pm i\frac{\pi}{3}}\\
y_{1,2,3}(x)e^{\mp i\frac{\pi}{3}}&=z\label{Ceq3.2}
\end{align}
We expect the zeros of the determinant located in the angle-region between $-\pi$ and $-\frac{2\pi}{3}$. The $\omega$'s in $y_{1,2,3}(x)$ have a negative sign, so eventually, they appear as $\omega e^{\pm i\frac{2\pi}{3}}$. We see that it is safer to choose $e^{i\frac{2\pi}{3}}$, because the zeros then fall near the 0 angle branch. So the argument for $\mathrm{Bi}$ are
\begin{align}
\mathrm{Bi}\left(y_{1,2,3}^B\right)&\approx\sqrt{\frac{2}{\pi}}e^{i\frac{\pi}{6}}\left(y_{1,2,3}^B\right)^{-\frac{1}{4}}\frac{1}{2i}\left(e^{i\left(\frac{2}{3}\left(y_{1,2,3}^B\right)^\frac{3}{2}+\frac{\pi}{4}-\frac{i}{2}\ln2\right)}\right.\nonumber\\
&\left.-e^{-i\left(\frac{2}{3}\left(y_{1,2,3}^B\right)^\frac{3}{2}+\frac{\pi}{4}-\frac{i}{2}\ln2\right)}\right)\label{Ceq3.3}\\
y_1^B(x)&=\gamma(\varepsilon x+\omega)e^{i\frac{2\pi}{3}}\label{Ceq3.4}\\
y_2^B(x)&=\gamma(\varepsilon x+V_0+\omega)e^{i\frac{2\pi}{3}}\label{Ceq3.5}\\
y_3^B(x)&=\gamma\left(i\varepsilon x+\varepsilon x_c(1-i)+\omega\right)e^{i\frac{2\pi}{3}}\label{Ceq3.6}
\end{align}
where we indicated with an upperscript $B$ that these go into Airy $Bi$ function. The Airy $\mathrm{Ai}$ does not have such problems, because the formula
\begin{align}
\mathrm{Ai}\left(y_{1,2,3}^A\right)&\approx\frac{1}{2}\pi^{-\frac{1}{2}}\left(y_{1,2,3}^A\right)^{-\frac{1}{4}}e^{-\frac{2}{3}\left(y_{1,2,3}^A\right)^\frac{3}{2}}\label{Ceq3.7}\\
\end{align}
where
\begin{align}
y_1^A(x)&=\gamma(-\varepsilon x+\tilde{\omega})\label{Ceq3.8}\\
y_2^A(x)&=\gamma(-\varepsilon x-V_0+\tilde{\omega})\label{Ceq3.9}\\
y_3^A(x)&=\gamma\left(-i\varepsilon x-\varepsilon x_c(1-i)+\tilde{\omega}\right)\label{Ceq3.10}
\end{align}
works for every argument and we denoted
\begin{align}
\omega(\xi)&=-\tilde{\omega}(\xi)\label{Ceq3}
\end{align}
The transformation from the $\omega$-plane into $\xi$-plane is then
\begin{align}
\omega(\xi)&=\gamma^{-1}\left(\frac{3\pi}{2}\xi\right)^\frac{2}{3}e^{-i\frac{2\pi}{3}}\label{Ceq4}\\
\tilde{\omega}(\xi)&=-\gamma^{-1}\left(\frac{3\pi}{2}\xi\right)^\frac{2}{3}e^{-i\frac{2\pi}{3}}\label{Ceq4.1}
\end{align}

\subsection{Asymptotic C-series determinant}
The asymptotic form of the determinant is obtained in a similar way as for A-series, we only have in this case different asymptotic Airy functions to work with. We provide an overview over the key terms used to obtain the final form. AS we already know the determinant has the form
\begin{align}
\det\mathbf{M}(\omega)=(A_0A'_1-A'_0A_1)(B_2C'_3-B'_2C_3)-(A_0B'_1-A'_0B_1)(A_2C'_3-A'_2C_3)\label{Ceq4.2}
\end{align}
together with
\begin{align}
\begin{array}{l l}
A_0=\mathrm{Ai}\left(-2\nu\left(-d\varepsilon+\omega\right)\right) & A_1=\mathrm{Ai}\left(-2\nu\left(-d\varepsilon+\omega+V_0\right)\right)\\
A_2=\mathrm{Ai}\left(-2\nu\left(d\varepsilon+\omega+V_0\right)\right) & B_2=\mathrm{Bi}\left(-2\nu\left(d\varepsilon+\omega+V_0\right)\right)\\
B_1=\mathrm{Bi}\left(-2\nu\left(-d\varepsilon+\omega+V_0\right)\right) & C_3=\mathrm{Ci}^{+}\left(-2\nu\left(d\varepsilon+\omega\right)\right)
\end{array}\label{Ceq4.3}
\end{align}
where $\nu=(2\varepsilon)^{-\frac{2}{3}}$. We refer to the formulas presented in this section. We express first the exponents and some other terms. The complexity in this case is given by the fact that we use different arguments for $\mathrm{Ai}$ and $\mathrm{Bi}$.
\begin{align}
\zeta^{\pm}_{1A}&=\frac{2}{3}(y_1^A(\pm d))^\frac{3}{2}\approx\frac{2}{3}(\gamma\tilde{\omega})^\frac{3}{2}\mp\gamma\varepsilon d(\gamma\tilde{\omega})^\frac{1}{2}\nonumber\\
\zeta^\pm_{2A}&=\frac{2}{3}(y_2^A(\pm d))^\frac{3}{2}\approx\frac{2}{3}(\gamma\tilde{\omega})^\frac{3}{2}-\gamma(\varepsilon d\pm V_0)(\gamma\tilde{\omega})^\frac{1}{2}\label{Ceq4.4}\\
\zeta^{\pm}_{1B}&=\frac{2}{3}(y_1^B(\pm d))^\frac{3}{2}\approx-\frac{2}{3}(\gamma\omega)^\frac{3}{2}\mp\gamma\varepsilon d(\gamma\omega)^\frac{1}{2}\nonumber\\
\zeta^\pm_{2B}&=\frac{2}{3}(y_2^B(\pm d))^\frac{3}{2}\approx-\frac{2}{3}(\gamma\omega)^\frac{3}{2}-\gamma(\varepsilon d\pm V_0)(\gamma\omega)^\frac{1}{2}\label{Ceq4.5}
\end{align}
With the same argumentation in A-series, if we look at the structure of the determinant, the following simplifications can be used.
\begin{align}
y_1^A(\pm d)^{-\frac{1}{4}}y_2^A(\pm d)^{\frac{1}{4}}&\approx1-\frac{V_0}{4\tilde{\omega}} & y_1^A(\pm d)^{\frac{1}{4}}y_2^A(\pm d)^{-\frac{1}{4}}&\approx1+\frac{V_0}{4\tilde{\omega}}\label{Ceq4.6}\\
y_1^B(\pm d)^{-\frac{1}{4}}y_2^B(\pm d)^{\frac{1}{4}}&\approx1+\frac{V_0}{4\omega} & y_1^B(\pm d)^{\frac{1}{4}}y_2^B(\pm d)^{-\frac{1}{4}}&\approx1-\frac{V_0}{4\tilde{\omega}}\label{Ceq4.7}
\end{align}
We can see from the form of the determinant (\ref{Ceq4.2}) that are also going to need the mixture of two different arguments.
{\medmuskip=0mu
\thinmuskip=0mu
\thickmuskip=0mu
\begin{align}
y_1^A(\pm d)^{-\frac{1}{4}}y_2^B(\pm d)^{\frac{1}{4}}&\approx e^{-i\frac{\pi}{12}}\left(1+\frac{V_0}{4\omega}\right) & y_1^A(\pm d)^{\frac{1}{4}}y_2^B(\pm d)^{-\frac{1}{4}}&\approx e^{i\frac{\pi}{12}}\left(1-\frac{V_0}{4\omega}\right)\label{Ceq4.8}\\
y_1^B(\pm d)^{-\frac{1}{4}}y_2^A(\pm d)^{\frac{1}{4}}&\approx e^{i\frac{\pi}{12}}\left(1+\frac{V_0}{4\omega}\right) & y_1^B(\pm d)^{\frac{1}{4}}y_2^A(\pm d)^{-\frac{1}{4}}&\approx e^{-i\frac{\pi}{12}}\left(1-\frac{V_0}{4\omega}\right)\label{Ceq4.9}
\end{align}}
After some tedious calculations the final result is
{\medmuskip=0mu
\thinmuskip=0mu
\thickmuskip=0mu
\begin{align}
\det\mathbf{M}(\omega,\tilde{\omega})\approx&-\frac{1}{\pi^2}+\frac{V_0}{4\pi^2\omega}\left(\sin\left(2\gamma\varepsilon d\sqrt{\gamma\omega}\right)e^{-i\frac{2}{3}(\gamma\omega)^\frac{3}{2}-2(\gamma\tilde{\omega})^\frac{3}{2}}\right.\nonumber\\
&\left.+\sin\left(\gamma V_0\sqrt{\gamma\omega}\right)e^{-\frac{4}{3}(\gamma\tilde{\omega})^\frac{3}{2}+(\gamma\tilde{\omega})^\frac{1}{2}\gamma(V_0-2d\varepsilon)}\right.\nonumber\\
&\left.+\sin\left((\gamma\omega)^\frac{1}{2}\gamma(2d\varepsilon-V_0)\right)e^{-\frac{4}{3}(\gamma\tilde{\omega})^\frac{3}{2}+(\gamma\tilde{\omega})^\frac{1}{2}\gamma V_0}\right)
\end{align}}
Using the transformations (\ref{Ceq4}) and (\ref{Ceq4.1}) and neglecting the decaying terms, the determinant in this case becomes
\begin{align}
\det\mathbf{M}(\xi)=-\frac{1}{\pi^2}+c_1\xi^{-\frac{2}{3}}e^{i\frac{\pi}{6}}e^{-2i\pi\xi}e^{c_2\xi^\frac{1}{3}e^{i\frac{\pi}{6}}}\label{Ceq5.1}
\end{align}
Here we use the fact that $\xi=\xi_p=p+s$, where $|s|\ll p$ to simplify the powers using Taylor expansion and solve the equation $\det\mathbf{M}(\xi)=0$. The asymptotic formula for the zeros of the determinant then is
\begin{align}
\xi_p=p+\frac{i}{3\pi}\ln(p)-i\frac{1}{2\pi}\ln\left(\pi^2c_1\right)-ic_2p^\frac{1}{3}-\frac{1}{2\pi}h(p)\label{Ceq6}
\end{align}
where
\begin{align}
c_1&=\frac{V_0}{4\pi^2}\gamma\left(\frac{3\pi}{2}\right)^{-\frac{2}{3}}\label{Ceq7}\\
c_2&=\frac{\sqrt{3}\gamma d\varepsilon}{2\pi}\left(\frac{3\pi}{2}\right)^{\frac{1}{3}}\label{Ceq7.1}\\
h(p)&=\arg\left(e^{-i\left(4\pi c_2p^\frac{1}{3}\frac{1}{2\sqrt{3}}+\frac{\pi}{6}\right)}\right)\label{Ceq7.2}
\end{align}

\subsection{\texorpdfstring{Coefficients $b_p(0)$}{Cbp}}
In this chapter we look in more details into the coefficients $b_p(0)$. They are computed as
\begin{align}
b_p(0)&=\psi_{\omega_p}(0)\int_{-a}^a\psi_{\omega_p}(x)\mathrm{d}x\nonumber\\
&=\left(a_2\mathrm{Ai}\left( y_2^A(0)\right)+a_3\mathrm{Bi}\left( y_2^B(0)\right)\right)\int_{-a}^a\left(a_2\mathrm{Ai}\left( y_2^A(x)\right)+a_3\mathrm{Bi}\left( y_2^B(x)\right)\right)\mathrm{d}x\label{Ceq2.3}
\end{align}
The asymptotic formulas for the Airy functions were set up in the beginning of this Appendix and we know that the arguments to them are different. Using the $\xi$-plane transformation we can write the arguments as
\begin{align}
z_1^B(\xi_p)&=y_1^B(x)|_{\omega=\omega(\xi_p)}=\gamma(\varepsilon x+\omega)e^{i\frac{2\pi}{3}}\nonumber\\
&=\gamma\varepsilon xe^{i\frac{2\pi}{3}}+\gamma\gamma^{-1}\left(\frac{3\pi}{2}\xi_p\right)^\frac{2}{3}e^{-i\frac{2\pi}{3}}e^{i\frac{2\pi}{3}}=\gamma\varepsilon xe^{i\frac{2\pi}{3}}+\beta\xi_p^\frac{2}{3}\label{Ceq14}\\
z_2^B(\xi_p)&=y_2^B(x)|_{\omega=\omega(\xi_p)}=\gamma(\varepsilon x+V_0)e^{i\frac{2\pi}{3}}+\beta\xi_p^\frac{2}{3}\label{Ceq15}\\
z_3^B(\xi_p)&=y_3^B(x)|_{\omega=\omega(\xi_p)}=\gamma(i\varepsilon x+\varepsilon x_c(1-i))e^{i\frac{2\pi}{3}}+\beta\xi_p^\frac{2}{3}\label{Ceq16}
\end{align}
where
\begin{align}
\beta=\left(\frac{3\pi}{2}\right)^\frac{2}{3}\label{Ceq17}
\end{align}
and
\begin{align}
z_1^A(\xi_p)&=y_1^A(x)|_{\tilde{\omega}=\tilde{\omega}(\xi_p)}=\gamma(-\varepsilon x+\tilde{\omega})=-\gamma\varepsilon x-\gamma\gamma^{-1}\left(\frac{3\pi}{2}\xi_p\right)^\frac{2}{3}e^{-i\frac{2\pi}{3}}\nonumber\\
&=-\gamma\varepsilon x+\mu\xi_p^\frac{2}{3}\label{Ceq18}\\
z_2^A(\xi_p)&=y_2^A(x)|_{\tilde{\omega}=\tilde{\omega}(\xi_p)}=-\gamma(\varepsilon x+V_0)+\mu\xi_p^\frac{2}{3}\label{Ceq19}\\
z_3^A(\xi_p)&=y_3^A(x)|_{\tilde{\omega}=\tilde{\omega}(\xi_p)}=-\gamma(i\varepsilon x+\varepsilon x_c(1-i))+\mu\xi_p^\frac{2}{3}\label{Ceq20}
\end{align}
where
\begin{align}
\mu=-\beta e^{-i\frac{2\pi}{3}}=\beta e^{i\frac{\pi}{3}}\label{Ceq21}
\end{align}
The coefficients $b_p(0)$ then become
{\medmuskip=0mu
\thinmuskip=0mu
\thickmuskip=0mu
\begin{align}
b_p(0)=&\left(a_2\mathrm{Ai}\left(z_2^A(\xi_p)|_{x=0}\right)+a_3\mathrm{Bi}\left(z_2^B(\xi_p)|_{x=0}\right)\right)\nonumber\\
&\int_{-a}^a\left(a_2\mathrm{Ai}\left(y_2^A(x)\right)+a_3\mathrm{Bi}\left(y_2(x)^B\right)\right)\mathrm{d}x\nonumber\\
=&\left(a_2\frac{1}{2}\pi^{-\frac{1}{2}}\left(z_2^A(\xi_p)|_{x=0}\right)^{-\frac{1}{4}}e^{-\frac{2}{3}\left(z_2^A(\xi_p)|_{x=0}\right)^\frac{3}{2}}+a_3\sqrt{\frac{2}{\pi}}e^{i\frac{\pi}{6}}\left(z_2^B(\xi_p)|_{x=0}\right)^{-\frac{1}{4}}\right.\nonumber\\
&\left.\frac{1}{2i}\left(e^{i\left(\frac{2}{3}\left(z_2^B(\xi_p)|_{x=0}\right)^\frac{3}{2}+\frac{\pi}{4}-\frac{i}{2}\ln2\right)}-e^{-i\left(\frac{2}{3}\left(z_2^B(\xi_p)|_{x=0}\right)^\frac{3}{2}+\frac{\pi}{4}-\frac{i}{2}\ln2\right)}\right)\right)\nonumber\\
&\int_{-a}^a\left(a_2\mathrm{Ai}\left(y_2^A(x)\right)+a_3\mathrm{Bi}\left(y_2(x)^B\right)\right)\mathrm{d}x\label{Ceq22}
\end{align}}
We are going to need the following arguments to express $\left(z_2^{A,B}(\xi_p)\right)^{-\frac{1}{4}}$ and $\frac{2}{3}\left(z_2^{A,B}(\xi_p)\right)^\frac{3}{2}$.
\begin{align}
\left(z_2^{A}(\xi_p)\right)^{-\frac{1}{4}}&=\left(-\gamma(\varepsilon x+V_0)+\mu\xi_p^\frac{2}{3}\right)^{-\frac{1}{4}}\approx\mu^{-\frac{1}{4}}p^{-\frac{1}{6}}\label{Ceq23}\\
\left(z_2^{B}(\xi_p)\right)^{-\frac{1}{4}}&=\left(\gamma(\varepsilon x+V_0)e^{i\frac{2\pi}{3}}+\beta\xi_p^\frac{2}{3}\right)^{-\frac{1}{4}}\approx\beta^{-\frac{1}{4}}p^{-\frac{1}{6}}\label{Ceq24}
\end{align}
where we used $\xi_p^{-\frac{1}{6}}\approx p^{-\frac{1}{6}}$. And the exponents
\begin{align}
\frac{2}{3}\left(z_2^{A}(\xi_p)\right)^\frac{3}{2}&=\frac{2}{3}\left(-\gamma(\varepsilon x+V_0)+\mu\xi_p^\frac{2}{3}\right)^\frac{3}{2}\approx\frac{2}{3}\mu^\frac{3}{2}\xi_p-\gamma(\varepsilon x+V_0)\mu^\frac{1}{2}p^\frac{1}{3}\nonumber\\
&=s_A(p)+xt_A(p)\label{Ceq25}\\
\frac{2}{3}\left(z_2^{B}(\xi_p)\right)^\frac{3}{2}&=\frac{2}{3}\left(\gamma(\varepsilon x+V_0)e^{i\frac{2\pi}{3}}+\beta\xi_p^\frac{2}{3}\right)^\frac{3}{2}\approx\frac{2}{3}\beta^\frac{3}{2}\xi_p+\gamma(\varepsilon x+V_0)e^{i\frac{2\pi}{3}}\beta^\frac{1}{2}p^\frac{1}{3}\nonumber\\
&=s_B(p)+xt_B(p)\label{Ceq26}
\end{align}
where we used $\xi_p^\frac{1}{3}\approx p^\frac{1}{3}$ and denoted
\begin{align}
s_A(p)&=\frac{2}{3}\mu^\frac{3}{2}\xi_p-\gamma V_0\mu^\frac{1}{2}p^\frac{1}{3}=i\pi\xi_p-\gamma V_0\mu^\frac{1}{2}p^\frac{1}{3} & t_A(p)&=-\gamma \varepsilon\mu^\frac{1}{2}p^\frac{1}{3}\label{Ceq28}\\
s_B(p)&=\frac{2}{3}\beta^\frac{3}{2}\xi_p+\gamma V_0e^{i\frac{2\pi}{3}}\beta^\frac{1}{2}p^\frac{1}{3}=\pi\xi_p+\gamma V_0e^{i\frac{2\pi}{3}}\beta^\frac{1}{2}p^\frac{1}{3} & t_B(p)&=\gamma\varepsilon e^{i\frac{2\pi}{3}}\beta^\frac{1}{2}p^\frac{1}{3}\label{Ceq30}
\end{align}
Let us use these simplifications and write (\ref{Ceq22}) as
{\medmuskip=0mu
\thinmuskip=0mu
\thickmuskip=0mu
\begin{align}
b_p(0)\approx&\left(a_2\frac{1}{2}\pi^{-\frac{1}{2}}\mu^{-\frac{1}{4}}p^{-\frac{1}{6}}e^{-s_A(p)}+a_3\sqrt{\frac{2}{\pi}}e^{i\frac{\pi}{6}}\beta^{-\frac{1}{4}}p^{-\frac{1}{6}}\sin\left(s_B(p)+\frac{\pi}{4}-\frac{i}{2}\ln2\right)\right)\nonumber\\
&\int_{-a}^a\left(a_2\frac{1}{2}\pi^{-\frac{1}{2}}\mu^{-\frac{1}{4}}p^{-\frac{1}{6}}e^{-(s_A(p)+xt_A(p))}\right.\nonumber\\
&\left.+a_3\sqrt{\frac{2}{\pi}}e^{i\frac{\pi}{6}}\beta^{-\frac{1}{4}}p^{-\frac{1}{6}}\sin\left(s_B(p)+xt_B(p)+\frac{\pi}{4}-\frac{i}{2}\ln2\right)\right)\mathrm{d}x\label{Ceq31}
\end{align}}
The integral should be easy to compute. We perform two substitutions. For the first term it is $y_A=s_A(p)+xt_A(p)\Rightarrow \mathrm{d}x=\frac{\mathrm{d}y_A}{t_A(p)}$ and for the other one we get $y_B=s_B(p)+xt_B(p)+\frac{\pi}{4}-\frac{i}{2}\ln2\Rightarrow\mathrm{d}x=\frac{\mathrm{d}y_B}{t_B(p)}$. The coefficients then become
{\medmuskip=0mu
\thinmuskip=0mu
\thickmuskip=0mu
\begin{align}
b_p(0)&=\left(a_2\frac{1}{2}\pi^{-\frac{1}{2}}\mu^{-\frac{1}{4}}p^{-\frac{1}{6}}e^{-s_A(p)}+a_3\sqrt{\frac{2}{\pi}}e^{i\frac{\pi}{6}}\beta^{-\frac{1}{4}}p^{-\frac{1}{6}}\sin\left(s_B(p)+\frac{\pi}{4}-\frac{i}{2}\ln2\right)\right)\nonumber\\
&\left(\frac{1}{t_A(p)}a_2\frac{1}{2}\pi^{-\frac{1}{2}}\mu^{-\frac{1}{4}}p^{-\frac{1}{6}}e^{-s_A(p)}\left(e^{at_A(p)}-e^{-at_A(p)}\right)\right.\nonumber\\
&\left.+\frac{2}{t_B(p)}a_3\sqrt{\frac{2}{\pi}}e^{i\frac{\pi}{6}}\beta^{-\frac{1}{4}}p^{-\frac{1}{6}}\sin\left(s_B(p)+\frac{\pi}{4}-\frac{i}{2}\ln2\right)\sin\left(at_B(p)\right)\right)\nonumber\\
=&a_2\frac{1}{2}\pi^{-\frac{1}{2}}\mu^{-\frac{1}{4}}p^{-\frac{1}{6}}e^{-s_A(p)}\frac{1}{t_A(p)}a_2\frac{1}{2}\pi^{-\frac{1}{2}}\mu^{-\frac{1}{4}}p^{-\frac{1}{6}}e^{-s_A(p)}\left(e^{at_A(p)}-e^{-at_A(p)}\right)\nonumber\\
&+a_3\sqrt{\frac{2}{\pi}}e^{i\frac{\pi}{6}}\beta^{-\frac{1}{4}}p^{-\frac{1}{6}}\sin\left(s_B(p)+\frac{\pi}{4}-\frac{i}{2}\ln2\right)\frac{1}{t_A(p)}a_2\frac{1}{2}\pi^{-\frac{1}{2}}\mu^{-\frac{1}{4}}p^{-\frac{1}{6}}e^{-s_A(p)}\nonumber\\
&\left(e^{at_A(p)}-e^{-at_A(p)}\right)+a_2\frac{1}{2}\pi^{-\frac{1}{2}}\mu^{-\frac{1}{4}}p^{-\frac{1}{6}}e^{-s_A(p)}\frac{2}{t_B(p)}a_3\sqrt{\frac{2}{\pi}}e^{i\frac{\pi}{6}}\beta^{-\frac{1}{4}}p^{-\frac{1}{6}}\nonumber\\
&\sin\left(s_B(p)+\frac{\pi}{4}-\frac{i}{2}\ln2\right)\sin\left(at_B(p)\right)+a_3\sqrt{\frac{2}{\pi}}e^{i\frac{\pi}{6}}\beta^{-\frac{1}{4}}p^{-\frac{1}{6}}\nonumber\\
&\sin\left(s_B(p)+\frac{\pi}{4}-\frac{i}{2}\ln2\right)\frac{2}{t_B(p)}a_3\sqrt{\frac{2}{\pi}}e^{i\frac{\pi}{6}}\beta^{-\frac{1}{4}}p^{-\frac{1}{6}}\sin\left(s_B(p)+\frac{\pi}{4}-\frac{i}{2}\ln2\right)\nonumber\\
&\sin\left(at_B(p)\right)\nonumber\\
=&a_2^2\frac{1}{4}\pi^{-1}\mu^{-\frac{1}{2}}p^{-\frac{1}{3}}e^{-2s_A(p)}\frac{1}{t_A(p)}\left(e^{at_A(p)}-e^{-at_A(p)}\right)\nonumber\\
&+a_3a_2\sqrt{\frac{1}{2}}e^{i\frac{\pi}{12}}\beta^{-\frac{1}{2}}\frac{1}{t_A(p)}\pi^{-1}p^{-\frac{1}{3}}\sin\left(s_B(p)+\frac{\pi}{4}-\frac{i}{2}\ln2\right)e^{-s_A(p)}\nonumber\\
&\left(e^{at_A(p)}-e^{-at_A(p)}\right)+a_2a_3\pi^{-1}p^{-\frac{1}{3}}e^{-s_A(p)}\frac{2}{t_B(p)}\sqrt{\frac{1}{2}}e^{i\frac{\pi}{12}}\beta^{-\frac{1}{2}}\nonumber\\
&\sin\left(s_B(p)+\frac{\pi}{4}-\frac{i}{2}\ln2\right)\sin\left(at_B(p)\right)+a_3^2\frac{4}{\pi}e^{i\frac{\pi}{3}}\beta^{-\frac{1}{2}}p^{-\frac{1}{3}}\sin^2\left(s_B(p)+\frac{\pi}{4}-\frac{i}{2}\ln2\right)\nonumber\\
&\frac{1}{t_B(p)}\sin\left(at_B(p)\right)\label{Ceq33}
\end{align}}
Let us now expand the exponentials and trigonometric terms.
\begin{align}
e^{-s_A(p)}&=(-1)^pp^{\frac{1}{3}}\left(\pi^2c_1\right)^{-\frac{1}{2}}e^{\frac{\sqrt{3}\gamma}{2}\left(\frac{3\pi}{2}\right)^{\frac{1}{3}}p^\frac{1}{3}(V_0-d\varepsilon)+\frac{i}{2}h(p)+\gamma V_0\left(\frac{3\pi}{2}\right)^{\frac{1}{3}}\frac{1}{2}i p^\frac{1}{3}}\label{Ceq34}\\
e^{at_A(p)}&=\approx 0\label{Ceq35}\\
e^{-at_A(p)}&= e^{a\gamma \varepsilon\left(\frac{3\pi}{2}\right)^{\frac{1}{3}}\frac{\sqrt{3}}{2}p^\frac{1}{3}+a\gamma \varepsilon\left(\frac{3\pi}{2}\right)^{\frac{1}{3}}\frac{1}{2}ip^\frac{1}{3}}\label{Ceq35.1}
\end{align}
and
\begin{align}
\sin\left(s_B(p)+\frac{\pi}{4}-\frac{i}{2}\ln2\right)&=\frac{1}{2i}\left((-1)^pp^{-\frac{1}{3}}\left(\pi^2c_1\right)^{\frac{1}{2}}2^{\frac{1}{2}} \right.\nonumber\\
&\left.e^{\frac{\sqrt{3}\gamma}{2}\left(\frac{3\pi}{2}\right)^{\frac{1}{3}}p^\frac{1}{3}(d\varepsilon-V_0)-i\frac{1}{2}h(p)-\gamma V_0\frac{1}{2}i \left(\frac{3\pi}{2}\right)^{\frac{1}{3}}p^\frac{1}{3}+i\frac{\pi}{4}}\right.\nonumber\\
&\left.-(-1)^pp^{\frac{1}{3}}\left(\pi^2c_1\right)^{-\frac{1}{2}}2^{-\frac{1}{2}} \right.\nonumber\\
&\left.e^{-\frac{\sqrt{3}\gamma}{2}\left(\frac{3\pi}{2}\right)^{\frac{1}{3}}p^\frac{1}{3}(d\varepsilon-V_0)+i\frac{1}{2}h(p)+\gamma V_0\frac{1}{2}i \left(\frac{3\pi}{2}\right)^{\frac{1}{3}}p^\frac{1}{3}-i\frac{\pi}{4}}\right)\label{Ceq36}\\
\sin\left(at_B(p)\right)&\approx-\frac{1}{2i}e^{ia\gamma\varepsilon \frac{1}{2}\left(\frac{3\pi}{2}\right)^{\frac{1}{3}}p^\frac{1}{3}+a\gamma\varepsilon \frac{\sqrt{3}}{2}\left(\frac{3\pi}{2}\right)^{\frac{1}{3}}p^\frac{1}{3}}\label{Ceq37}
\end{align}
Observe, that the second and third term in (\ref{Ceq33}) are the same taking into account the simplifications made above. The expression (\ref{Ceq33}) is complicated, but after some long calculations we find that
\begin{align}
b_p(0)&=e^{a\gamma\varepsilon \frac{\sqrt{3}}{2}\left(\frac{3\pi}{2}\right)^{\frac{1}{3}}p^\frac{1}{3}+ia\gamma\varepsilon \frac{1}{2}\left(\frac{3\pi}{2}\right)^{\frac{1}{3}}p^\frac{1}{3}}\left(-a_3^2\frac{1}{\beta\gamma\varepsilon\pi}e^{-i\frac{\pi}{3}}\right)\nonumber\\
&+e^{\sqrt{3}\gamma\left(\frac{3\pi}{2}\right)^{\frac{1}{3}}p^\frac{1}{3}\left( V_0-d\varepsilon+\frac{a\varepsilon}{2}\right)+ih(p)+\gamma\left(\frac{3\pi}{2}\right)^{\frac{1}{3}}i p^\frac{1}{3}\left( V_0+\frac{a\varepsilon}{2}\right)}\left(4\pi^3c_1\beta\gamma\varepsilon\right)^{-1}\nonumber\\
&\left(a_2^2e^{-i\frac{\pi}{3}}-2a_2a_3e^{-i\frac{5\pi}{6}}-a_3^2e^{-i\frac{\pi}{3}}\right)\nonumber\\
&+e^{\sqrt{3}\gamma\left(\frac{3\pi}{2}\right)^{\frac{1}{3}}p^\frac{1}{3}\left( d\varepsilon-V_0+\frac{a\varepsilon}{2}\right)-ih(p)+\gamma i \left(\frac{3\pi}{2}\right)^{\frac{1}{3}}p^\frac{1}{3}\left(-V_0+\frac{a\varepsilon}{2}\right)}\left(a_3^2\frac{\pi c_1}{\gamma\varepsilon\beta}e^{-i\frac{\pi}{3}}p^{-\frac{4}{3}}\right)\label{Ceq43}
\end{align}

\subsection{Normalization}
In this section we compute the normalization coefficients $N_p$ in (\ref{Ceq0}) in the following way
\begin{align}
N_p=\int_\mathcal{L}\psi_{\omega_p}^2(z)\mathrm{d}z\label{Ceq44}
\end{align}
where $\mathcal{L}$ is the curve
\begin{align}
\mathcal{L}=u(x)=\left\{\begin{array}{c c}
x & x<x_c\\
x_c+i(x-x_c) & x_c<x
\end{array}\right.\label{Ceq45}
\end{align}
on which the resonant states decay on both ends. We split the integral into parts and use the corresponding parts of $\psi_{\omega_p}(x)$ as
\begin{align}
N_p&=\int_{-\infty}^{-d}a_1^2\mathrm{Ai}^2(y_1(x))\mathrm{d}x+\int_{-d}^d\left(a_2\mathrm{Ai}(y_2(x))+a_3\mathrm{Bi}(y_2(x))\right)^2\mathrm{d}x\nonumber\\
&+\int_{d}^{x_c}a_4^2\left(\mathrm{Ci}^+(y_1(x))\right)^2\mathrm{d}x+i\int_{x_c}^\infty a_4^2\left(\mathrm{Ci}^+(y_3(x))\right)^2\mathrm{d}x=I_1+I_2+I_3+I_4\label{Ceq46}
\end{align}
where we denoted the integrals as $I_1,\ldots,I_4$ in their respective order and as before we have
\begin{align}
y_1(x)&=-\gamma(\varepsilon x+\omega_p)\label{Ceq47}\\
y_2(x)&=-\gamma(\varepsilon x+V_0+\omega_p)\label{Ceq48}\\
y_3(x)&=-\gamma\left(i\varepsilon x+\varepsilon x_c(1-i)+\omega_p\right)\label{Ceq49}
\end{align}
To compute the integrals we use a known formula. Let $\mathrm{A}(x)$ and $\mathrm{B}(x)$ be any linear combination of Airy functions and $\mu,\nu$ be any numbers, then
\begin{align}
\int\mathrm{A}(\mu(x+\nu))&\mathrm{B}(\mu(x+\nu))\mathrm{d}x\nonumber\\
&=(x+\nu)\mathrm{A}(\mu(x+\nu))\mathrm{B}(\mu(x+\nu))-\frac{1}{\mu}\mathrm{A}'(\mu(x+\nu))\mathrm{B}'(\mu(x+\nu))\label{Ceq50}
\end{align}
or in our case we can write
\begin{align}
\int\mathrm{A}^2(\mu(x+\nu))\mathrm{d}x=(x+\nu)\mathrm{A}^2(\mu(x+\nu))-\frac{1}{\mu}\left[\mathrm{A}'(\mu(x+\nu))\right]^2\label{Ceq50.1}
\end{align}
however, as we will see later, this can be used only for the integral $I_1$. Let us denote the integrals in (\ref{Ceq46}) in order as they appear $I_1,\ldots,I_4$. We can use a lot of formulas computed in the previous section. The arguments to the Airy functions will be
\begin{align}
z_1^B(\xi_p)&=y_1^B(x)|_{\omega=\omega(\xi_p)}=\gamma\varepsilon xe^{i\frac{2\pi}{3}}+\beta\xi_p^\frac{2}{3}\label{Ceq51}\\
z_2^B(\xi_p)&=y_2^B(x)|_{\omega=\omega(\xi_p)}=\gamma(\varepsilon x+V_0)e^{i\frac{2\pi}{3}}+\beta\xi_p^\frac{2}{3}\label{Ceq52}\\
z_3^B(\xi_p)&=y_3^B(x)|_{\omega=\omega(\xi_p)}=\gamma(i\varepsilon x+\varepsilon x_c(1-i))e^{i\frac{2\pi}{3}}+\beta\xi_p^\frac{2}{3}\label{Ceq53}
\end{align}
where
\begin{align}
\beta=\left(\frac{3\pi}{2}\right)^\frac{2}{3}\label{Ceq54}
\end{align}
and
\begin{align}
z_1^A(\xi_p)&=y_1^A(x)|_{\tilde{\omega}=\tilde{\omega}(\xi_p)}=-\gamma\varepsilon x+\mu\xi_p^\frac{2}{3}\label{Ceq55}\\
z_2^A(\xi_p)&=y_2^A(x)|_{\tilde{\omega}=\tilde{\omega}(\xi_p)}=-\gamma(\varepsilon x+V_0)+\mu\xi_p^\frac{2}{3}\label{Ceq56}\\
z_3^A(\xi_p)&=y_3^A(x)|_{\tilde{\omega}=\tilde{\omega}(\xi_p)}=-\gamma(i\varepsilon x+\varepsilon x_c(1-i))+\mu\xi_p^\frac{2}{3}\label{Ceq57}
\end{align}
where
\begin{align}
\mu=-\beta e^{-i\frac{2\pi}{3}}=\beta e^{i\frac{\pi}{3}}\label{Ceq58}
\end{align}
and $\xi_p$ defined as
\begin{align}
\xi_p=p+\frac{i}{3\pi}\ln(p)-i\frac{1}{2\pi}\ln\left(\pi^2c_1\right)-ic_2p^\frac{1}{3}-\frac{1}{2\pi}h(p)\label{Ceq59}
\end{align}
where
\begin{align}
c_1&=\frac{V_0}{4\pi^2}\gamma\left(\frac{3\pi}{2}\right)^{-\frac{2}{3}}\label{Ceq60}\\
c_2&=\frac{\sqrt{3}\gamma d\varepsilon}{2\pi}\left(\frac{3\pi}{2}\right)^{\frac{1}{3}}\label{Ceq61}\\
h(p)&=\arg\left(e^{-i\left(4\pi c_2p^\frac{1}{3}\frac{1}{2\sqrt{3}}+\frac{\pi}{6}\right)}\right)\label{Ceq62}
\end{align}
We rewrite the asymptotic Airy functions using the previously computed (\ref{Ceq28}), (\ref{Ceq30}) but with slightly different names as
\begin{align}
\mathrm{Ai}\left(z_{1,2,3}^A\right)&\approx\frac{1}{2}\pi^{-\frac{1}{2}}\left(z_{1,2,3}^A\right)^{-\frac{1}{4}}e^{-\left(s_{1,2,3}^A+xt_{12,3}^A\right)}\label{Ceq69}\\
\mathrm{Ai'}\left(z_{1,2,3}^A\right)&\approx-\frac{1}{2}\pi^{-\frac{1}{2}}\left(z_{1,2,3}^A\right)^{\frac{1}{4}}e^{-\left(s_{1,2,3}^A+xt_{12,3}^A\right)}\label{Ceq70}\\
\mathrm{Bi}\left(z_{1,2,3}^B\right)&\approx\sqrt{\frac{2}{\pi}}e^{i\frac{\pi}{6}}\left(z_{1,2,3}^B\right)^{-\frac{1}{4}}\sin\left(\left(s_{1,2,3}^B+xt_{12,3}^B\right)+\frac{\pi}{4}-\frac{i}{2}\ln2\right)\label{Ceq71}\\
\mathrm{Bi'}\left(z_{1,2,3}^B\right)&\approx\sqrt{\frac{2}{\pi}}e^{-i\frac{\pi}{6}}\left(z_{1,2,3}^B\right)^{\frac{1}{4}}\cos\left(\left(s_{1,2,3}^B+xt_{12,3}^B\right)+\frac{\pi}{4}-\frac{i}{2}\ln2\right)\label{Ceq72}
\end{align}
where
{\medmuskip=0mu
\thinmuskip=0mu
\thickmuskip=0mu
\begin{align}
s_1^A(p)&=i\pi\xi_p & s_1^B(p)&=\pi\xi_p\label{Ceq73}\\
s_2^A(p)&=i\pi\xi_p-\gamma V_0\beta^\frac{1}{2}e^{i\frac{\pi}{6}}p^\frac{1}{3} &\nonumber\\
s_2^B(p)&=\pi\xi_p+\gamma V_0e^{i\frac{2\pi}{3}}\beta^\frac{1}{2}p^\frac{1}{3} &\label{Ceq77}\\
t_{12}^A(p)&=-\gamma \varepsilon\beta^\frac{1}{2}e^{i\frac{\pi}{6}}p^\frac{1}{3} & t_{12}^B(p)&=\gamma\varepsilon e^{i\frac{2\pi}{3}}\beta^\frac{1}{2}p^\frac{1}{3}\label{Ceq74}\\
\frac{2}{3}\left(z_3^{A}(\xi_p)\right)^\frac{3}{2}&\approx s_3^A(p)+xt_3^A(p) & \frac{2}{3}\left(z_3^{B}(\xi_p)\right)^\frac{3}{2}&\approx s_3^B(p)+xt_3^B(p)\label{Ceq79}\\
s_3^A(p)&=i\pi\xi_p-\gamma\varepsilon x_c(1-i)\beta^\frac{1}{2}e^{i\frac{\pi}{6}}p^\frac{1}{3} & \nonumber\\
s_3^B(p)&=\pi\xi_p+\gamma\varepsilon x_c(1-i)e^{i\frac{2\pi}{3}}\beta^\frac{1}{2}p^\frac{1}{3} & \label{Ceq80}\\
t_3^A(p)&=-\gamma i\varepsilon\mu^\frac{1}{2}p^\frac{1}{3}=-\gamma i\varepsilon\beta^\frac{1}{2}e^{i\frac{\pi}{6}}p^\frac{1}{3} & t_3^B&=\gamma i\varepsilon e^{i\frac{2\pi}{3}}\beta^\frac{1}{2}p^\frac{1}{3}\label{Ceq81}\\
\end{align}}
Observe, that we have
\begin{align}
s_{1,2,3}^A(p)&=is_{1,2,3}^B(p)\label{Ceq84.1}\\
t_{12,3}^A(p)&=it_{12,3}^B(p)\label{Ceq84.2}
\end{align}
From the form of the primitive in (\ref{Ceq50.1}) and the fact that we are integrating the squares of Airy functions, we see that we need the expansions of the following expressions as well.
\begin{align}
\left(z_{1}^A\right)^{-\frac{1}{2}}&\approx\beta^{-\frac{1}{2}}e^{-i\frac{\pi}{6}}p^{-\frac{1}{3}} & \left(z_{2}^A\right)^{-\frac{1}{2}}&\approx\beta^{-\frac{1}{2}}e^{-i\frac{\pi}{6}}p^{-\frac{1}{3}} & \left(z_{3}^A\right)^{-\frac{1}{2}}&\approx\beta^{-\frac{1}{2}}e^{-i\frac{\pi}{6}}p^{-\frac{1}{3}}\label{Ceq85}\\
\left(z_{1}^A\right)^{\frac{1}{2}}&\approx\beta^{\frac{1}{2}}e^{i\frac{\pi}{6}}p^{\frac{1}{3}} & \left(z_{2}^A\right)^{\frac{1}{2}}&\approx\beta^{\frac{1}{2}}e^{i\frac{\pi}{6}}p^{\frac{1}{3}} & \left(z_{3}^A\right)^{\frac{1}{2}}&\approx\beta^{\frac{1}{2}}e^{i\frac{\pi}{6}}p^{\frac{1}{3}}\label{Ceq86}\\
\left(z_{1}^B\right)^{-\frac{1}{2}}&\approx\beta^{-\frac{1}{2}}p^{-\frac{1}{3}} & \left(z_{2}^B\right)^{-\frac{1}{2}}&\approx\beta^{-\frac{1}{2}}p^{-\frac{1}{3}} & \left(z_{3}^B\right)^{-\frac{1}{2}}&\approx\beta^{-\frac{1}{2}}p^{-\frac{1}{3}}\label{Ceq87}\\
\left(z_{1}^B\right)^{\frac{1}{2}}&\approx\beta^{\frac{1}{2}}p^{\frac{1}{3}} & \left(z_{2}^B\right)^{\frac{1}{2}}&\approx\beta^{\frac{1}{2}}p^{\frac{1}{3}} & \left(z_{3}^B\right)^{\frac{1}{2}}&\approx\beta^{\frac{1}{2}}p^{\frac{1}{3}}\label{Ceq88}
\end{align}
In the next subchapter we take the integrals $I_1,\ldots,I_4$ and compute their asymptotic forms.

\subsubsection{\texorpdfstring{$I_1$}{CI1}}
This section is about the first integral $I_1$ in (\ref{Ceq46}). It has the form
\begin{align}
I_1&=\int_{-\infty}^{-d}a_1^2\mathrm{Ai}^2\left( z_1^A(x)\right)\mathrm{d}x=\int_{-\infty}^{-d}a_1^2\mathrm{Ai}^2\left( -\gamma\varepsilon x+\mu\xi_p^\frac{2}{3}\right)\mathrm{d}x\nonumber\\
&=\int_{-\infty}^{-d}a_1^2\mathrm{Ai}^2\left( -\gamma\varepsilon\left( x-\frac{\mu\xi_p^\frac{2}{3}}{\gamma\varepsilon}\right)\right)\mathrm{d}x\nonumber\\
&=\left( -d-\frac{\mu\xi_p^\frac{2}{3}}{\gamma\varepsilon}\right)a_1^2\mathrm{Ai}^2\left( -\gamma\varepsilon\left( -d-\frac{\mu\xi_p^\frac{2}{3}}{\gamma\varepsilon}\right)\right)\nonumber\\
&+\frac{a_1^2}{\gamma\varepsilon}\left[\mathrm{Ai'}\left( -\gamma\varepsilon\left( -d-\frac{\mu\xi_p^\frac{2}{3}}{\gamma\varepsilon}\right)\right)\right]^2\nonumber\\
&=\left( -d-\frac{\mu\xi_p^\frac{2}{3}}{\gamma\varepsilon}\right)a_1^2\mathrm{Ai}^2\left( \gamma\varepsilon\left( d+\frac{\mu\xi_p^\frac{2}{3}}{\gamma\varepsilon}\right)\right)+\frac{a_1^2}{\gamma\varepsilon}\left[\mathrm{Ai'}\left( \gamma\varepsilon\left( d+\frac{\mu\xi_p^\frac{2}{3}}{\gamma\varepsilon}\right)\right)\right]^2\label{Ceq97}
\end{align}
First, we approximate $\xi_p^{\frac{2}{3}}\approx p^\frac{2}{3}$ and use the framework we set up earlier.
\begin{align}
I_1&=\left( -d-\frac{\mu\xi_p^\frac{2}{3}}{\gamma\varepsilon}\right)a_1^2\mathrm{Ai}^2\left( \gamma\varepsilon\left( d+\frac{\mu\xi_p^\frac{2}{3}}{\gamma\varepsilon}\right)\right)+\frac{a_1^2}{\gamma\varepsilon}\left[\mathrm{Ai'}\left( \gamma\varepsilon\left( d+\frac{\mu\xi_p^\frac{2}{3}}{\gamma\varepsilon}\right)\right)\right]^2\nonumber\\
&\approx\left( -d-\frac{\mu p^\frac{2}{3}}{\gamma\varepsilon}\right)a_1^2\frac{1}{4}\pi^{-1}\beta^{-\frac{1}{2}}e^{-i\frac{\pi}{6}}p^{-\frac{1}{3}}e^{-2\left(s_1^A(p)-dt_{12}^A(p)\right)}\nonumber\\
&+\frac{a_1^2}{\gamma\varepsilon}\frac{1}{4}\pi^{-1}\beta^{\frac{1}{2}}e^{i\frac{\pi}{6}}p^{\frac{1}{3}} e^{-2\left(s_1^A(p)-dt_{12}^A(p)\right)}\label{Ceq99}
\end{align}
We express the exponentials
\begin{align}
&e^{-2\left(s_1^A(p)-dt_{12}^A(p)\right)}=p^\frac{2}{3}\left(\pi^2c_1\right)^{-1} e^{-2\sqrt{3}\gamma d\varepsilon\left(\frac{3\pi}{2}\right)^{\frac{1}{3}}p^\frac{1}{3}+ih(p)-d\gamma \varepsilon\left(\frac{3\pi}{2}\right)^{\frac{1}{3}}i p^\frac{1}{3}}\label{Ceq100}
\end{align}
and see that the expression (\ref{Ceq99}) decays exponentially.

\subsubsection{\texorpdfstring{$I_2$}{CI2}}
The second integral in (\ref{Ceq46}) is
\begin{align}
I_2&=\int_{-d}^d\left(a_2\mathrm{Ai}(z_2^A(x))+a_3\mathrm{Bi}(z_2^B(x))\right)^2\mathrm{d}x\nonumber\\
&=\int_{-d}^d\left(a_2\mathrm{Ai}\left(-\gamma(\varepsilon x+V_0)+\mu\xi_p^\frac{2}{3}\right)+a_3\mathrm{Bi}\left(\gamma(\varepsilon x+V_0)e^{i\frac{2\pi}{3}}+\beta\xi_p^\frac{2}{3}\right)\right)^2\mathrm{d}x\nonumber\\
&=\int_{-d}^d\left(a_2\mathrm{Ai}\left(-\gamma\varepsilon\left( x +\frac{\gamma V_0-\mu\xi_p^\frac{2}{3}}{\gamma\varepsilon}\right)\right)\right.\nonumber\\
&\left.+a_3\mathrm{Bi}\left(\gamma\varepsilon e^{i\frac{2\pi}{3}}\left( x+\frac{\gamma V_0e^{i\frac{2\pi}{3}}+\beta\xi_p^\frac{2}{3}}{\gamma\varepsilon e^{i\frac{2\pi}{3}}}\right)\right)\right)^2\mathrm{d}x\label{Ceq102}
\end{align}
For this integral we can not use the primitive (\ref{Ceq50}) because of the different arguments. Instead, we rewrite the integrand in its asymptotic form and try to integrate that way.
{\medmuskip=0mu
\thinmuskip=0mu
\thickmuskip=0mu
\begin{align}
I_2=&\int_{-d}^da_2^2\left(\frac{1}{2}\pi^{-\frac{1}{2}}\left(z_{2}^A\right)^{-\frac{1}{4}}e^{-\left(s_{2}^A+xt_{12}^A\right)}\right)^2\nonumber\\
&+a_3^2\left(\sqrt{\frac{2}{\pi}}e^{i\frac{\pi}{6}}\left(z_{2}^B\right)^{-\frac{1}{4}}\sin\left(\left(s_{2}^B+xt_{12}^B\right)+\frac{\pi}{4}-\frac{i}{2}\ln2\right)\right)^2+2a_2a_3\frac{1}{2}\pi^{-\frac{1}{2}}\left(z_{2}^A\right)^{-\frac{1}{4}}\nonumber\\
&e^{-\left(s_{2}^A+xt_{12}^A\right)}\sqrt{\frac{2}{\pi}}e^{i\frac{\pi}{6}}\left(z_{2}^B\right)^{-\frac{1}{4}}\sin\left(\left(s_{2}^B+xt_{12}^B\right)+\frac{\pi}{4}-\frac{i}{2}\ln2\right)\mathrm{d}x\nonumber\\
\approx&\int_{-d}^da_2^2\frac{1}{4}\pi^{-1}\beta^{-\frac{1}{2}}e^{-i\frac{\pi}{6}}p^{-\frac{1}{3}}e^{-2\left(s_{2}^A+xt_{12}^A\right)}+a_3^2\frac{2}{\pi}e^{i\frac{\pi}{3}}\beta^{-\frac{1}{2}}p^{-\frac{1}{3}}\nonumber\\
&\sin^2\left(\left(s_{2}^B+xt_{12}^B\right)+\frac{\pi}{4}-\frac{i}{2}\ln2\right)+2a_2a_3\frac{1}{2}\pi^{-\frac{1}{2}}\mu^{-\frac{1}{4}}p^{-\frac{1}{6}} e^{-i\left(s_{2}^B+xt_{12}^B\right)}\nonumber\\
&\sqrt{\frac{2}{\pi}}e^{i\frac{\pi}{6}}\beta^{-\frac{1}{4}}p^{-\frac{1}{6}}\sin\left(\left(s_{2}^B+xt_{12}^B\right)+\frac{\pi}{4}-\frac{i}{2}\ln2\right)\mathrm{d}x=i_1+i_3+i_2\label{Ceq103}
\end{align}}
where we used (\ref{Ceq84.1}), (\ref{Ceq84.2}) and split the big integral into three smaller, each of them including one term in the sum and denoted them $i_1,i_3,i_2$. We start with $i_1$ and use a substitution $2s_2^A(p)+x2t_{12}^A(p)=y\Rightarrow\mathrm{d}x=\frac{\mathrm{d}y}{2t_{12}^A(p)}$.
\begin{align}
i_1&=\int_{-d}^da_2^2\frac{1}{4}\pi^{-1}\beta^{-\frac{1}{2}}e^{-i\frac{\pi}{6}}p^{-\frac{1}{3}}e^{-2\left(s_{2}^A+xt_{12}^A\right)}\mathrm{d}x\nonumber\\
&=\frac{1}{2t_{12}^A(p)}a_2^2\frac{1}{4}\pi^{-1}\beta^{-\frac{1}{2}}e^{-i\frac{\pi}{6}}p^{-\frac{1}{3}}\int_{2s_2^A(p)-d2t_{12}^A(p)}^{2s_2^A(p)+d2t_{12}^A(p)}e^{-y}\mathrm{d}y\nonumber\\
&=-\frac{a_2^2\beta^{-\frac{1}{2}}e^{-i\frac{\pi}{6}}p^{-\frac{1}{3}}}{8\pi t_{12}^A(p)}e^{-y}\Big|_{y=2s_2^A(p)-d2t_{12}^A(p)}^{y=2s_2^A(p)+d2t_{12}^A(p)}\nonumber\\
&=-\frac{a_2^2\beta^{-\frac{1}{2}}e^{-i\frac{\pi}{6}}p^{-\frac{1}{3}}}{8\pi t_{12}^A(p)}\left(e^{-2\left(s_2^A(p)+dt_{12}^A(p)\right)}-e^{-2\left(s_2^A(p)-dt_{12}^A(p)\right)}\right)\label{Ceq104}
\end{align}
From (\ref{Ceq100}) we can deduce that
\begin{align}
&e^{-2\left(s_2^A(p)-dt_{12}^A(p)\right)}=p^\frac{2}{3}\left(\pi^2c_1\right)^{-1} e^{ih(p)+\gamma V_0\beta^\frac{1}{2}\sqrt{3}p^\frac{1}{3}+\gamma\left(\frac{3\pi}{2}\right)^{\frac{1}{3}}i p^\frac{1}{3}(V_0+d\varepsilon)}\label{Ceq106}
\end{align}
so that (\ref{Ceq104}) becomes
{\medmuskip=0mu
\thinmuskip=0mu
\thickmuskip=0mu
\begin{align}
&i_1=\frac{a_2^2\beta^{-\frac{1}{2}}e^{-i\frac{\pi}{6}}p^{-\frac{1}{3}}}{8\pi\gamma \varepsilon\beta^\frac{1}{2}e^{i\frac{\pi}{6}}p^\frac{1}{3}}\left(p^\frac{2}{3}\left(\pi^2c_1\right)^{-1} e^{ih(p)+\gamma V_0\beta^\frac{1}{2}\sqrt{3}p^\frac{1}{3}+\gamma\left(\frac{3\pi}{2}\right)^{\frac{1}{3}}i p^\frac{1}{3}(V_0+d\varepsilon)}\right.\nonumber\\
&\left.-p^\frac{2}{3}\left(\pi^2c_1\right)^{-1} e^{ih(p)+\gamma\beta^\frac{1}{2}\sqrt{3} p^\frac{1}{3}\left(V_0-2d\varepsilon\right)+\gamma \left(\frac{3\pi}{2}\right)^{\frac{1}{3}}i p^\frac{1}{3}(V_0-d\varepsilon)}\right)\nonumber\\
&=\frac{a_2^2}{8\pi^3c_1\gamma \varepsilon\beta e^{i\frac{\pi}{3}}}e^{ih(p)+\gamma V_0\beta^\frac{1}{2}\sqrt{3}p^\frac{1}{3}}\left( e^{\gamma\left(\frac{3\pi}{2}\right)^{\frac{1}{3}}i p^\frac{1}{3}(V_0+d\varepsilon)}\right.\nonumber\\
&\left.-e^{-\gamma\beta^\frac{1}{2}\sqrt{3} p^\frac{1}{3}2d\varepsilon+\gamma \left(\frac{3\pi}{2}\right)^{\frac{1}{3}}i p^\frac{1}{3}(V_0-d\varepsilon)}\right)\nonumber\\
&\approx\frac{a_2^2}{8\pi^3c_1\gamma \varepsilon\beta e^{i\frac{\pi}{3}}}e^{ih(p)+\gamma V_0\beta^\frac{1}{2}\sqrt{3}p^\frac{1}{3}+\gamma\left(\frac{3\pi}{2}\right)^{\frac{1}{3}}i p^\frac{1}{3}(V_0+d\varepsilon)}\label{Ceq107}
\end{align}}
The integral $i_2$ in (\ref{Ceq103}) is
{\medmuskip=0mu
\thinmuskip=0mu
\thickmuskip=0mu
\begin{align}
i_2=&\int_{-d}^d2a_2a_3\frac{1}{2}\pi^{-\frac{1}{2}}\mu^{-\frac{1}{4}}p^{-\frac{1}{6}} e^{-i\left(s_{2}^B+xt_{12}^B\right)}\sqrt{\frac{2}{\pi}}e^{i\frac{\pi}{6}}\beta^{-\frac{1}{4}}p^{-\frac{1}{6}}\nonumber\\
&\sin\left(\left(s_{2}^B+xt_{12}^B\right)+\frac{\pi}{4}-\frac{i}{2}\ln2\right)\mathrm{d}x\nonumber\\
=&a_2a_3\frac{\sqrt{2}}{\pi}\beta^{-\frac{1}{2}}e^{i\frac{\pi}{12}}p^{-\frac{1}{3}}\int_{-d}^d e^{-i\left(s_{2}^B+xt_{12}^B\right)}\sin\left(\left(s_{2}^B+xt_{12}^B\right)+\frac{\pi}{4}-\frac{i}{2}\ln2\right)\mathrm{d}x\nonumber\\
=&a_2a_3\frac{\sqrt{2}}{\pi}\beta^{-\frac{1}{2}}e^{i\frac{\pi}{12}}p^{-\frac{1}{3}}\frac{1}{2i}\int_{-d}^d e^{-i\left(s_{2}^B+xt_{12}^B\right)}\left(e^{i\left(s_{2}^B+xt_{12}^B\right)+i\frac{\pi}{4}+\frac{1}{2}\ln2}\right.\nonumber\\
&\left.-e^{-i\left(s_{2}^B+xt_{12}^B\right)-i\frac{\pi}{4}-\frac{1}{2}\ln2}\right)\mathrm{d}x\nonumber\\
=&a_2a_3\frac{\sqrt{2}}{\pi}\beta^{-\frac{1}{2}}e^{i\frac{\pi}{12}}p^{-\frac{1}{3}}\frac{1}{2i}\int_{-d}^d \left(e^{i\frac{\pi}{4}+\frac{1}{2}\ln2}-e^{-i2\left(s_{2}^B+xt_{12}^B\right)-i\frac{\pi}{4}-\frac{1}{2}\ln2}\right)\mathrm{d}x\nonumber\\
=&a_2a_3\frac{\sqrt{2}}{\pi}\beta^{-\frac{1}{2}}e^{i\frac{\pi}{12}}p^{-\frac{1}{3}}\frac{1}{2i}\sqrt{2} e^{i\frac{\pi}{4}}2d-a_2a_3\frac{\sqrt{2}}{\pi}\beta^{-\frac{1}{2}}e^{i\frac{\pi}{12}}p^{-\frac{1}{3}}\frac{1}{2i}e^{-i\frac{\pi}{4}}2^{-\frac{1}{2}}\nonumber\\
&\int_{-d}^d e^{-i2\left(s_{2}^B+xt_{12}^B\right)}\mathrm{d}x\nonumber\\
\approx& -a_2a_3\frac{1}{2\pi}\beta^{-\frac{1}{2}}e^{-i\frac{2\pi}{3}}p^{-\frac{1}{3}}\int_{-d}^d e^{-i2\left(s_{2}^B+xt_{12}^B\right)}\mathrm{d}x\label{Ceq108}
\end{align}}
The transformation we use here will be $2\left(s_{2}^B+xt_{12}^B\right)=y\Rightarrow\mathrm{d}x=\frac{\mathrm{d}y}{2t_{12}^B(p)}$.
\begin{align}
i_2=&-a_2a_3\frac{1}{2\pi}\beta^{-\frac{1}{2}}e^{-i\frac{2\pi}{3}}p^{-\frac{1}{3}}\int_{-d}^d e^{-i2\left(s_{2}^B+xt_{12}^B\right)}\mathrm{d}x\nonumber\\
&=-a_2a_3\frac{1}{2\pi}\beta^{-\frac{1}{2}}e^{-i\frac{2\pi}{3}}p^{-\frac{1}{3}}\frac{1}{2t_{12}^B(p)}\int_{2\left(s_{2}^B-dt_{12}^B\right)}^{2\left(s_{2}^B+dt_{12}^B\right)} e^{-iy}\mathrm{d}y\nonumber\\
=&a_2a_3\frac{1}{2\pi}\beta^{-\frac{1}{2}}e^{-i\frac{2\pi}{3}}p^{-\frac{1}{3}}\frac{1}{i2t_{12}^B(p)}\left( e^{-i2\left(s_{2}^B+dt_{12}^B\right)}-e^{-i2\left(s_{2}^B-dt_{12}^B\right)}\right)\label{Ceq109}
\end{align}
We are going to need the exponentials
\begin{align}
e^{-i2\left(s_{2}^B+dt_{12}^B\right)}&=p^\frac{2}{3}\left(\pi^2c_1\right)^{-1} e^{ih(p)+\gamma\sqrt{3} \beta^\frac{1}{2}p^\frac{1}{3}V_0+\gamma i \beta^\frac{1}{2}p^\frac{1}{3}(V_0+d\varepsilon)}\label{Ceq110}\\
e^{-i2\left(s_{2}^B-dt_{12}^B\right)}&=p^\frac{2}{3}\left(\pi^2c_1\right)^{-1} e^{ih(p)+\gamma\sqrt{3} \beta^\frac{1}{2}p^\frac{1}{3}(V_0-2d\varepsilon)+\gamma i \beta^\frac{1}{2}p^\frac{1}{3}(V_0-d\varepsilon)}\label{Ceq111}
\end{align}
The integral $i_2$ then is
\begin{align}
i_2&=a_2a_3\frac{1}{2\pi}\frac{\beta^{-\frac{1}{2}}e^{-i\frac{2\pi}{3}}p^{-\frac{1}{3}}}{i2\gamma\varepsilon e^{i\frac{2\pi}{3}}\beta^\frac{1}{2}p^\frac{1}{3}}\left( p^\frac{2}{3}\left(\pi^2c_1\right)^{-1} e^{ih(p)+\gamma\sqrt{3} \beta^\frac{1}{2}p^\frac{1}{3}V_0+\gamma i \beta^\frac{1}{2}p^\frac{1}{3}(V_0+d\varepsilon)}\right.\nonumber\\
&\left.-p^\frac{2}{3}\left(\pi^2c_1\right)^{-1} e^{ih(p)+\gamma\sqrt{3} \beta^\frac{1}{2}p^\frac{1}{3}(V_0-2d\varepsilon)+\gamma i \beta^\frac{1}{2}p^\frac{1}{3}(V_0-d\varepsilon)}\right)\nonumber\\
=&a_2a_3\frac{1}{4\pi^3c_1\gamma\varepsilon e^{-i\frac{\pi}{6}}\beta }e^{ih(p)+\gamma\sqrt{3} \beta^\frac{1}{2}p^\frac{1}{3}V_0}\left( e^{\gamma i \beta^\frac{1}{2}p^\frac{1}{3}(V_0+d\varepsilon)}\right.\nonumber\\
&\left.-e^{-\gamma\sqrt{3} \beta^\frac{1}{2}p^\frac{1}{3}2d\varepsilon+\gamma i \beta^\frac{1}{2}p^\frac{1}{3}(V_0-d\varepsilon)}\right)\nonumber\\
\approx& a_2a_3\frac{1}{4\pi^3c_1\gamma\varepsilon e^{-i\frac{\pi}{6}}\beta }e^{ih(p)+\gamma\sqrt{3} \beta^\frac{1}{2}p^\frac{1}{3}V_0+\gamma i \beta^\frac{1}{2}p^\frac{1}{3}(V_0+d\varepsilon)}\label{Ceq112}
\end{align}
Finally the third integral in (\ref{Ceq103}).
\begin{align}
&i_3=a_3^2\frac{2}{\pi}e^{i\frac{\pi}{3}}\beta^{-\frac{1}{2}}p^{-\frac{1}{3}}\int_{-d}^d\sin^2\left(s_{2}^B+xt_{12}^B+\frac{\pi}{4}-\frac{i}{2}\ln2\right)\mathrm{d}x\label{Ceq113}
\end{align}
We perform a transformation $s_{2}^B+xt_{12}^B+\frac{\pi}{4}-\frac{i}{2}\ln2=y\Rightarrow\mathrm{d}x=\frac{\mathrm{d}y}{t_{12}^B(p)}$.
\begin{align}
i_3=&a_3^2\frac{2}{\pi}e^{i\frac{\pi}{3}}\beta^{-\frac{1}{2}}p^{-\frac{1}{3}}\frac{1}{t_{12}^B(p)}\int_{s_{2}^B-dt_{12}^B+\frac{\pi}{4}-\frac{i}{2}\ln2}^{s_{2}^B+dt_{12}^B+\frac{\pi}{4}-\frac{i}{2}\ln2}\sin^2\left(y\right)\mathrm{d}y\nonumber\\
=&a_3^2\frac{2}{\pi}e^{i\frac{\pi}{3}}\beta^{-\frac{1}{2}}p^{-\frac{1}{3}}\frac{1}{t_{12}^B(p)}\left(\frac{y}{2}-\frac{\sin(2y)}{4}\right)\Big|_{y=s_{2}^B-dt_{12}^B+\frac{\pi}{4}-\frac{i}{2}\ln2}^{y=s_{2}^B+dt_{12}^B+\frac{\pi}{4}-\frac{i}{2}\ln2}\nonumber\\
=&a_3^2\frac{2}{\pi}e^{i\frac{\pi}{3}}\beta^{-\frac{1}{2}}p^{-\frac{1}{3}}\frac{1}{t_{12}^B(p)}\frac{1}{2}\left(s_{2}^B+dt_{12}^B+\frac{\pi}{4}-\frac{i}{2}\ln2\right.\nonumber\\
&\left.-\frac{\sin\left(2\left(s_{2}^B+dt_{12}^B+\frac{\pi}{4}-\frac{i}{2}\ln2\right)\right)}{2}-\left( s_{2}^B-dt_{12}^B+\frac{\pi}{4}-\frac{i}{2}\ln2\right)\right.\nonumber\\
&\left.+\frac{\sin\left(2\left(s_{2}^B-dt_{12}^B+\frac{\pi}{4}-\frac{i}{2}\ln2\right)\right)}{2}\right)\nonumber\\
=&a_3^2\frac{1}{\pi}e^{i\frac{\pi}{3}}\beta^{-\frac{1}{2}}p^{-\frac{1}{3}}\frac{1}{t_{12}^B(p)}\left(2dt_{12}^B-\frac{\sin\left(2s_{2}^B+2dt_{12}^B+\frac{\pi}{2}-i\ln2\right)}{2}\right.\nonumber\\
&\left.+\frac{\sin\left(2s_{2}^B-2dt_{12}^B+\frac{\pi}{2}-i\ln2\right)}{2}\right)\label{Ceq114}
\end{align}
We investigate the sines.
{\medmuskip=0mu
\thinmuskip=0mu
\thickmuskip=0mu
\begin{align}
\sin&\left(2s_{2}^B+2dt_{12}^B+\frac{\pi}{2}-i\ln2\right)\approx\frac{1}{4\pi^2c_1}p^{\frac{2}{3}}e^{ih(p)+\gamma\sqrt{3}\beta^\frac{1}{2}p^\frac{1}{3}V_0+\gamma i \beta^\frac{1}{2}p^\frac{1}{3}(V_0+d\varepsilon)}\label{Ceq115}\\
\sin&\left(2s_{2}^B-2dt_{12}^B+\frac{\pi}{2}-i\ln2\right)\nonumber\\
&=\frac{1}{2i}\left(p^{-\frac{2}{3}}\left(\pi^2c_1\right)2ie^{-ih(p)-\gamma\sqrt{3}\beta^\frac{1}{2}p^\frac{1}{3}(V_0-2d\varepsilon)-\gamma i \beta^\frac{1}{2}p^\frac{1}{3}(V_0+d\varepsilon)}\right.\nonumber\\
&\left.-p^{\frac{2}{3}}\left(2i\pi^2c_1\right)^{-1}e^{ih(p)+\gamma\sqrt{3}\beta^\frac{1}{2}p^\frac{1}{3}(V_0-2d\varepsilon)+\gamma i \beta^\frac{1}{2}p^\frac{1}{3}(V_0+d\varepsilon)}\right)\label{Ceq116}
\end{align}}
Using these expressions in (\ref{Ceq114}), we get
{\medmuskip=0mu
\thinmuskip=0mu
\thickmuskip=0mu
\begin{align}
i_3&=a_3^2\frac{1}{\pi}e^{i\frac{\pi}{3}}\beta^{-\frac{1}{2}}p^{-\frac{1}{3}}\frac{1}{t_{12}^B(p)}\nonumber\\
&\left(2dt_{12}^B-\frac{\sin\left(2s_{2}^B+2dt_{12}^B+\frac{\pi}{2}-i\ln2\right)}{2}+\frac{\sin\left(2s_{2}^B-2dt_{12}^B+\frac{\pi}{2}-i\ln2\right)}{2}\right)\nonumber\\
=&a_3^2\frac{2}{\pi}e^{i\frac{\pi}{3}}\beta^{-\frac{1}{2}}p^{-\frac{1}{3}}-a_3^2\frac{e^{i\frac{\pi}{3}}\beta^{-\frac{1}{2}}p^{-\frac{1}{3}}}{\pi\gamma\varepsilon e^{i\frac{2\pi}{3}}\beta^\frac{1}{2}p^\frac{1}{3}}\frac{\sin\left(2s_{2}^B+2dt_{12}^B+\frac{\pi}{2}-i\ln2\right)}{2}\nonumber\\
&+a_3^2\frac{e^{i\frac{\pi}{3}}\beta^{-\frac{1}{2}}p^{-\frac{1}{3}}}{\pi\gamma\varepsilon e^{i\frac{2\pi}{3}}\beta^\frac{1}{2}p^\frac{1}{3}}\frac{\sin\left(2s_{2}^B-2dt_{12}^B+\frac{\pi}{2}-i\ln2\right)}{2}\nonumber\\
\approx& a_3^2\frac{1}{2\pi\gamma\varepsilon e^{i\frac{\pi}{3}}\beta p^\frac{2}{3}}\left(-\frac{1}{4\pi^2c_1}p^{\frac{2}{3}}e^{ih(p)+\gamma\sqrt{3}\beta^\frac{1}{2}p^\frac{1}{3}V_0+\gamma i \beta^\frac{1}{2}p^\frac{1}{3}(V_0+d\varepsilon)}\right.\nonumber\\
&\left.+\frac{1}{2i}\left(p^{-\frac{2}{3}}\left(\pi^2c_1\right)2ie^{-ih(p)-\gamma\sqrt{3}\beta^\frac{1}{2}p^\frac{1}{3}(V_0-2d\varepsilon)-\gamma i \beta^\frac{1}{2}p^\frac{1}{3}(V_0+d\varepsilon)}\right.\right.\nonumber\\
&\left.\left.-p^{\frac{2}{3}}\left(2i\pi^2c_1\right)^{-1}e^{ih(p)+\gamma\sqrt{3}\beta^\frac{1}{2}p^\frac{1}{3}(V_0-2d\varepsilon)+\gamma i \beta^\frac{1}{2}p^\frac{1}{3}(V_0+d\varepsilon)}\right)\right)\nonumber\\
=& a_3^2\frac{1}{2\pi\gamma\varepsilon e^{i\frac{\pi}{3}}\beta}\left(-\frac{1}{4\pi^2c_1}e^{ih(p)+\gamma\sqrt{3}\beta^\frac{1}{2}p^\frac{1}{3}V_0+\gamma i \beta^\frac{1}{2}p^\frac{1}{3}(V_0+d\varepsilon)}\right.\nonumber\\
&\left.+p^{-\frac{4}{3}}\left(\pi^2c_1\right)e^{-ih(p)-\gamma\sqrt{3}\beta^\frac{1}{2}p^\frac{1}{3}(V_0-2d\varepsilon)-\gamma i \beta^\frac{1}{2}p^\frac{1}{3}(V_0+d\varepsilon)}\right.\nonumber\\
&\left.+\left(4\pi^2c_1\right)^{-1}e^{ih(p)+\gamma\sqrt{3}\beta^\frac{1}{2}p^\frac{1}{3}(V_0-2d\varepsilon)+\gamma i \beta^\frac{1}{2}p^\frac{1}{3}(V_0+d\varepsilon)}\right)\label{Ceq117}
\end{align}}
It is not possible to simplify more this expression since we do not know the sign of $V_0-2d\varepsilon$, but we can say for sure, that it grows exponentially either way. The overall integral $I_2$ is then
{\medmuskip=0mu
\thinmuskip=0mu
\thickmuskip=0mu
\begin{align}
I_2&=i_1+i_2+i_3\nonumber\\
&=\frac{a_2^2}{8\pi^3c_1\gamma \varepsilon\beta e^{i\frac{\pi}{3}}}e^{ih(p)+\gamma V_0\beta^\frac{1}{2}\sqrt{3}p^\frac{1}{3}+\gamma\left(\frac{3\pi}{2}\right)^{\frac{1}{3}}i p^\frac{1}{3}(V_0+d\varepsilon)}\nonumber\\
&+a_2a_3\frac{1}{4\pi^3c_1\gamma\varepsilon e^{-i\frac{\pi}{6}}\beta }e^{ih(p)+\gamma\sqrt{3} \beta^\frac{1}{2}p^\frac{1}{3}V_0+\gamma i \beta^\frac{1}{2}p^\frac{1}{3}(V_0+d\varepsilon)}\nonumber\\
&+a_3^2\frac{1}{2\pi\gamma\varepsilon e^{i\frac{\pi}{3}}\beta}\left(-\frac{1}{4\pi^2c_1}e^{ih(p)+\gamma\sqrt{3}\beta^\frac{1}{2}p^\frac{1}{3}V_0+\gamma i \beta^\frac{1}{2}p^\frac{1}{3}(V_0+d\varepsilon)}\right.\nonumber\\
&\left.+p^{-\frac{4}{3}}\left(\pi^2c_1\right)e^{-ih(p)-\gamma\sqrt{3}\beta^\frac{1}{2}p^\frac{1}{3}(V_0-2d\varepsilon)-\gamma i \beta^\frac{1}{2}p^\frac{1}{3}(V_0+d\varepsilon)}\right.\nonumber\\
&\left.+\left(4\pi^2c_1\right)^{-1}e^{ih(p)+\gamma\sqrt{3}\beta^\frac{1}{2}p^\frac{1}{3}(V_0-2d\varepsilon)+\gamma i \beta^\frac{1}{2}p^\frac{1}{3}(V_0+d\varepsilon)}\right)\label{Ceq118}
\end{align}}

\subsubsection{\texorpdfstring{$I_3$}{CI3}}
The third integral in (\ref{Ceq46}) $I_3$ is very similar to the previous one, since we have a linear combination of $\mathrm{Ai}$ and $\mathrm{Bi}$. Sadly, we can not use the formula here either.
\begin{align}
I_3&=\int_{d}^{x_c}a_4^2\left(\mathrm{Ci}^+(y_1(x))\right)^2\mathrm{d}x=a_4^2\int_{d}^{x_c}\left(\mathrm{Bi}\left( z_1^B(x)\right)+i\mathrm{Ai}\left( z_1^A(x)\right)\right)^2\mathrm{d}x\nonumber\\
&=a_4^2\int_{d}^{x_c}\left(\mathrm{Bi}\left( \gamma\varepsilon xe^{i\frac{2\pi}{3}}+\beta\xi_p^\frac{2}{3}\right)+i\mathrm{Ai}\left( -\gamma\varepsilon x+\mu\xi_p^\frac{2}{3}\right)\right)^2\mathrm{d}x\label{Ceq119}
\end{align}
We can divide this integral into three smaller integrals $i_1,i_2,i_3$ as before.
\begin{align}
i_1&=-a_4^2\int_{d}^{x_c}\mathrm{Ai}^2\left( -\gamma\varepsilon x+\mu\xi_p^\frac{2}{3}\right)\mathrm{d}x\label{Ceq120}\\
i_2&=2ia_4^2\int_{d}^{x_c}\mathrm{Bi}\left( \gamma\varepsilon xe^{i\frac{2\pi}{3}}+\beta\xi_p^\frac{2}{3}\right)\mathrm{Ai}\left( -\gamma\varepsilon x+\mu\xi_p^\frac{2}{3}\right)\mathrm{d}x\label{Ceq121}\\
i_3&=a_4^2\int_{d}^{x_c}\mathrm{Bi}^2\left( \gamma\varepsilon xe^{i\frac{2\pi}{3}}+\beta\xi_p^\frac{2}{3}\right)\mathrm{d}x\label{Ceq122}
\end{align}
We can see that $i_1$ is the same as (\ref{Ceq104}) up to the exponentials, where we have different boundary values but do not have any $V_0$.
\begin{align}
i_1&=\frac{a_4^2\beta^{-\frac{1}{2}}e^{-i\frac{\pi}{6}}p^{-\frac{1}{3}}}{8\pi t_{12}^A(p)}e^{-y}\Big|_{y=2s_2^A(p)-d2t_{12}^A(p)}^{y=2s_2^A(p)+d2t_{12}^A(p)}\nonumber\\
&=\frac{a_4^2\beta^{-\frac{1}{2}}e^{-i\frac{\pi}{6}}p^{-\frac{1}{3}}}{8\pi t_{12}^A(p)}\left(e^{-2\left(s_1^A(p)+x_ct_{12}^A(p)\right)}-e^{-2\left(s_1^A(p)+dt_{12}^A(p)\right)}\right)\label{Ceq123}
\end{align}
From (\ref{Ceq100}) we have that the exponentials are
\begin{align}
e^{-2\left(s_1^A(p)+dt_{12}^A(p)\right)}&=p^\frac{2}{3}\left(\pi^2c_1\right)^{-1} e^{ih(p)+d\gamma \varepsilon\left(\frac{3\pi}{2}\right)^{\frac{1}{3}}i p^\frac{1}{3}}\label{Ceq124}\\
e^{-2\left(s_1^A(p)+x_ct_{12}^A(p)\right)}&=p^\frac{2}{3}\left(\pi^2c_1\right)^{-1} e^{ih(p)+\sqrt{3}\gamma\varepsilon\left(\frac{3\pi}{2}\right)^{\frac{1}{3}}p^\frac{1}{3}(x_c-d)+x_c\gamma \varepsilon\left(\frac{3\pi}{2}\right)^{\frac{1}{3}}i p^\frac{1}{3}}\label{Ceq125}
\end{align}
so that (\ref{Ceq120}) becomes
{\medmuskip=0mu
\thinmuskip=0mu
\thickmuskip=0mu
\begin{align}
i_1&=-\frac{a_4^2}{8\pi^3c_1\gamma \varepsilon\beta e^{i\frac{\pi}{3}}}\nonumber\\
&\left(e^{ih(p)+\sqrt{3}\gamma\varepsilon\left(\frac{3\pi}{2}\right)^{\frac{1}{3}}p^\frac{1}{3}(x_c-d)+x_c\gamma \varepsilon\left(\frac{3\pi}{2}\right)^{\frac{1}{3}}i p^\frac{1}{3}}-e^{ih(p)+d\gamma \varepsilon\left(\frac{3\pi}{2}\right)^{\frac{1}{3}}i p^\frac{1}{3}}\right)\label{Ceq126}
\end{align}}
For the second smaller integral $i_2$ we can use (\ref{Ceq109}).
\begin{align}
&i_2=ia_4^2\frac{1}{2\pi}\beta^{-\frac{1}{2}}e^{-i\frac{2\pi}{3}}p^{-\frac{1}{3}}\frac{1}{i2t_{12}^B(p)}\left( e^{-i2\left(s_{1}^B+x_ct_{12}^B\right)}-e^{-i2\left(s_{1}^B+dt_{12}^B\right)}\right)\label{Ceq127}
\end{align}
To express the exponential terms we use (\ref{Ceq110}).
\begin{align}
e^{-i2\left(s_{1}^B+dt_{12}^B\right)}&=p^\frac{2}{3}\left(\pi^2c_1\right)^{-1} e^{ih(p)+\gamma i \beta^\frac{1}{2}p^\frac{1}{3}d\varepsilon}\label{Ceq128}\\
e^{-i2\left(s_{1}^B+x_ct_{12}^B\right)}&=p^\frac{2}{3}\left(\pi^2c_1\right)^{-1} e^{ih(p)+\gamma\sqrt{3} \beta^\frac{1}{2}p^\frac{1}{3}\varepsilon(x_c-d)+\gamma i \beta^\frac{1}{2}p^\frac{1}{3}x_c\varepsilon}\label{Ceq129}
\end{align}
and (\ref{Ceq127}) can be rewritten as
\begin{align}
i_2&=ia_4^2\frac{1}{2\pi}\beta^{-\frac{1}{2}}e^{-i\frac{2\pi}{3}}p^{-\frac{1}{3}}\frac{1}{i2t_{12}^B(p)}\left( e^{-i2\left(s_{1}^B+x_ct_{12}^B\right)}-e^{-i2\left(s_{1}^B+dt_{12}^B\right)}\right)\nonumber\\
&=a_4^2\frac{1}{4\pi^3c_1\gamma\varepsilon e^{-i\frac{2\pi}{3}}\beta }\left( e^{ih(p)+\gamma\sqrt{3} \beta^\frac{1}{2}p^\frac{1}{3}\varepsilon(x_c-d)+\gamma i \beta^\frac{1}{2}p^\frac{1}{3}x_c\varepsilon}-e^{ih(p)+\gamma i \beta^\frac{1}{2}p^\frac{1}{3}d\varepsilon}\right)\label{Ceq130}
\end{align}
The third small integral $i_3$ is again very similar to (\ref{Ceq114}).
{\medmuskip=0mu
\thinmuskip=0mu
\thickmuskip=0mu
\begin{align}
i_3&=a_4^2\frac{1}{\pi}e^{i\frac{\pi}{3}}\beta^{-\frac{1}{2}}p^{-\frac{1}{3}}\frac{1}{t_{12}^B(p)}\nonumber\\
&\left((x_c+d)t_{12}^B-\frac{\sin\left(2s_{1}^B+2x_ct_{12}^B+\frac{\pi}{2}-i\ln2\right)}{2}+\frac{\sin\left(2s_{1}^B+2dt_{12}^B+\frac{\pi}{2}-i\ln2\right)}{2}\right)\label{Ceq131}
\end{align}}
As before, we express the sinuses with help from (\ref{Ceq115}).
\begin{align}
\sin\left(2s_{2}^B+2dt_{12}^B+\frac{\pi}{2}-i\ln2\right)&\approx\frac{1}{4\pi^2c_1}p^{\frac{2}{3}}e^{ih(p)+\gamma i \beta^\frac{1}{2}p^\frac{1}{3}d\varepsilon}\label{Ceq132}\\
\sin\left(2s_{2}^B+2x_ct_{12}^B+\frac{\pi}{2}-i\ln2\right)&\approx\frac{1}{4\pi^2c_1}p^{\frac{2}{3}}e^{ih(p)+\gamma\varepsilon\sqrt{3}\beta^\frac{1}{2}p^\frac{1}{3}(x_c-d)+\gamma i \beta^\frac{1}{2}p^\frac{1}{3}x_c\varepsilon}\label{Ceq133}
\end{align}
So $i_3$ turns to
{\medmuskip=0mu
\thinmuskip=0mu
\thickmuskip=0mu
\begin{align}
i_3&=a_4^2\frac{1}{\pi}e^{i\frac{\pi}{3}}\beta^{-\frac{1}{2}}p^{-\frac{1}{3}}\frac{1}{t_{12}^B(p)}\nonumber\\
&\left((x_c+d)t_{12}^B-\frac{\sin\left(2s_{1}^B+2x_ct_{12}^B+\frac{\pi}{2}-i\ln2\right)}{2}+\frac{\sin\left(2s_{1}^B+2dt_{12}^B+\frac{\pi}{2}-i\ln2\right)}{2}\right)\nonumber\\
&\approx a_4^2\frac{1}{2\pi\gamma\varepsilon e^{i\frac{\pi}{3}}\beta}\nonumber\\
&\left(-\frac{1}{4\pi^2c_1}e^{ih(p)+\gamma\varepsilon\sqrt{3}\beta^\frac{1}{2}p^\frac{1}{3}(x_c-d)+\gamma i \beta^\frac{1}{2}p^\frac{1}{3}x_c\varepsilon}+\frac{1}{4\pi^2c_1}e^{ih(p)+\gamma i \beta^\frac{1}{2}p^\frac{1}{3}d\varepsilon}\right)\nonumber\\
&=a_4^2\frac{1}{8\pi^3c_1\gamma\varepsilon e^{i\frac{\pi}{3}}\beta}\left(-e^{ih(p)+\gamma\varepsilon\sqrt{3}\beta^\frac{1}{2}p^\frac{1}{3}(x_c-d)+\gamma i \beta^\frac{1}{2}p^\frac{1}{3}x_c\varepsilon}+e^{ih(p)+\gamma i \beta^\frac{1}{2}p^\frac{1}{3}d\varepsilon}\right)\label{Ceq134}
\end{align}}
and the overall integral $I_3$ is
{\medmuskip=0mu
\thinmuskip=0mu
\thickmuskip=0mu
\begin{align}
I_3&=i_1+i_2+i_3\nonumber\\
&=-\frac{a_4^2}{8\pi^3c_1\gamma \varepsilon\beta e^{i\frac{\pi}{3}}}\nonumber\\
&\left(e^{ih(p)+\sqrt{3}\gamma\varepsilon\left(\frac{3\pi}{2}\right)^{\frac{1}{3}}p^\frac{1}{3}(x_c-d)+x_c\gamma \varepsilon\left(\frac{3\pi}{2}\right)^{\frac{1}{3}}i p^\frac{1}{3}}-e^{ih(p)+d\gamma \varepsilon\left(\frac{3\pi}{2}\right)^{\frac{1}{3}}i p^\frac{1}{3}}\right)\nonumber\\
&+a_4^2\frac{1}{4\pi^3c_1\gamma\varepsilon e^{-i\frac{2\pi}{3}}\beta }\left( e^{ih(p)+\gamma\sqrt{3} \beta^\frac{1}{2}p^\frac{1}{3}\varepsilon(x_c-d)+\gamma i \beta^\frac{1}{2}p^\frac{1}{3}x_c\varepsilon}-e^{ih(p)+\gamma i \beta^\frac{1}{2}p^\frac{1}{3}d\varepsilon}\right)\nonumber\\
&+a_4^2\frac{1}{8\pi^3c_1\gamma\varepsilon e^{i\frac{\pi}{3}}\beta}\left(-e^{ih(p)+\gamma\varepsilon\sqrt{3}\beta^\frac{1}{2}p^\frac{1}{3}(x_c-d)+\gamma i \beta^\frac{1}{2}p^\frac{1}{3}x_c\varepsilon}+e^{ih(p)+\gamma i \beta^\frac{1}{2}p^\frac{1}{3}d\varepsilon}\right)\nonumber\\
&=-\frac{a_4^2}{4\pi^3c_1\gamma \varepsilon\beta e^{i\frac{\pi}{3}}}\nonumber\\
&\left(e^{ih(p)+\sqrt{3}\gamma\varepsilon\left(\frac{3\pi}{2}\right)^{\frac{1}{3}}p^\frac{1}{3}(x_c-d)+x_c\gamma \varepsilon\left(\frac{3\pi}{2}\right)^{\frac{1}{3}}i p^\frac{1}{3}}-e^{ih(p)+d\gamma \varepsilon\left(\frac{3\pi}{2}\right)^{\frac{1}{3}}i p^\frac{1}{3}}\right)\nonumber\\
&+a_4^2\frac{1}{4\pi^3c_1\gamma\varepsilon e^{-i\frac{2\pi}{3}}\beta }\left( e^{ih(p)+\gamma\sqrt{3} \beta^\frac{1}{2}p^\frac{1}{3}\varepsilon(x_c-d)+\gamma i \beta^\frac{1}{2}p^\frac{1}{3}x_c\varepsilon}-e^{ih(p)+\gamma i \beta^\frac{1}{2}p^\frac{1}{3}d\varepsilon}\right)\nonumber\\
&=\frac{a_4^2}{4\pi^3c_1\gamma\varepsilon\beta }\left(e^{ih(p)+\sqrt{3}\gamma\varepsilon\left(\frac{3\pi}{2}\right)^{\frac{1}{3}}p^\frac{1}{3}(x_c-d)+x_c\gamma \varepsilon\left(\frac{3\pi}{2}\right)^{\frac{1}{3}}i p^\frac{1}{3}}-e^{ih(p)+d\gamma \varepsilon\left(\frac{3\pi}{2}\right)^{\frac{1}{3}}i p^\frac{1}{3}}\right)\nonumber\\
&\left(-\frac{1}{e^{i\frac{\pi}{3}}}+\frac{1}{e^{-i\frac{2\pi}{3}}}\right)\nonumber\\
&=\frac{a_4^2e^{i\frac{2\pi}{3}}}{2\pi^3c_1\gamma\varepsilon\beta }\left(e^{ih(p)+\sqrt{3}\gamma\varepsilon\left(\frac{3\pi}{2}\right)^{\frac{1}{3}}p^\frac{1}{3}(x_c-d)+x_c\gamma \varepsilon\left(\frac{3\pi}{2}\right)^{\frac{1}{3}}i p^\frac{1}{3}}-e^{ih(p)+d\gamma \varepsilon\left(\frac{3\pi}{2}\right)^{\frac{1}{3}}i p^\frac{1}{3}}\right)\nonumber\\
&\approx\frac{a_4^2e^{i\frac{2\pi}{3}}}{2\pi^3c_1\gamma\varepsilon\beta }e^{ih(p)+\sqrt{3}\gamma\varepsilon\left(\frac{3\pi}{2}\right)^{\frac{1}{3}}p^\frac{1}{3}(x_c-d)+x_c\gamma \varepsilon\left(\frac{3\pi}{2}\right)^{\frac{1}{3}}i p^\frac{1}{3}}\label{Ceq135}
\end{align}}

\subsubsection{\texorpdfstring{$I_4$}{CI4}}
Finally the last integral in (\ref{Ceq46}).
\begin{align}
I_4&=i\int_{x_c}^\infty a_4^2\left(\mathrm{Ci}^+(z_3(x))\right)^2\mathrm{d}x=ia_4^2\int_{x_c}^\infty \left(\mathrm{Bi}\left( z_3^B(x)\right)+i\mathrm{Ai}\left( z_3^A(x)\right)\right)^2\mathrm{d}x
\nonumber\\
&=ia_4^2\int_{x_c}^\infty \left(\mathrm{Bi}\left( \gamma(i\varepsilon x+\varepsilon x_c(1-i))e^{i\frac{2\pi}{3}}+\beta\xi_p^\frac{2}{3}\right)\right.\nonumber\\
&\left.+i\mathrm{Ai}\left( -\gamma(i\varepsilon x+\varepsilon x_c(1-i))+\mu\xi_p^\frac{2}{3}\right)\right)^2\mathrm{d}x\label{Ceq136}
\end{align}
We do not get to use the formula again, but at least we have that the integrand vanishes at infinity. We divide the integral into $i_1,i_2,i_3$ again.
{\medmuskip=0mu
\thinmuskip=0mu
\thickmuskip=0mu
\begin{align}
i_1&=-ia_4^2\int_{d}^{x_c}\mathrm{Ai}^2\left( -\gamma(i\varepsilon x+\varepsilon x_c(1-i))+\mu\xi_p^\frac{2}{3}\right)\mathrm{d}x\label{Ceq137}\\
i_2&=-2a_4^2\int_{d}^{x_c}\mathrm{Bi}\left( \gamma(i\varepsilon x+\varepsilon x_c(1-i))e^{i\frac{2\pi}{3}}+\beta\xi_p^\frac{2}{3}\right)\mathrm{Ai}\left( -\gamma(i\varepsilon x+\varepsilon x_c(1-i))+\mu\xi_p^\frac{2}{3}\right)\mathrm{d}x\label{Ceq138}\\
i_3&=ia_4^2\int_{d}^{x_c}\mathrm{Bi}^2\left( \gamma(i\varepsilon x+\varepsilon x_c(1-i))e^{i\frac{2\pi}{3}}+\beta\xi_p^\frac{2}{3}\right)\mathrm{d}x\label{Ceq139}
\end{align}}
For the first $i_1$ we can use (\ref{Ceq104}) as before but without the upper boundary.
\begin{align}
i_1&=\frac{ia_4^2\beta^{-\frac{1}{2}}e^{-i\frac{\pi}{6}}p^{-\frac{1}{3}}}{8\pi t_{3}^A(p)}\left(-e^{-2\left(s_3^A(p)+x_ct_{3}^A(p)\right)}\right)\nonumber\\
&=-\frac{ia_4^2\beta^{-\frac{1}{2}}e^{-i\frac{\pi}{6}}p^{-\frac{1}{3}}}{8\pi t_{3}^A(p)}e^{-2\left(s_3^A(p)+x_ct_{3}^A(p)\right)}\label{Ceq140}
\end{align}
This exponential is somewhat different from the other so we need to compute it.
\begin{align}
&e^{-2\left(s_3^A(p)+x_ct_{3}^A(p)\right)}=p^{\frac{2}{3}}\left(\pi^2c_1\right)^{-1}e^{\sqrt{3}\gamma \varepsilon\left(\frac{3\pi}{2}\right)^{\frac{1}{3}}p^\frac{1}{3}(x_c-d)+ih(p)+x_c\gamma\varepsilon\beta^\frac{1}{2}ip^\frac{1}{3}}\label{Ceq141}
\end{align}
We can write (\ref{Ceq140}) as
{\medmuskip=0mu
\thinmuskip=0mu
\thickmuskip=0mu
\begin{align}
i_1&=-\frac{ia_4^2\beta^{-\frac{1}{2}}e^{-i\frac{\pi}{6}}p^{-\frac{1}{3}}}{8\pi t_{3}^A(p)}e^{-2\left(s_3^A(p)+x_ct_{3}^A(p)\right)}\nonumber\\
&=\frac{ia_4^2\beta^{-\frac{1}{2}}e^{-i\frac{\pi}{6}}p^{-\frac{1}{3}}}{8\pi \gamma i\varepsilon\beta^\frac{1}{2}e^{i\frac{\pi}{6}}p^\frac{1}{3}}p^{\frac{2}{3}}\frac{1}{\pi^2c_1}e^{\sqrt{3}\gamma \varepsilon\left(\frac{3\pi}{2}\right)^{\frac{1}{3}}p^\frac{1}{3}(x_c-d)+ih(p)+x_c\gamma\varepsilon\beta^\frac{1}{2}ip^\frac{1}{3}}\nonumber\\
&=\frac{a_4^2}{8\pi^3c_1 \gamma \varepsilon\beta e^{i\frac{\pi}{3}}}e^{\sqrt{3}\gamma \varepsilon\left(\frac{3\pi}{2}\right)^{\frac{1}{3}}p^\frac{1}{3}(x_c-d)+ih(p)+x_c\gamma\varepsilon\beta^\frac{1}{2}ip^\frac{1}{3}}\label{Ceq142}
\end{align}}
We take as help (\ref{Ceq109}) for the next integral $i_2$.
\begin{align}
i_2&=-a_4^2\frac{1}{2\pi}\beta^{-\frac{1}{2}}e^{-i\frac{2\pi}{3}}p^{-\frac{1}{3}}\frac{1}{i2t_{3}^B(p)}\left(-e^{-i2\left(s_{3}^B+x_ct_{3}^B\right)}\right)\nonumber\\
&=a_4^2\frac{1}{2\pi}\beta^{-\frac{1}{2}}e^{-i\frac{2\pi}{3}}p^{-\frac{1}{3}}\frac{1}{i2t_{3}^B(p)}e^{-i2\left(s_{3}^B+x_ct_{3}^B\right)}\label{Ceq143}
\end{align}
Let us expand the exponential. We can use the fact (\ref{Ceq84.1}), (\ref{Ceq84.2}).
\begin{align}
e^{-i2\left(s_{3}^B+x_ct_{3}^B\right)}&=e^{-2\left(s_3^A(p)+x_ct_{3}^A(p)\right)}\nonumber\\
&=p^{\frac{2}{3}}\left(\pi^2c_1\right)^{-1}e^{\sqrt{3}\gamma \varepsilon\left(\frac{3\pi}{2}\right)^{\frac{1}{3}}p^\frac{1}{3}(x_c-d)+ih(p)+x_c\gamma\varepsilon\beta^\frac{1}{2}ip^\frac{1}{3}}\label{Ceq144}
\end{align}
Hence, the integral $i_2$ becomes
\begin{align}
i_2&=a_4^2\frac{1}{2\pi}\beta^{-\frac{1}{2}}e^{-i\frac{2\pi}{3}}p^{-\frac{1}{3}}\frac{1}{i2t_{3}^B(p)}e^{-i2\left(s_{3}^B+x_ct_{3}^B\right)}\nonumber\\
&=\frac{a_4^2\beta^{-\frac{1}{2}}e^{-i\frac{2\pi}{3}}p^{-\frac{1}{3}}}{2\pi i2\gamma i\varepsilon e^{i\frac{2\pi}{3}}\beta^\frac{1}{2}p^\frac{1}{3}}p^{\frac{2}{3}}\left(\pi^2c_1\right)^{-1}e^{\sqrt{3}\gamma \varepsilon\left(\frac{3\pi}{2}\right)^{\frac{1}{3}}p^\frac{1}{3}(x_c-d)+ih(p)+x_c\gamma\varepsilon\beta^\frac{1}{2}ip^\frac{1}{3}}\nonumber\\
&=-\frac{a_4^2}{4\pi^3c_1\gamma\varepsilon e^{i\frac{4\pi}{3}}\beta}e^{\sqrt{3}\gamma \varepsilon\left(\frac{3\pi}{2}\right)^{\frac{1}{3}}p^\frac{1}{3}(x_c-d)+ih(p)+x_c\gamma\varepsilon\beta^\frac{1}{2}ip^\frac{1}{3}}\label{Ceq145}
\end{align}
At last, the final integral $i_3$. We use (\ref{Ceq114}) as a starting point.
\begin{align}
i_3&=ia_4^2\frac{1}{\pi}e^{i\frac{\pi}{3}}\beta^{-\frac{1}{2}}p^{-\frac{1}{3}}\frac{1}{t_{3}^B(p)}\nonumber\\
&\left(-s_{3}^B-x_ct_{3}^B-\frac{\pi}{4}+\frac{i}{2}\ln2+\frac{\sin\left(2s_{3}^B+2x_ct_{3}^B+\frac{\pi}{2}-i\ln2\right)}{2}\right)\nonumber\\
&\approx \frac{ia_4^2e^{i\frac{\pi}{3}}\beta^{-\frac{1}{2}}p^{-\frac{1}{3}}}{2\pi t_{3}^B(p)}\sin\left(2s_{3}^B+2x_ct_{3}^B+\frac{\pi}{2}-i\ln2\right)\label{Ceq146}
\end{align}
The sinus in this expression is
{\medmuskip=0mu
\thinmuskip=0mu
\thickmuskip=0mu
\begin{align}
&\sin\left(2s_{3}^B+2x_ct_{3}^B+\frac{\pi}{2}-i\ln2\right)\approx\frac{1}{4\pi^2c_1}p^{\frac{2}{3}}e^{\sqrt{3}\gamma d\varepsilon\left(\frac{3\pi}{2}\right)^{\frac{1}{3}}p^\frac{1}{3}(x_c-d)+ih(p)+i\gamma\varepsilon x_c\beta^\frac{1}{2}p^\frac{1}{3}}\label{Ceq147}
\end{align}}
The expression (\ref{Ceq146}) then becomes
\begin{align}
&i_3=\frac{ia_4^2e^{i\frac{\pi}{3}}\beta^{-\frac{1}{2}}p^{-\frac{1}{3}}}{2\pi t_{3}^B(p)}\sin\left(2s_{3}^B+2x_ct_{3}^B+\frac{\pi}{2}-i\ln2\right)\nonumber\\
&=\frac{ia_4^2e^{i\frac{\pi}{3}}\beta^{-\frac{1}{2}}p^{-\frac{1}{3}}}{2\pi \gamma i\varepsilon e^{i\frac{2\pi}{3}}\beta^\frac{1}{2}p^\frac{1}{3}}\frac{1}{4\pi^2c_1}p^{\frac{2}{3}}e^{\sqrt{3}\gamma d\varepsilon\left(\frac{3\pi}{2}\right)^{\frac{1}{3}}p^\frac{1}{3}(x_c-d)+ih(p)+i\gamma\varepsilon x_c\beta^\frac{1}{2}p^\frac{1}{3}}\nonumber\\
&=\frac{a_4^2}{8\pi^3c_1 \gamma \varepsilon e^{i\frac{\pi}{3}}\beta}e^{\sqrt{3}\gamma d\varepsilon\left(\frac{3\pi}{2}\right)^{\frac{1}{3}}p^\frac{1}{3}(x_c-d)+ih(p)+i\gamma\varepsilon x_c\beta^\frac{1}{2}p^\frac{1}{3}}\label{Ceq148}
\end{align}
So the overall integral $I_4$ is
{\medmuskip=0mu
\thinmuskip=0mu
\thickmuskip=0mu
\begin{align}
I_4=&i_1+i_2+i_3\nonumber\\
=&\frac{a_4^2}{8\pi^3c_1 \gamma \varepsilon\beta e^{i\frac{\pi}{3}}}e^{\sqrt{3}\gamma \varepsilon\left(\frac{3\pi}{2}\right)^{\frac{1}{3}}p^\frac{1}{3}(x_c-d)+ih(p)+x_c\gamma\varepsilon\beta^\frac{1}{2}ip^\frac{1}{3}}\nonumber\\
&-\frac{a_4^2}{4\pi^3c_1\gamma\varepsilon e^{i\frac{4\pi}{3}}\beta}e^{\sqrt{3}\gamma \varepsilon\left(\frac{3\pi}{2}\right)^{\frac{1}{3}}p^\frac{1}{3}(x_c-d)+ih(p)+x_c\gamma\varepsilon\beta^\frac{1}{2}ip^\frac{1}{3}}\nonumber\\
&+\frac{a_4^2}{8\pi^3c_1 \gamma \varepsilon e^{i\frac{\pi}{3}}\beta}e^{\sqrt{3}\gamma d\varepsilon\left(\frac{3\pi}{2}\right)^{\frac{1}{3}}p^\frac{1}{3}(x_c-d)+ih(p)+i\gamma\varepsilon x_c\beta^\frac{1}{2}p^\frac{1}{3}}\nonumber\\
=&\frac{a_4^2}{4\pi^3c_1 \gamma \varepsilon\beta e^{i\frac{\pi}{3}}}e^{\sqrt{3}\gamma \varepsilon\left(\frac{3\pi}{2}\right)^{\frac{1}{3}}p^\frac{1}{3}(x_c-d)+ih(p)+x_c\gamma\varepsilon\beta^\frac{1}{2}ip^\frac{1}{3}}\nonumber\\
&+\frac{a_4^2}{4\pi^3c_1\gamma\varepsilon e^{i\frac{\pi}{3}}\beta}e^{\sqrt{3}\gamma \varepsilon\left(\frac{3\pi}{2}\right)^{\frac{1}{3}}p^\frac{1}{3}(x_c-d)+ih(p)+x_c\gamma\varepsilon\beta^\frac{1}{2}ip^\frac{1}{3}}\nonumber\\
=&\frac{a_4^2}{2\pi^3c_1 \gamma \varepsilon\beta e^{i\frac{\pi}{3}}}e^{\sqrt{3}\gamma \varepsilon\left(\frac{3\pi}{2}\right)^{\frac{1}{3}}p^\frac{1}{3}(x_c-d)+ih(p)+x_c\gamma\varepsilon\beta^\frac{1}{2}ip^\frac{1}{3}}\label{Ceq149}
\end{align}}

\subsection{Normalization coefficients}
The integrals we calculated appear as a sum $I_1+I_2+I_3+I_4$. As we computed, the integral $I_1$ is decaying so we can neglect it and we realize that $I_3=-I_4$ so these two cancel. Using (\ref{Ceq118}), the final normalisation coefficients are
{\medmuskip=0mu
\thinmuskip=0mu
\thickmuskip=0mu
\begin{align}
N_p&\approx\frac{a_2^2}{8\pi^3c_1\gamma \varepsilon\beta e^{i\frac{\pi}{3}}}e^{ih(p)+\gamma V_0\beta^\frac{1}{2}\sqrt{3}p^\frac{1}{3}+\gamma\left(\frac{3\pi}{2}\right)^{\frac{1}{3}}i p^\frac{1}{3}(V_0+d\varepsilon)}\nonumber\\
&+a_2a_3\frac{1}{4\pi^3c_1\gamma\varepsilon e^{-i\frac{\pi}{6}}\beta }e^{ih(p)+\gamma\sqrt{3} \beta^\frac{1}{2}p^\frac{1}{3}V_0+\gamma i \beta^\frac{1}{2}p^\frac{1}{3}(V_0+d\varepsilon)}\nonumber\\
&+a_3^2\frac{1}{2\pi\gamma\varepsilon e^{i\frac{\pi}{3}}\beta}\left(-\frac{1}{4\pi^2c_1}e^{ih(p)+\gamma\sqrt{3}\beta^\frac{1}{2}p^\frac{1}{3}V_0+\gamma i \beta^\frac{1}{2}p^\frac{1}{3}(V_0+d\varepsilon)}\right.\nonumber\\
&\left.+p^{-\frac{4}{3}}\left(\pi^2c_1\right)e^{-ih(p)-\gamma\sqrt{3}\beta^\frac{1}{2}p^\frac{1}{3}(V_0-2d\varepsilon)-\gamma i \beta^\frac{1}{2}p^\frac{1}{3}(V_0+d\varepsilon)}\right.\nonumber\\
&\left.+\left(4\pi^2c_1\right)^{-1}e^{ih(p)+\gamma\sqrt{3}\beta^\frac{1}{2}p^\frac{1}{3}(V_0-2d\varepsilon)+\gamma i \beta^\frac{1}{2}p^\frac{1}{3}(V_0+d\varepsilon)}\right)\label{Ceq150}
\end{align}}

\subsection{\texorpdfstring{C-series expansion coefficients $c_p$}{Ccp}}
We are after all interested in the ratio of (\ref{Ceq43}) and (\ref{Ceq150}) so let us write what the numerator and denominator is.
The numerator
{\medmuskip=0mu
\thinmuskip=0mu
\thickmuskip=0mu
\begin{align}
b_p&(0)=e^{a\gamma\varepsilon \frac{\sqrt{3}}{2}\left(\frac{3\pi}{2}\right)^{\frac{1}{3}}p^\frac{1}{3}+ia\gamma\varepsilon \frac{1}{2}\left(\frac{3\pi}{2}\right)^{\frac{1}{3}}p^\frac{1}{3}}\left(-a_3^2\frac{1}{\beta\gamma\varepsilon\pi}e^{-i\frac{\pi}{3}}\right)\nonumber\\
&+e^{\sqrt{3}\gamma\left(\frac{3\pi}{2}\right)^{\frac{1}{3}}p^\frac{1}{3}\left( V_0-d\varepsilon+\frac{a\varepsilon}{2}\right)+ih(p)+\gamma\left(\frac{3\pi}{2}\right)^{\frac{1}{3}}i p^\frac{1}{3}\left( V_0+\frac{a\varepsilon}{2}\right)}\nonumber\\
&\left(4\pi^3c_1\beta\gamma\varepsilon\right)^{-1}\left(a_2^2e^{-i\frac{\pi}{3}}-2a_2a_3e^{-i\frac{5\pi}{6}}-a_3^2e^{-i\frac{\pi}{3}}\right)\nonumber\\
&+e^{\sqrt{3}\gamma\left(\frac{3\pi}{2}\right)^{\frac{1}{3}}p^\frac{1}{3}\left( d\varepsilon-V_0+\frac{a\varepsilon}{2}\right)-ih(p)+\gamma i \left(\frac{3\pi}{2}\right)^{\frac{1}{3}}p^\frac{1}{3}\left(-V_0+\frac{a\varepsilon}{2}\right)}\left(a_3^2\frac{\pi c_1}{\gamma\varepsilon\beta}e^{-i\frac{\pi}{3}}p^{-\frac{4}{3}}\right)\label{Ceq151}
\end{align}}
and the denominator from (\ref{Ceq150}) that can be simplifies into
\begin{align}
N_p&\approx\frac{1}{4\pi^3c_1\gamma\varepsilon\beta }e^{ih(p)+\gamma\sqrt{3} \beta^\frac{1}{2}p^\frac{1}{3}V_0+\gamma i \beta^\frac{1}{2}p^\frac{1}{3}(V_0+d\varepsilon)}\left(\frac{a_2^2}{2e^{i\frac{\pi}{3}}}+\frac{a_2a_3}{e^{-i\frac{\pi}{6}}}\right)\nonumber\\
&+a_3^2\frac{1}{2\pi\gamma\varepsilon e^{i\frac{\pi}{3}}\beta}e^{\gamma\sqrt{3}\beta^\frac{1}{2}p^\frac{1}{3}V_0}\left(-\frac{1}{4\pi^2c_1}e^{ih(p)+\gamma i \beta^\frac{1}{2}p^\frac{1}{3}(V_0+d\varepsilon)}\right.\nonumber\\
&\left.+p^{-\frac{4}{3}}\left(\pi^2c_1\right)e^{-ih(p)-2\gamma\sqrt{3}\beta^\frac{1}{2}p^\frac{1}{3}(V_0-d\varepsilon)-\gamma i \beta^\frac{1}{2}p^\frac{1}{3}(V_0+d\varepsilon)}\right)\label{Ceq152}
\end{align}
Observe, that both are growing exponentially, but there are some conditions under which some of the terms vanish. In the case of $b_p(0)$ if $V_0-d\varepsilon<-\frac{a\varepsilon}{2}$ then the second term vanishes and the third grows. If $-\frac{a\varepsilon}{2}<V_0-d\varepsilon<\frac{a\varepsilon}{2}$, then both terms grow and if $\frac{a\varepsilon}{2}<V_0-d\varepsilon$ then the second grows and the third vanishes. We look into each case separately.

\subsubsection{\texorpdfstring{$V_0-d\varepsilon<-\frac{a\varepsilon}{2}$}{Case 1}}
We have the first case when $V_0-d\varepsilon<-\frac{a\varepsilon}{2}$. Then (\ref{Ceq151}) turns to
\begin{align}
b_p(0)&\approx e^{a\gamma\varepsilon \frac{\sqrt{3}}{2}\left(\frac{3\pi}{2}\right)^{\frac{1}{3}}p^\frac{1}{3}+ia\gamma\varepsilon \frac{1}{2}\left(\frac{3\pi}{2}\right)^{\frac{1}{3}}p^\frac{1}{3}}\left(-a_3^2\frac{1}{\beta\gamma\varepsilon\pi}e^{-i\frac{\pi}{3}}\right)\nonumber\\
&+e^{\sqrt{3}\gamma\left(\frac{3\pi}{2}\right)^{\frac{1}{3}}p^\frac{1}{3}\left( d\varepsilon-V_0+\frac{a\varepsilon}{2}\right)-ih(p)+\gamma i \left(\frac{3\pi}{2}\right)^{\frac{1}{3}}p^\frac{1}{3}\left(-V_0+\frac{a\varepsilon}{2}\right)}\left(a_3^2\frac{\pi c_1}{\gamma\varepsilon\beta}e^{-i\frac{\pi}{3}}p^{-\frac{4}{3}}\right)\label{Ceq153}
\end{align}
From the initial assumption implies that $V_0<d\varepsilon$. Then $d\varepsilon-V_0+\frac{a\varepsilon}{2}>a\varepsilon\Rightarrow d\varepsilon-V_0>0\Rightarrow d\varepsilon>V_0$ which is true, so the second term in (\ref{Ceq153}) grows faster than the first and we write
\begin{align}
b_p(0)&\approx e^{\sqrt{3}\gamma\left(\frac{3\pi}{2}\right)^{\frac{1}{3}}p^\frac{1}{3}\left( d\varepsilon-V_0+\frac{a\varepsilon}{2}\right)-ih(p)+\gamma i \left(\frac{3\pi}{2}\right)^{\frac{1}{3}}p^\frac{1}{3}\left(-V_0+\frac{a\varepsilon}{2}\right)}\left(a_3^2\frac{\pi c_1}{\gamma\varepsilon\beta}e^{-i\frac{\pi}{3}}p^{-\frac{4}{3}}\right)\label{Ceq154}
\end{align}
The coefficients $N_p$ are
\begin{align}
N_p&=\frac{1}{4\pi^3c_1\gamma\varepsilon\beta }e^{ih(p)+\gamma\sqrt{3} \beta^\frac{1}{2}p^\frac{1}{3}V_0+\gamma i \beta^\frac{1}{2}p^\frac{1}{3}(V_0+d\varepsilon)}\left(\frac{a_2^2}{2e^{i\frac{\pi}{3}}}+\frac{a_2a_3}{e^{-i\frac{\pi}{6}}}\right)\nonumber\\
&+a_3^2\frac{1}{2\pi\gamma\varepsilon e^{i\frac{\pi}{3}}\beta}\left(-\frac{1}{4\pi^2c_1}e^{ih(p)+\gamma\sqrt{3}\beta^\frac{1}{2}p^\frac{1}{3}V_0+\gamma i \beta^\frac{1}{2}p^\frac{1}{3}(V_0+d\varepsilon)}\right.\nonumber\\
&\left.+p^{-\frac{4}{3}}\left(\pi^2c_1\right)e^{-ih(p)+\gamma\sqrt{3}\beta^\frac{1}{2}p^\frac{1}{3}(2d\varepsilon-V_0)-\gamma i \beta^\frac{1}{2}p^\frac{1}{3}(V_0+d\varepsilon)}\right)\label{Ceq155}
\end{align}
and it is true that $2d\varepsilon-V_0>V_0\Rightarrow2d\varepsilon>2V_0\Rightarrow d\varepsilon>V_0$, so the third term in (\ref{Ceq155}) is growing faster than the others, so we can write
\begin{align}
N_p&\approx a_3^2\frac{1}{2\pi\gamma\varepsilon e^{i\frac{\pi}{3}}\beta}p^{-\frac{4}{3}}\left(\pi^2c_1\right)e^{-ih(p)+\gamma\sqrt{3}\beta^\frac{1}{2}p^\frac{1}{3}(2d\varepsilon-V_0)-\gamma i \beta^\frac{1}{2}p^\frac{1}{3}(V_0+d\varepsilon)}\label{Ceq156}
\end{align}
At this stage, we can divide.
{\medmuskip=-1mu
\thinmuskip=-1mu
\thickmuskip=-1mu
\begin{align}
\frac{b_p(0)}{N_p}=&e^{\sqrt{3}\gamma\left(\frac{3\pi}{2}\right)^{\frac{1}{3}}p^\frac{1}{3}\left( d\varepsilon-V_0+\frac{a\varepsilon}{2}\right)-ih(p)+\gamma i \left(\frac{3\pi}{2}\right)^{\frac{1}{3}}p^\frac{1}{3}\left(-V_0+\frac{a\varepsilon}{2}\right)}\left(a_3^2\frac{\pi c_1}{\gamma\varepsilon\beta}e^{-i\frac{\pi}{3}}p^{-\frac{4}{3}}\right)\nonumber\\
&\frac{1}{a_3^2\frac{1}{2\pi\gamma\varepsilon e^{i\frac{\pi}{3}}\beta}p^{-\frac{4}{3}}\left(\pi^2c_1\right)}e^{ih(p)-\gamma\sqrt{3}\beta^\frac{1}{2}p^\frac{1}{3}(2d\varepsilon-V_0)+\gamma i \beta^\frac{1}{2}p^\frac{1}{3}(V_0+d\varepsilon)}\nonumber\\
=&\frac{e^{-i\frac{\pi}{3}}p^{-\frac{4}{3}}a_3^22\pi\gamma\varepsilon e^{i\frac{\pi}{3}}\beta\pi c_1}{a_3^2p^{-\frac{4}{3}}\pi^2c_1\gamma\varepsilon\beta}e^{\sqrt{3}\gamma\left(\frac{3\pi}{2}\right)^{\frac{1}{3}}p^\frac{1}{3}\left( d\varepsilon-V_0+\frac{a\varepsilon}{2}\right)-\gamma\sqrt{3}\beta^\frac{1}{2}p^\frac{1}{3}(2d\varepsilon-V_0)}\nonumber\\
&e^{-ih(p)+\gamma i \left(\frac{3\pi}{2}\right)^{\frac{1}{3}}p^\frac{1}{3}\left(-V_0+\frac{a\varepsilon}{2}\right)+ih(p)+\gamma i \beta^\frac{1}{2}p^\frac{1}{3}(V_0+d\varepsilon)}\nonumber\\
=&2e^{\sqrt{3}\gamma\left(\frac{3\pi}{2}\right)^{\frac{1}{3}}p^\frac{1}{3}\left( d\varepsilon-V_0+\frac{a\varepsilon}{2}-2d\varepsilon+V_0\right)+\gamma i \left(\frac{3\pi}{2}\right)^{\frac{1}{3}}p^\frac{1}{3}\left(-V_0+\frac{a\varepsilon}{2}+V_0+d\varepsilon\right)}\nonumber\\
=&2e^{\sqrt{3}\gamma\left(\frac{3\pi}{2}\right)^{\frac{1}{3}}p^\frac{1}{3}\left( -d\varepsilon+\frac{a\varepsilon}{2}\right)+\gamma i \left(\frac{3\pi}{2}\right)^{\frac{1}{3}}p^\frac{1}{3}\left(\frac{a\varepsilon}{2}+d\varepsilon\right)}\label{Ceq157}
\end{align}}
Since $a<d$, then $-d\varepsilon+\frac{a\varepsilon}{2}<0$ which makes that this expression is decaying.

The second case is when $-\frac{a\varepsilon}{2}<V_0-d\varepsilon<\frac{a\varepsilon}{2}$. In this case we do not know the relationship between $V_0$ and $d\varepsilon$, but we can split this condition into $-\frac{a\varepsilon}{2}<V_0-d\varepsilon<0$ and $0<V_0-d\varepsilon<\frac{a\varepsilon}{2}$.

\subsubsection{\texorpdfstring{$-\frac{a\varepsilon}{2}<V_0-d\varepsilon<0$}{Case 2}}
For the next case we have $-\frac{a\varepsilon}{2}<V_0-d\varepsilon<0$, so $V_0<d\varepsilon$. In the coefficients $b_p(0)$ all the terms grow, but we pick only the fastest. Let us compare the exponents in the second and the third term in (\ref{Ceq151}).
\begin{align}
V_0-d\varepsilon+\frac{a\varepsilon}{2}&<d\varepsilon-V_0+\frac{a\varepsilon}{2}\nonumber\\
V_0<d\varepsilon\label{Ceq158}
\end{align}
which is true. Consequently we have that $d\varepsilon-V_0+\frac{a\varepsilon}{2}>\frac{a\varepsilon}{2}$, so the third terms is the fastest.
\begin{align}
b_p(0)&\approx e^{\sqrt{3}\gamma\left(\frac{3\pi}{2}\right)^{\frac{1}{3}}p^\frac{1}{3}\left( d\varepsilon-V_0+\frac{a\varepsilon}{2}\right)-ih(p)+\gamma i \left(\frac{3\pi}{2}\right)^{\frac{1}{3}}p^\frac{1}{3}\left(-V_0+\frac{a\varepsilon}{2}\right)}\left(a_3^2\frac{\pi c_1}{\gamma\varepsilon\beta}e^{-i\frac{\pi}{3}}p^{-\frac{4}{3}}\right)\label{Ceq159}
\end{align}
$N_p$ simplify to the same expression as in the previous case.
\begin{align}
N_p&\approx a_3^2\frac{1}{2\pi\gamma\varepsilon e^{i\frac{\pi}{3}}\beta}p^{-\frac{4}{3}}\left(\pi^2c_1\right)e^{-ih(p)+\gamma\sqrt{3}\beta^\frac{1}{2}p^\frac{1}{3}(2d\varepsilon-V_0)-\gamma i \beta^\frac{1}{2}p^\frac{1}{3}(V_0+d\varepsilon)}\label{Ceq160}
\end{align}
We got the same conclusion as last time, so the division $c_p=b_p(0)/N_p$ decays exponentially.

\subsubsection{\texorpdfstring{$0<V_0-d\varepsilon<\frac{a\varepsilon}{2}$}{Case 3}}
In this case we have $0<V_0-d\varepsilon<\frac{a\varepsilon}{2}$ from which we deduce that $d\varepsilon<V_0$. In (\ref{Ceq151}) all terms grow, but based on the comparing we did in (\ref{Ceq158}) we conclude that the second term grows faster than the third. If we compare it with the first in (\ref{Ceq151}) we get
\begin{align}
V_0-d\varepsilon+\frac{a\varepsilon}{2}&>\frac{a\varepsilon}{2}\nonumber\\
V_0>d\varepsilon\label{Ceq161}
\end{align}
which confirms that the second term wins and we write
\begin{align}
b_p(0)&\approx e^{\sqrt{3}\gamma\left(\frac{3\pi}{2}\right)^{\frac{1}{3}}p^\frac{1}{3}\left( V_0-d\varepsilon+\frac{a\varepsilon}{2}\right)+ih(p)+\gamma\left(\frac{3\pi}{2}\right)^{\frac{1}{3}}i p^\frac{1}{3}\left( V_0+\frac{a\varepsilon}{2}\right)}\nonumber\\
&\left(4\pi^3c_1\beta\gamma\varepsilon\right)^{-1}\left(a_2^2e^{-i\frac{\pi}{3}}-2a_2a_3e^{-i\frac{5\pi}{6}}-a_3^2e^{-i\frac{\pi}{3}}\right)\label{Ceq162}
\end{align}
The general coefficients $N_p$ are
\begin{align}
N_p&=\frac{1}{4\pi^3c_1\gamma\varepsilon\beta }e^{ih(p)+\gamma\sqrt{3} \beta^\frac{1}{2}p^\frac{1}{3}V_0+\gamma i \beta^\frac{1}{2}p^\frac{1}{3}(V_0+d\varepsilon)}\left(\frac{a_2^2}{2e^{i\frac{\pi}{3}}}+\frac{a_2a_3}{e^{-i\frac{\pi}{6}}}\right)\nonumber\\
&+a_3^2\frac{1}{2\pi\gamma\varepsilon e^{i\frac{\pi}{3}}\beta}e^{\gamma\sqrt{3}\beta^\frac{1}{2}p^\frac{1}{3}V_0}\left(-\frac{1}{4\pi^2c_1}e^{ih(p)+\gamma i \beta^\frac{1}{2}p^\frac{1}{3}(V_0+d\varepsilon)}\right.\nonumber\\
&\left.+p^{-\frac{4}{3}}\left(\pi^2c_1\right)e^{-ih(p)-2\gamma\sqrt{3}\beta^\frac{1}{2}p^\frac{1}{3}(V_0-d\varepsilon)-\gamma i \beta^\frac{1}{2}p^\frac{1}{3}(V_0+d\varepsilon)}\right)\label{Ceq163}
\end{align}
but since $V_0<d\varepsilon$, the last term vanishes and we get
\begin{align}
N_p&\approx\frac{1}{4\pi^3c_1\gamma\varepsilon\beta }e^{ih(p)+\gamma\sqrt{3} \beta^\frac{1}{2}p^\frac{1}{3}V_0+\gamma i \beta^\frac{1}{2}p^\frac{1}{3}(V_0+d\varepsilon)}\left(\frac{a_2^2}{2e^{i\frac{\pi}{3}}}+\frac{a_2a_3}{e^{-i\frac{\pi}{6}}}\right)\nonumber\\
&+a_3^2\frac{1}{2\pi\gamma\varepsilon e^{i\frac{\pi}{3}}\beta}e^{\gamma\sqrt{3}\beta^\frac{1}{2}p^\frac{1}{3}V_0+ih(p)+\gamma i \beta^\frac{1}{2}p^\frac{1}{3}(V_0+d\varepsilon)}\left(-\frac{1}{4\pi^2c_1}\right)\nonumber\\
&=e^{\gamma\sqrt{3}\beta^\frac{1}{2}p^\frac{1}{3}V_0+ih(p)+\gamma i \beta^\frac{1}{2}p^\frac{1}{3}(V_0+d\varepsilon)}\nonumber\\
&\left(\frac{1}{4\pi^3c_1\gamma\varepsilon\beta }\left(\frac{a_2^2}{2e^{i\frac{\pi}{3}}}+\frac{a_2a_3}{e^{-i\frac{\pi}{6}}}\right)-a_3^2\frac{1}{2\pi\gamma\varepsilon e^{i\frac{\pi}{3}}\beta}\frac{1}{4\pi^2c_1}\right)\nonumber\\
&=e^{\gamma\sqrt{3}\beta^\frac{1}{2}p^\frac{1}{3}V_0+ih(p)+\gamma i \beta^\frac{1}{2}p^\frac{1}{3}(V_0+d\varepsilon)}\frac{1}{4\pi^3c_1\gamma\varepsilon\beta }\left(\frac{a_2^2}{2e^{i\frac{\pi}{3}} }+\frac{a_2a_3}{e^{-i\frac{\pi}{6}}}-\frac{a_3^2}{2e^{i\frac{\pi}{3}}}\right)\label{Ceq164}
\end{align}
Dividing (\ref{Ceq162}) by (\ref{Ceq164}) we get
{\medmuskip=-1mu
\thinmuskip=-1mu
\thickmuskip=-1mu
\begin{align}
c_p&=\frac{b_p(0)}{N_p}=e^{\sqrt{3}\gamma\left(\frac{3\pi}{2}\right)^{\frac{1}{3}}p^\frac{1}{3}\left( V_0-d\varepsilon+\frac{a\varepsilon}{2}\right)+ih(p)+\gamma\left(\frac{3\pi}{2}\right)^{\frac{1}{3}}i p^\frac{1}{3}\left( V_0+\frac{a\varepsilon}{2}\right)}\left(4\pi^3c_1\beta\gamma\varepsilon\right)^{-1}\nonumber\\
&\left(a_2^2e^{-i\frac{\pi}{3}}-2a_2a_3e^{-i\frac{5\pi}{6}}-a_3^2e^{-i\frac{\pi}{3}}\right)4\pi^3c_1\gamma\varepsilon\beta\left(\frac{a_2^2}{2e^{i\frac{\pi}{3}} }+\frac{a_2a_3}{e^{-i\frac{\pi}{6}}}-\frac{a_3^2}{2e^{i\frac{\pi}{3}}}\right)^{-1}\nonumber\\
&e^{-\gamma\sqrt{3}\beta^\frac{1}{2}p^\frac{1}{3}V_0-ih(p)-\gamma i \beta^\frac{1}{2}p^\frac{1}{3}(V_0+d\varepsilon)}\nonumber\\
&=\left(a_2^2e^{-i\frac{\pi}{3}}-2a_2a_3e^{-i\frac{5\pi}{6}}-a_3^2e^{-i\frac{\pi}{3}}\right)\left(\frac{a_2^2}{2e^{i\frac{\pi}{3}} }+\frac{a_2a_3}{e^{-i\frac{\pi}{6}}}-\frac{a_3^2}{2e^{i\frac{\pi}{3}}}\right)^{-1}\nonumber\\
&e^{\sqrt{3}\gamma\left(\frac{3\pi}{2}\right)^{\frac{1}{3}}p^\frac{1}{3}\left( V_0-d\varepsilon+\frac{a\varepsilon}{2}\right)-\gamma\sqrt{3}\beta^\frac{1}{2}p^\frac{1}{3}V_0}\nonumber\\
&e^{ih(p)+\gamma\left(\frac{3\pi}{2}\right)^{\frac{1}{3}}i p^\frac{1}{3}\left( V_0+\frac{a\varepsilon}{2}\right)-ih(p)-\gamma i \beta^\frac{1}{2}p^\frac{1}{3}(V_0+d\varepsilon)}\nonumber\\
&=\left(a_2^2e^{-i\frac{\pi}{3}}-2a_2a_3e^{-i\frac{5\pi}{6}}-a_3^2e^{-i\frac{\pi}{3}}\right)\left(\frac{a_2^2}{2e^{i\frac{\pi}{3}} }+\frac{a_2a_3}{e^{-i\frac{\pi}{6}}}-\frac{a_3^2}{2e^{i\frac{\pi}{3}}}\right)^{-1}\nonumber\\
&e^{\sqrt{3}\gamma\left(\frac{3\pi}{2}\right)^{\frac{1}{3}}p^\frac{1}{3}\left( V_0-d\varepsilon+\frac{a\varepsilon}{2}-V_0\right)+\gamma\left(\frac{3\pi}{2}\right)^{\frac{1}{3}}i p^\frac{1}{3}\left( V_0+\frac{a\varepsilon}{2}-V_0-d\varepsilon\right)}\nonumber\\
&=\left(a_2^2e^{-i\frac{\pi}{3}}-2a_2a_3e^{-i\frac{5\pi}{6}}-a_3^2e^{-i\frac{\pi}{3}}\right)\left(\frac{a_2^2}{2e^{i\frac{\pi}{3}} }+\frac{a_2a_3}{e^{-i\frac{\pi}{6}}}-\frac{a_3^2}{2e^{i\frac{\pi}{3}}}\right)^{-1}\nonumber\\
&e^{\sqrt{3}\gamma\left(\frac{3\pi}{2}\right)^{\frac{1}{3}}p^\frac{1}{3}\left(-d\varepsilon+\frac{a\varepsilon}{2}\right)+\gamma\left(\frac{3\pi}{2}\right)^{\frac{1}{3}}i p^\frac{1}{3}\left(\frac{a\varepsilon}{2}-d\varepsilon\right)}\label{Ceq165}
\end{align}}
This decays always, since $d>a$.

\subsubsection{\texorpdfstring{$\frac{a\varepsilon}{2}<V_0-d\varepsilon$}{Case 4}}
Lastly, we have again that $\frac{a\varepsilon}{2}<V_0-d\varepsilon$ and $V_0<d\varepsilon$. This leaves us with the second and first term in (\ref{Ceq151}), but as we saw last time, the second is faster. There is neither a change for $N_p$ comparing to the last case, so everything stays the same and we have an exponential decay for this case as well.

\subsection{\texorpdfstring{Continuity coefficients}{Cai}}
Just as in the case of A-series, we assumed that the continuity coefficients $a_1,\ldots,a_4$ throughout the computations did not have any effect on the result. Let us actually confirm that it is really so. The continuity matrix we use to compute the coefficients is
\begin{align}
\mathbf{M}=
\begin{pmatrix}
A_0 & -A_1 & -B_1 & 0\\
A'_0 & -A'_1 & -B'_1 & 0\\
0 & A_2 & B_2 & -C_3\\
0 & A'_2 & B'_2 & -C'_3
\end{pmatrix}\label{Ceq21.1}
\end{align}
where
\begin{align}
A_0&=\mathrm{Ai}(y_1(-d)) & A_1&=\mathrm{Ai}(y_2(-d)) & B_1&=\mathrm{Bi}(y_2(-d))\label{Ceq21.2}\\
A_2&=\mathrm{Ai}(y_2(d)) & B_2&=\mathrm{Bi}(y_2(d)) & C_3&=\mathrm{Ci}^+(y_1(d))\label{Ceq21.3}
\end{align}
Using the asymptotic expressions
\begin{align}
\mathrm{Ai}\left(z_{1,2,3}^A\right)&\approx\frac{1}{2}\pi^{-\frac{1}{2}}\left(z_{1,2,3}^A\right)^{-\frac{1}{4}}e^{-\frac{2}{3}\left(z_{1,2,3}^A\right)^\frac{3}{2}}\label{Ceq21.4}\\
\mathrm{Ai'}\left(z_{1,2,3}^A\right)&\approx-\frac{1}{2}\pi^{-\frac{1}{2}}\left(z_{1,2,3}^A\right)^{\frac{1}{4}}e^{-\frac{2}{3}\left(z_{1,2,3}^A\right)^\frac{3}{2}}\label{Ceq21.5}\\
\mathrm{Bi}\left(z_{1,2,3}^B\right)&\approx\sqrt{\frac{2}{\pi}}e^{i\frac{\pi}{6}}\left(z_{1,2,3}^B\right)^{-\frac{1}{4}}\sin\left(\frac{2}{3}\left(z_{1,2,3}^B\right)^\frac{3}{2}+\frac{\pi}{4}-\frac{i}{2}\ln2\right)\label{Ceq21.6}\\
\mathrm{Bi'}\left(z_{1,2,3}^B\right)&\approx\sqrt{\frac{2}{\pi}}e^{-i\frac{\pi}{6}}\left(z_{1,2,3}^B\right)^{\frac{1}{4}}\cos\left(\frac{2}{3}\left(z_{1,2,3}^B\right)^\frac{3}{2}+\frac{\pi}{4}-\frac{i}{2}\ln2\right)\label{Ceq21.7}\\
\mathrm{Ci}^+&=\mathrm{Bi}\left(z_{1,2,3}^B\right)+i\mathrm{Ai}\left(z_{1,2,3}^A\right)\label{Ceq21.8}\\
\mathrm{Ci'}^+&=\mathrm{Bi'}\left(z_{1,2,3}^B\right)+i\mathrm{Ai'}\left(z_{1,2,3}^A\right)\label{Ceq21.9}
\end{align}
and the knowledge we gained throughout this Appendix, we write down the terms in the matrix $\mathbf{M}$ as
\begin{align}
A_0&\approx\frac{1}{2}\pi^{-\frac{1}{2}}\mu^{-\frac{1}{4}}p^{-\frac{1}{6}}(-1)^pp^{\frac{1}{3}}\left(\pi^2c_1\right)^{-\frac{1}{2}}e^{-\gamma\left(\frac{3\pi}{2}\right)^{\frac{1}{3}}\frac{\sqrt{3}}{2} p^\frac{1}{3}2d\varepsilon-\gamma\left(\frac{3\pi}{2}\right)^{\frac{1}{3}}\frac{1}{2}i p^\frac{1}{3}d\varepsilon}\approx 0\label{Ceq21.14}\\
A_1&\approx\frac{1}{2}\pi^{-\frac{1}{2}}\mu^{-\frac{1}{4}}p^{\frac{1}{6}}(-1)^p\left(\pi^2c_1\right)^{-\frac{1}{2}}e^{\gamma\left(\frac{3\pi}{2}\right)^{\frac{1}{3}}\frac{\sqrt{3}}{2} p^\frac{1}{3}(-2d\varepsilon+V_0)+\gamma\left(\frac{3\pi}{2}\right)^{\frac{1}{3}}\frac{1}{2}i p^\frac{1}{3}(-d\varepsilon+V_0)}\label{Ceq21.15}\\
A_2&\approx\frac{1}{2}\pi^{-\frac{1}{2}}\mu^{-\frac{1}{4}}p^{\frac{1}{6}}(-1)^p\left(\pi^2c_1\right)^{-\frac{1}{2}}e^{\gamma\left(\frac{3\pi}{2}\right)^{\frac{1}{3}}\frac{\sqrt{3}}{2} p^\frac{1}{3}V_0+\gamma\left(\frac{3\pi}{2}\right)^{\frac{1}{3}}\frac{1}{2}i p^\frac{1}{3}(d\varepsilon+V_0)}\label{Ceq21.16}\\
A_3&\approx\frac{1}{2}\pi^{-\frac{1}{2}}\mu^{-\frac{1}{4}}p^{\frac{1}{6}}(-1)^p\left(\pi^2c_1\right)^{-\frac{1}{2}}e^{\gamma\left(\frac{3\pi}{2}\right)^{\frac{1}{3}}\frac{1}{2}i p^\frac{1}{3}d\varepsilon}\label{Ceq21.17}\\
A'_0&\approx-\frac{1}{2}\pi^{-\frac{1}{2}}\mu^{\frac{1}{4}}p^{\frac{1}{2}}(-1)^p\left(\pi^2c_1\right)^{-\frac{1}{2}}e^{-\gamma\left(\frac{3\pi}{2}\right)^{\frac{1}{3}}\frac{\sqrt{3}}{2} p^\frac{1}{3}2d\varepsilon-\gamma\left(\frac{3\pi}{2}\right)^{\frac{1}{3}}\frac{1}{2}i p^\frac{1}{3}d\varepsilon}\approx 0\label{Ceq21.18}\\
A'_1&\approx-\frac{1}{2}\pi^{-\frac{1}{2}}\mu^{\frac{1}{4}}p^{\frac{1}{2}}(-1)^p\left(\pi^2c_1\right)^{-\frac{1}{2}}e^{\gamma\left(\frac{3\pi}{2}\right)^{\frac{1}{3}}\frac{\sqrt{3}}{2} p^\frac{1}{3}(-2d\varepsilon+V_0)+\gamma\left(\frac{3\pi}{2}\right)^{\frac{1}{3}}\frac{1}{2}i p^\frac{1}{3}(-d\varepsilon+V_0)}\label{Ceq21.19}\\
A'_2&\approx-\frac{1}{2}\pi^{-\frac{1}{2}}\mu^{\frac{1}{4}}p^{\frac{1}{2}}(-1)^p\left(\pi^2c_1\right)^{-\frac{1}{2}}e^{\gamma\left(\frac{3\pi}{2}\right)^{\frac{1}{3}}\frac{\sqrt{3}}{2} p^\frac{1}{3}V_0+\gamma\left(\frac{3\pi}{2}\right)^{\frac{1}{3}}\frac{1}{2}i p^\frac{1}{3}(d\varepsilon+V_0)}\label{Ceq21.20}\\
A'_3&\approx-\frac{1}{2}\pi^{-\frac{1}{2}}\mu^{\frac{1}{4}}p^{\frac{1}{2}}(-1)^p\left(\pi^2c_1\right)^{-\frac{1}{2}}e^{\gamma\left(\frac{3\pi}{2}\right)^{\frac{1}{3}}\frac{1}{2}i p^\frac{1}{3}d\varepsilon}\label{Ceq21.21}
\end{align}
where $\mu=\beta e^{i\frac{\pi}{3}}, \beta=\left(\frac{3\pi}{2}\right)^\frac{2}{3}$ and the Airy $\mathrm{Bi}$ functions
{\medmuskip=0mu
\thinmuskip=0mu
\thickmuskip=0mu
\begin{align}
B_0\approx&\sqrt{\frac{2}{\pi}}e^{i\frac{\pi}{6}}\beta^{-\frac{1}{4}}p^{-\frac{1}{2}}\frac{1}{2i}(-1)^p\left(\pi^2c_1\right)^\frac{1}{2}\sqrt{2}e^{i\gamma\frac{1}{2}\left(\frac{3\pi}{2}\right)^{\frac{1}{3}}p^\frac{1}{3}d\varepsilon+\gamma\frac{\sqrt{3}}{2}\left(\frac{3\pi}{2}\right)^{\frac{1}{3}}p^\frac{1}{3}2d\varepsilon}\label{Ceq21.22}\\
B_1=&\sqrt{\frac{2}{\pi}}e^{i\frac{\pi}{6}}\beta^{-\frac{1}{4}}\frac{1}{2i}\nonumber\\
&\left((-1)^pp^{-\frac{1}{2}}\left(\pi^2c_1\right)^\frac{1}{2}\sqrt{2}e^{-i\gamma\frac{1}{2}\left(\frac{3\pi}{2}\right)^{\frac{1}{3}}p^\frac{1}{3}(V_0-d\varepsilon)-\gamma\frac{\sqrt{3}}{2}\left(\frac{3\pi}{2}\right)^{\frac{1}{3}}p^\frac{1}{3}(V_0-2d\varepsilon)}\right.\nonumber\\
&\left.-(-1)^pp^{\frac{1}{6}}\left(\pi^2c_1\right)^{-\frac{1}{2}}2^{-\frac{1}{2}} e^{i\gamma\frac{1}{2}\left(\frac{3\pi}{2}\right)^{\frac{1}{3}}p^\frac{1}{3}(V_0-d\varepsilon)+\gamma\frac{\sqrt{3}}{2}\left(\frac{3\pi}{2}\right)^{\frac{1}{3}}p^\frac{1}{3}(V_0-2d\varepsilon)}\right)\label{Ceq21.23}\\
B_2\approx&-\sqrt{\frac{2}{\pi}}e^{i\frac{\pi}{6}}\beta^{-\frac{1}{4}}p^{\frac{1}{6}}\frac{1}{2i}(-1)^p\left(\pi^2c_1\right)^{-\frac{1}{2}}2^{-\frac{1}{2}}e^{i\gamma\frac{1}{2}\left(\frac{3\pi}{2}\right)^{\frac{1}{3}}p^\frac{1}{3}(V_0+d\varepsilon)+\gamma\frac{\sqrt{3}}{2}\left(\frac{3\pi}{2}\right)^{\frac{1}{3}}p^\frac{1}{3}V_0}\label{Ceq21.24}\\
B_3\approx&-\sqrt{\frac{2}{\pi}}e^{i\frac{\pi}{6}}\beta^{-\frac{1}{4}}\frac{1}{2i}(-1)^pp^{\frac{1}{6}}\left(\pi^2c_1\right)^{-\frac{1}{2}}2^{-\frac{1}{2}}e^{i\gamma\frac{1}{2}\left(\frac{3\pi}{2}\right)^{\frac{1}{3}}p^\frac{1}{3}d\varepsilon})\label{Ceq21.25}\\
B'_0\approx&\sqrt{\frac{2}{\pi}}e^{-i\frac{\pi}{6}}\beta^{\frac{1}{4}}p^{-\frac{1}{6}}\frac{1}{2}(-1)^p\left(\pi^2c_1\right)^\frac{1}{2}\sqrt{2}e^{i\gamma\frac{1}{2}\left(\frac{3\pi}{2}\right)^{\frac{1}{3}}p^\frac{1}{3}d\varepsilon+\gamma\frac{\sqrt{3}}{2}\left(\frac{3\pi}{2}\right)^{\frac{1}{3}}p^\frac{1}{3}2d\varepsilon}\label{Ceq21.26}\\
B'_1\approx&\sqrt{\frac{2}{\pi}}e^{-i\frac{\pi}{6}}\beta^{\frac{1}{4}}\frac{1}{2}\nonumber\\
&\left((-1)^pp^{-\frac{1}{6}}\left(\pi^2c_1\right)^\frac{1}{2}\sqrt{2}e^{-i\gamma\frac{1}{2}\left(\frac{3\pi}{2}\right)^{\frac{1}{3}}p^\frac{1}{3}(V_0-d\varepsilon)-\gamma\frac{\sqrt{3}}{2}\left(\frac{3\pi}{2}\right)^{\frac{1}{3}}p^\frac{1}{3}(V_0-2d\varepsilon)}\right.\nonumber\\
&\left.+(-1)^pp^{\frac{1}{2}}\left(\pi^2c_1\right)^{-\frac{1}{2}}2^{-\frac{1}{2}} e^{i\gamma\frac{1}{2}\left(\frac{3\pi}{2}\right)^{\frac{1}{3}}p^\frac{1}{3}(V_0-d\varepsilon)+\gamma\frac{\sqrt{3}}{2}\left(\frac{3\pi}{2}\right)^{\frac{1}{3}}p^\frac{1}{3}(V_0-2d\varepsilon)}\right)\label{Ceq21.27}\\
B'_2\approx&\sqrt{\frac{2}{\pi}}e^{-i\frac{\pi}{6}}\beta^{\frac{1}{4}}\frac{1}{2}(-1)^pp^{\frac{1}{2}}\left(\pi^2c_1\right)^{-\frac{1}{2}}2^{-\frac{1}{2}}e^{i\gamma\frac{1}{2}\left(\frac{3\pi}{2}\right)^{\frac{1}{3}}p^\frac{1}{3}(V_0+d\varepsilon)+\gamma\frac{\sqrt{3}}{2}\left(\frac{3\pi}{2}\right)^{\frac{1}{3}}p^\frac{1}{3}V_0}\label{Ceq21.28}\\
B'_3\approx&\sqrt{\frac{2}{\pi}}e^{-i\frac{\pi}{6}}\beta^{\frac{1}{4}}\frac{1}{2}(-1)^pp^{\frac{1}{2}}\left(\pi^2c_1\right)^{-\frac{1}{2}}2^{-\frac{1}{2}}e^{i\gamma\frac{1}{2}\left(\frac{3\pi}{2}\right)^{\frac{1}{3}}p^\frac{1}{3}d\varepsilon}\label{Ceq21.29}
\end{align}}
Some of the terms grow exponentially ($exp.$), some algebraically ($algeb.$) and some terms grow exponentially under a condition that depends on the relationship between the parameters $V_0,d,\varepsilon$ ($cond.exp.$). Let us sum up, which term does which.
\begin{align}
\mathbf{M}&=\begin{pmatrix}
A_0 & -A_1 & -B_1 & 0\\
A'_0 & -A'_1 & -B'_1 & 0\\
0 & A_2 & B_2 & -C_3\\
0 & A'_2 & B'_2 & -C'_3
\end{pmatrix}\approx
\begin{pmatrix}
0 & cond. exp. & exp. & 0\\
0 & cond. exp. & exp. & 0\\
0 & exp. & exp. & algeb.\\
0 & exp. & exp & algeb.
\end{pmatrix}\nonumber\\
&\approx
\begin{pmatrix}
0 & cond. exp. & exp. & 0\\
0 & cond. exp. & exp. & 0\\
0 & exp. & exp. & 0\\
0 & exp. & exp & 0
\end{pmatrix}\label{Ceq21.30}
\end{align}
Only two terms are conditioned $A_1$ and $A'_1$. From (\ref{Ceq21.15}) and (\ref{Ceq21.19}) we have that both of them grow exponentially if $-2d\varepsilon+V_0>0$, so when $V_0>2d\varepsilon$. However, the terms $B_1$ and $B'_1$ also have conditions, but since they include both cases (see (\ref{Ceq21.23}), (\ref{Ceq21.27})), they will grow anyway. The important part is, that the same condition applies for them as well.

Let us look at the first case, when $V_0>2d\varepsilon$. Then we have the non-zero terms
{\medmuskip=0mu
\thinmuskip=0mu
\thickmuskip=0mu
\begin{align}
-A_1&=-\frac{1}{2}\pi^{-\frac{1}{2}}\mu^{-\frac{1}{4}}p^{\frac{1}{6}}(-1)^p\left(\pi^2c_1\right)^{-\frac{1}{2}}\nonumber\\
&e^{\gamma\left(\frac{3\pi}{2}\right)^{\frac{1}{3}}\frac{\sqrt{3}}{2} p^\frac{1}{3}(-2d\varepsilon+V_0)+\gamma\left(\frac{3\pi}{2}\right)^{\frac{1}{3}}\frac{1}{2}i p^\frac{1}{3}(-d\varepsilon+V_0)}\label{Ceq21.31}\\
-B_1&=\sqrt{\frac{2}{\pi}}e^{i\frac{\pi}{6}}\beta^{-\frac{1}{4}}\frac{1}{2i}(-1)^pp^{\frac{1}{6}}\left(\pi^2c_1\right)^{-\frac{1}{2}}2^{-\frac{1}{2}}\nonumber\\
& e^{i\gamma\frac{1}{2}\left(\frac{3\pi}{2}\right)^{\frac{1}{3}}p^\frac{1}{3}(V_0-d\varepsilon)+\gamma\frac{\sqrt{3}}{2}\left(\frac{3\pi}{2}\right)^{\frac{1}{3}}p^\frac{1}{3}(V_0-2d\varepsilon)}\label{Ceq21.32}\\
-A'_1&=\frac{1}{2}\pi^{-\frac{1}{2}}\mu^{\frac{1}{4}}p^{\frac{1}{2}}(-1)^p\left(\pi^2c_1\right)^{-\frac{1}{2}}\nonumber\\
&e^{\gamma\left(\frac{3\pi}{2}\right)^{\frac{1}{3}}\frac{\sqrt{3}}{2} p^\frac{1}{3}(-2d\varepsilon+V_0)+\gamma\left(\frac{3\pi}{2}\right)^{\frac{1}{3}}\frac{1}{2}i p^\frac{1}{3}(-d\varepsilon+V_0)}\label{Ceq21.33}\\
-B'_1&=-\sqrt{\frac{2}{\pi}}e^{-i\frac{\pi}{6}}\beta^{\frac{1}{4}}\frac{1}{2}(-1)^pp^{\frac{1}{2}}\left(\pi^2c_1\right)^{-\frac{1}{2}}2^{-\frac{1}{2}}\nonumber\\
& e^{i\gamma\frac{1}{2}\left(\frac{3\pi}{2}\right)^{\frac{1}{3}}p^\frac{1}{3}(V_0-d\varepsilon)+\gamma\frac{\sqrt{3}}{2}\left(\frac{3\pi}{2}\right)^{\frac{1}{3}}p^\frac{1}{3}(V_0-2d\varepsilon)}\label{Ceq21.34}\\
A_2&=\frac{1}{2}\pi^{-\frac{1}{2}}\mu^{-\frac{1}{4}}p^{\frac{1}{6}}(-1)^p\left(\pi^2c_1\right)^{-\frac{1}{2}}\nonumber\\
&e^{\gamma\left(\frac{3\pi}{2}\right)^{\frac{1}{3}}\frac{\sqrt{3}}{2} p^\frac{1}{3}V_0+\gamma\left(\frac{3\pi}{2}\right)^{\frac{1}{3}}\frac{1}{2}i p^\frac{1}{3}(d\varepsilon+V_0)}\label{Ceq21.35}\\
B_2&=-\sqrt{\frac{2}{\pi}}e^{i\frac{\pi}{6}}\beta^{-\frac{1}{4}}p^{\frac{1}{6}}\frac{1}{2i}(-1)^p\left(\pi^2c_1\right)^{-\frac{1}{2}}2^{-\frac{1}{2}}\nonumber\\
&e^{i\gamma\frac{1}{2}\left(\frac{3\pi}{2}\right)^{\frac{1}{3}}p^\frac{1}{3}(V_0+d\varepsilon)+\gamma\frac{\sqrt{3}}{2}\left(\frac{3\pi}{2}\right)^{\frac{1}{3}}p^\frac{1}{3}V_0}\label{Ceq21.36}\\
A'_2&=-\frac{1}{2}\pi^{-\frac{1}{2}}\mu^{\frac{1}{4}}p^{\frac{1}{2}}(-1)^p\left(\pi^2c_1\right)^{-\frac{1}{2}}e^{\gamma\left(\frac{3\pi}{2}\right)^{\frac{1}{3}}\frac{\sqrt{3}}{2} p^\frac{1}{3}V_0+\gamma\left(\frac{3\pi}{2}\right)^{\frac{1}{3}}\frac{1}{2}i p^\frac{1}{3}(d\varepsilon+V_0)}\label{Ceq21.37}\\
B'_2&=\sqrt{\frac{2}{\pi}}e^{-i\frac{\pi}{6}}\beta^{\frac{1}{4}}\frac{1}{2}(-1)^pp^{\frac{1}{2}}\left(\pi^2c_1\right)^{-\frac{1}{2}}2^{-\frac{1}{2}}e^{i\gamma\frac{1}{2}\left(\frac{3\pi}{2}\right)^{\frac{1}{3}}p^\frac{1}{3}(V_0+d\varepsilon)+\gamma\frac{\sqrt{3}}{2}\left(\frac{3\pi}{2}\right)^{\frac{1}{3}}p^\frac{1}{3}V_0}\label{Ceq21.38}
\end{align}}
We can now look for the common factors and we find that\\
$\frac{1}{2}\pi^{-\frac{1}{2}}(-1)^p\left(\pi^2c_1\right)^{-\frac{1}{2}}e^{\gamma\frac{\sqrt{3}}{2}\left(\frac{3\pi}{2}\right)^{\frac{1}{3}}p^\frac{1}{3}V_0}e^{i\gamma\frac{1}{2}\left(\frac{3\pi}{2}\right)^{\frac{1}{3}}p^\frac{1}{3}V_0}$ is one, so the expressions become
\begin{align}
-A_1&=-\mu^{-\frac{1}{4}}p^{\frac{1}{6}}e^{-\gamma\left(\frac{3\pi}{2}\right)^{\frac{1}{3}}\frac{\sqrt{3}}{2} p^\frac{1}{3}2d\varepsilon-\gamma\left(\frac{3\pi}{2}\right)^{\frac{1}{3}}\frac{1}{2}i p^\frac{1}{3}d\varepsilon}\approx 0\label{Ceq21.39}\\
-B_1&=e^{i\frac{\pi}{6}}\beta^{-\frac{1}{4}}\frac{1}{i}p^{\frac{1}{6}} e^{-i\gamma\frac{1}{2}\left(\frac{3\pi}{2}\right)^{\frac{1}{3}}p^\frac{1}{3}d\varepsilon-\gamma\frac{\sqrt{3}}{2}\left(\frac{3\pi}{2}\right)^{\frac{1}{3}}p^\frac{1}{3}2d\varepsilon}\approx 0\label{Ceq21.40}\\
-A'_1&=\mu^{\frac{1}{4}}p^{\frac{1}{2}}e^{-\gamma\left(\frac{3\pi}{2}\right)^{\frac{1}{3}}\frac{\sqrt{3}}{2} p^\frac{1}{3}2d\varepsilon-\gamma\left(\frac{3\pi}{2}\right)^{\frac{1}{3}}\frac{1}{2}i p^\frac{1}{3}d\varepsilon}\approx 0\label{Ceq21.41}\\
-B'_1&=-e^{-i\frac{\pi}{6}}\beta^{\frac{1}{4}}p^{\frac{1}{2}} e^{-i\gamma\frac{1}{2}\left(\frac{3\pi}{2}\right)^{\frac{1}{3}}p^\frac{1}{3}d\varepsilon-\gamma\frac{\sqrt{3}}{2}\left(\frac{3\pi}{2}\right)^{\frac{1}{3}}p^\frac{1}{3}2d\varepsilon}\approx 0\label{Ceq21.42}\\
A_2&=\mu^{-\frac{1}{4}}p^{\frac{1}{6}}e^{\gamma\left(\frac{3\pi}{2}\right)^{\frac{1}{3}}\frac{1}{2}i p^\frac{1}{3}d\varepsilon}\label{Ceq21.43}\\
B_2&=-e^{i\frac{\pi}{6}}\beta^{-\frac{1}{4}}p^{\frac{1}{6}}\frac{1}{i}e^{i\gamma\frac{1}{2}\left(\frac{3\pi}{2}\right)^{\frac{1}{3}}p^\frac{1}{3}d\varepsilon}\label{Ceq21.44}\\
A'_2&=-\mu^{\frac{1}{4}}p^{\frac{1}{2}}e^{\gamma\left(\frac{3\pi}{2}\right)^{\frac{1}{3}}\frac{1}{2}i p^\frac{1}{3}d\varepsilon}\label{Ceq21.45}\\
B'_2&=e^{-i\frac{\pi}{6}}\beta^{\frac{1}{4}}p^{\frac{1}{2}}e^{i\gamma\frac{1}{2}\left(\frac{3\pi}{2}\right)^{\frac{1}{3}}p^\frac{1}{3}d\varepsilon}\label{Ceq21.46}
\end{align}
After removing the common factors, we see that some of the terms decay and the rest has also a common factor $e^{i\gamma\frac{1}{2}\left(\frac{3\pi}{2}\right)^{\frac{1}{3}}p^\frac{1}{3}d\varepsilon}$. Cancelling this one, we get
\begin{align}
A_2&=\mu^{-\frac{1}{4}}p^{\frac{1}{6}}\label{Ceq21.47}\\
B_2&=-e^{i\frac{\pi}{6}}\beta^{-\frac{1}{4}}p^{\frac{1}{6}}\frac{1}{i}\label{Ceq21.48}\\
A'_2&=-\mu^{\frac{1}{4}}p^{\frac{1}{2}}\label{Ceq21.49}\\
B'_2&=e^{-i\frac{\pi}{6}}\beta^{\frac{1}{4}}p^{\frac{1}{2}}\label{Ceq21.50}
\end{align}
Having only these in the matrix gives us nullspace consisting of two vectors
\begin{align}
\mathbf{n}_1=
\begin{pmatrix}
1\\
0\\
0\\
0
\end{pmatrix},\quad 
\mathbf{n}_2=
\begin{pmatrix}
0\\
0\\
0\\
1
\end{pmatrix}\label{Ceq21.51}
\end{align}

The other case when $V_0<2d\varepsilon$ gives us that $A_1,A'_1$ decay and $-B_1$ will have different sign. One can remove the common factors and end up with the same non-zero terms as before and therefore the same nullspace. Just as in the A-series case, we see, that the coefficients $a_1,\ldots,a_4$ do not depend on the eigenvalue index $p$. Even if they did, it must have been exponential growth to beat the exponential decay of the coefficients $c_p$.

\bibliographystyle{unsrt}
\bibliography{RessonantModeExpansion}

\end{document}